\documentclass[conference]{IEEEtran}
\IEEEoverridecommandlockouts

\usepackage{cite}
\usepackage{amsmath,amssymb,amsfonts}
\usepackage{algorithmic}
\usepackage{graphicx}
\usepackage{textcomp}
\usepackage{xcolor}

\usepackage{algorithm}
\usepackage{algorithmic}
\usepackage{subfig}
\usepackage{amsthm}
\usepackage{amsmath}
\usepackage{array}
\usepackage{multirow}
\usepackage{booktabs}
\usepackage{dsfont}
\usepackage{comment}
\usepackage{endnotes}
\usepackage{makecell}
\usepackage{listings}
\usepackage{xcolor}
\usepackage{hyperref}
\usepackage{bibentry}
\usepackage{amsmath}
\usepackage{dsfont}
\usepackage{amsfonts}

\usepackage{amssymb}
\usepackage{bbm}
\usepackage{bbding}
\newtheorem{theorem}{Theorem}
\newtheorem{lemma}{Lemma}

\def\BibTeX{{\rm B\kern-.05em{\sc i\kern-.025em b}\kern-.08em
    T\kern-.1667em\lower.7ex\hbox{E}\kern-.125emX}}
\begin{document}

\title{DaRec: A Disentangled Alignment Framework for Large Language Model and Recommender System\\
\thanks{$^{\dagger}$: Corresponding author.  \textsuperscript{\rm 1}National University of Defense Technology, Changsha, China; \textsuperscript{\rm 2}Baidu Inc, Beijing, China; \textsuperscript{\rm 3}University of Science and Technology of China, Hefei, China; This work was done when Xihong Yang (xihong\_edu@163.com) was a research intern at Baidu Inc.}
}

\author{Xihong Yang\textsuperscript{\rm 1,2}, Heming Jing\textsuperscript{\rm 2}, Zixing Zhang\textsuperscript{\rm 2}, Jindong Wang\textsuperscript{\rm 2}, Huakang Niu\textsuperscript{\rm 2}, Shuaiqiang Wang\textsuperscript{\rm 2}, Yu Lu\textbf{\textsuperscript{\rm 2}}, \\  Junfeng Wang\textsuperscript{\rm 2}, Dawei Yin\textsuperscript{\rm 2}, Xinwang Liu\textsuperscript{\rm 1}, En Zhu\textsuperscript{\rm 1}, Defu Lian\textsuperscript{\rm 3}, Erxue Min\textsuperscript{\rm 2}$^{\dagger}$\\

%

}

\maketitle
\begin{abstract}
Benefiting from the strong reasoning capabilities, Large language models (LLMs) have demonstrated remarkable performance in recommender systems. 
Various efforts have been made to distill knowledge from LLMs to enhance collaborative models, employing techniques like contrastive learning for representation alignment. In this work, we prove that directly aligning the representations of LLMs and collaborative models is sub-optimal for enhancing downstream recommendation tasks performance, based on the information theorem. Consequently, the challenge of effectively aligning semantic representations between collaborative models and LLMs remains unresolved. Inspired by this viewpoint, we propose a novel plug-and-play alignment framework for LLMs and collaborative models. Specifically, we first disentangle the latent representations of both LLMs and collaborative models into specific and shared components via projection layers and representation regularization. Subsequently, we perform both global and local structure alignment on the shared representations to facilitate knowledge transfer. Additionally, we theoretically prove that the specific and shared representations contain more pertinent and less irrelevant information, which can enhance the effectiveness of downstream recommendation tasks. Extensive experimental results on benchmark datasets demonstrate that our method is superior to existing state-of-the-art algorithms.
\end{abstract}


\section{Introduction}
Recommender systems have become a hot spot recently, which play a crucial role in various applications, such as video streaming, social media, and e-commerce. Owing to the strong representation learning ability, deep neural network-based recommendation algorithms \cite{ICDE_rec,Scenario_min, Neighbour_min, Reneasyrec,Rensslrec,yindataset,yin2023apgl4sr} have demonstrated impressive capabilities. More recently, large language models (LLMs) have exhibited strong reasonable proficiency in many tasks, e.g., vision task \cite{visionllm1, liuyue_AdvLoRA}, natural language processing \cite{liuyue_FlipAttack, haonanLLM}, and graph \cite{LLMgraph}. Several works explore the application of LLMs in recommendation tasks, including semantic representation alignment \cite{RLMRec, controlrec, ctrl, RA-Rec, WWW_align_rec,luo2024kellmrec}, representation augmentation \cite{LLMRec, KAR, luo2024integrating}, ranking function \cite{recranker, tallrec, zhu2023collaborative}, etc.

\begin{figure}
\centering
\scalebox{0.25}{
\includegraphics{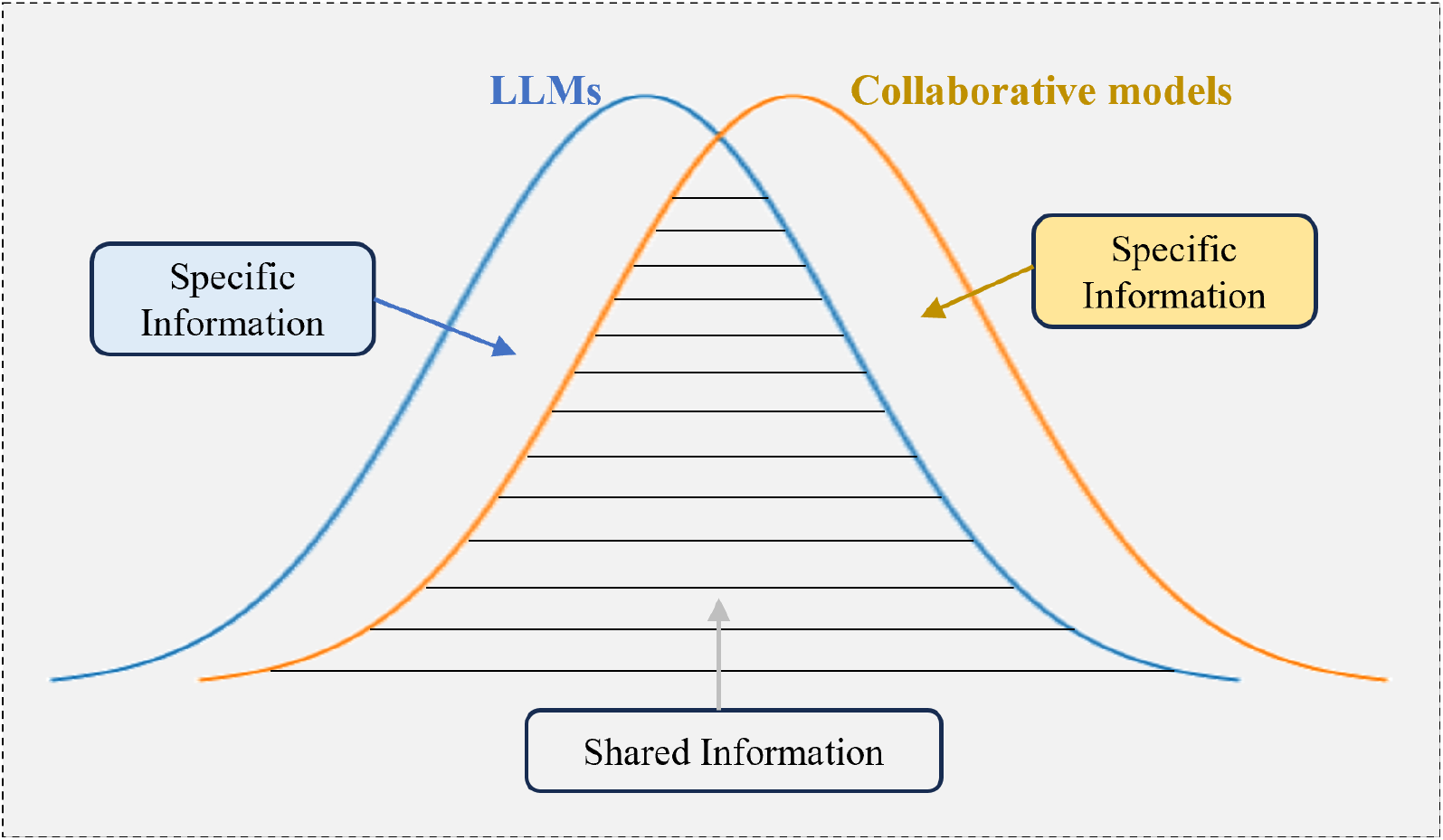}}
\caption{Illustration of the information gap between LLMs and collaborative models. The noisy signals within the specific information of each aspect impede the alignment of shared information, leading to a decline in the quality of representation.}
\label{motivation}
\end{figure}

Although various methods have explored the possibility of applying LLMs in recommender systems, most of them are hindered by two significant limitations: Firstly, LLMs have a huge number of parameters, it is quite arduous for LLMs to meet the low latency requirements for recommender systems. Secondly, LLMs always perform prediction with semantics ignoring the collaborative signal. Therefore, recent studies have explored semantic alignment methods \cite{RLMRec, controlrec, ctrl, RA-Rec} to transfer the semantic knowledge from LLMs to collaborative models by aligning their latent representations, aiming to improve the recommendation performance of existing collaborative models. 
However, due to the diverse nature of the interaction data employed in collaborative models compared to the nature language used for training LLMs, there exists a significant semantic gap between LLMs and recommendation tasks. Consequently, effectively aligning these two modalities poses a critical question. Some semantic alignment methods align the representations of collaborative models and LLMs via contrastive learning\cite{RLMRec, ctrl,controlrec}. Intuitively, alignment strategies like contrastive learning could reduce the gap by pulling the positive samples close. However, directly aligning the representation in latent space may be suboptimal due to the neglect of potential specific information inherent to each modality, as illustrated in Fig.\ref{motivation}. Inspired by this observation, we first theoretically investigate the representation gap in Theorem~\ref{motivation_the}, proving that when the gap is zero, which means exactly aligning two representations from collaborative models and LLMs, the downstream recommendation tasks have to pay a price for the performance. Simply mapping representations with a zero gap into the same latent space would introduce irrelevant noise from the specific representation, leading to a decline in recommendation tasks performance.

\begin{figure*}
\centering
\scalebox{0.6}{
\includegraphics{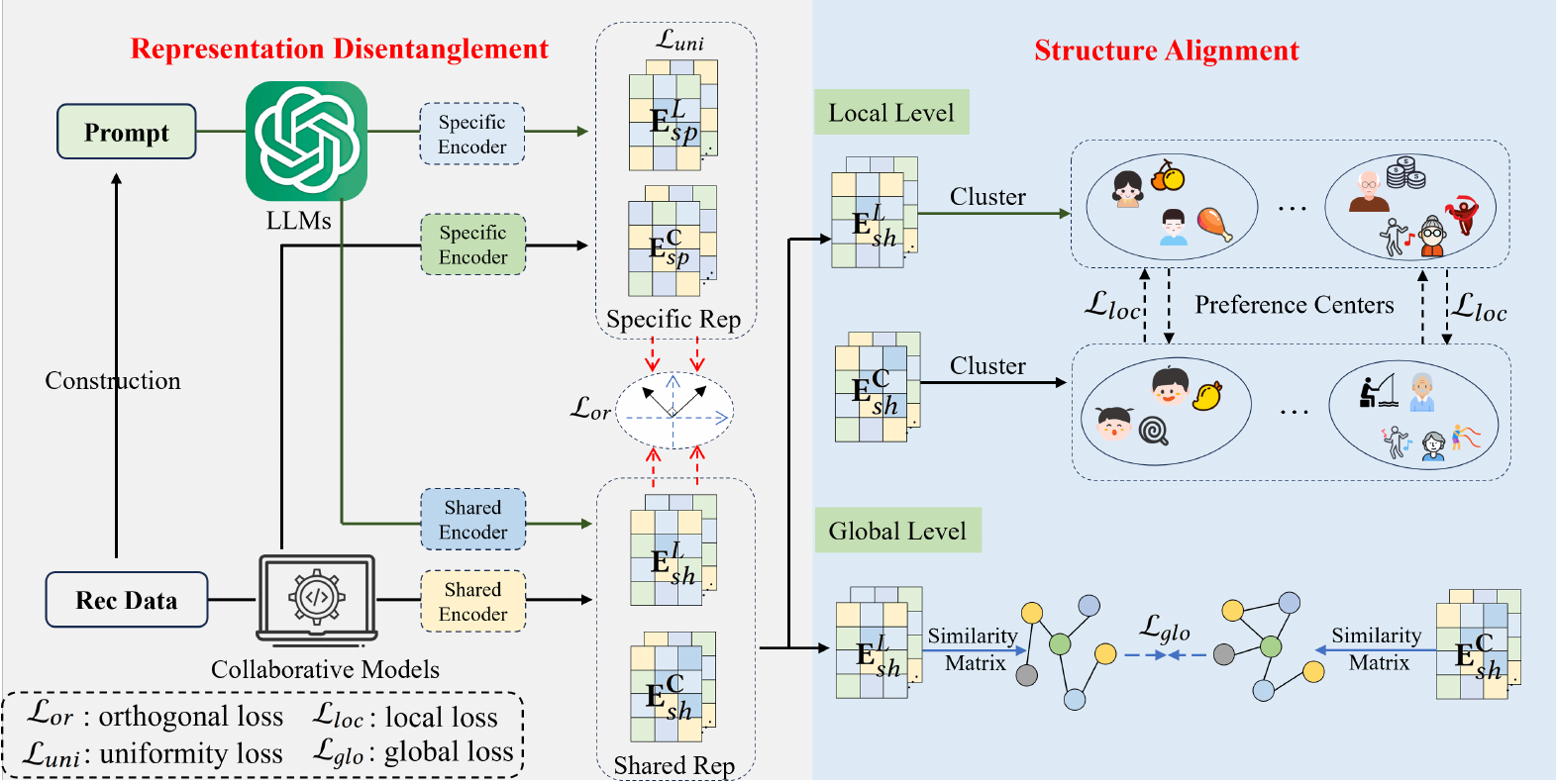}}
\caption{Illustration of our proposed disentangled alignment strategy. In our method, we first disentangle the representation into shared and specific components with two exclusive encoders and introduce orthogonal and uniformity loss to guarantee informative representations. Then, based on the shared representation, we devise a structure alignment strategy at both global and local levels to enhance the transfer of semantic knowledge from LLMs to collaborative models.}
\label{overall}
\vspace{-5pt}
\end{figure*}

Motivated by our theoretical findings,  we align the semantic knowledge of LLMs and collaborative models by disentangling the representations instead of exactly aligning all representations. We propose a novel plug-and-play representation \underline{\textbf{D}}isentangled \underline{\textbf{a}}lignment framework for \underline{\textbf{Rec}}ommendation model and LLMs, termed \textbf{DaRec}. To be specific, we first disentangle the representations into shared and specific components, reducing the negative impact of the specific information. Subsequently, the uniformity and orthogonal loss are designed to keep the informativity of representations.  Finally, we design a structure alignment strategy at both local and global levels to effectively transfer the semantic knowledge. Our method is shown to yield shared and specific representations that contain more relevant and less irrelevant information for the recommendation tasks, as supported by our theoretical analysis.

In summary, the main contributions of this work can be summarized as:
\begin{itemize}
    \item We provide a theoretical analysis to understand the impact of alignment strategy on recommendation performance. We prove that reducing the gap to zero between collaborative models and LLMs may not always benefit the performance when the gap between two models is large. To the best of our knowledge, this paper is the first work to demonstrate this phenomenon in mutual information perspective. 
    
    \item Motivated by our theorem, we disentangle the representations into two components, i.e., shared and specific representations, regularized by orthogonality and uniformity. Moreover, we design a global and local structure alignment strategy to better transfer the semantic knowledge from LLMs to collaborative models.
    
    \item We theoretically prove that the shared and specific representations by our method contain more relevant information and less irrelevant information to the recommendation tasks. Extensive experiments on the benchmark datasets have demonstrated the effectiveness and superiority of our designed algorithms with several state-of-the-art recommendation methods. 
\end{itemize}

\section{Preliminary}
This work proposes strategies to align the semantic representations of collaborative models and LLMs. Let $f_{\textbf{C}}(\cdot)$ and $f_{\textbf{L}}(\cdot)$ denote collaborative models and LLMs to obtain the corresponding representation in the latent space, respectively. Besides, $\textbf{D}$ and $\textbf{D'}$ are two types of input for collaborative models and LLMs, i.e., review data and prompt. We use $Y$ to indicate the target variable in the recommendation tasks. $h$ denotes the prediction function. The representation in LLMs and collaborative models can be denoted as $\textbf{E}^\textbf{L}$ and $\textbf{E}^\textbf{C}$, respectively.  Moreover, we define the mutual information between two representations as $I(\textbf{E}^{\textbf{C}};\textbf{E}^{\textbf{L}})$, and use $H(Y|\textbf{E}^{\textbf{C}},\textbf{E}^{\textbf{L}})$ to indicate the conditional entropy with two representations. $\ell_{CE}(\cdot)$ is the cross-entropy loss. The basic notations are summarized in Table \ref{notation_sum}.


\begin{table}[]
\centering
\caption{Notation Summary.}
\begin{tabular}{c|c}
\hline
\textbf{Notation} & \textbf{Meaning }                                   \\ \hline
$\textbf{D}$                 & The input for collaborative models         \\
$\textbf{D'}$              & The input for LLMs                         \\
$\textbf{E}^{\textbf{L}}$                 & The representations of LLMs                \\
$\textbf{E}^{\textbf{C}}$                 & The representations of collaborative models \\
$Y$                                & The target variable in the recommendation tasks \\
$I(\textbf{E}^{\textbf{C}};\textbf{E}^{\textbf{L}})$              & The mutual information between two representations                     \\
$H(Y|\textbf{E}^{\textbf{C}},\textbf{E}^{\textbf{L}})$                & The conditional entropy                    \\
$N_U$                & The number of users                        \\
$N_I$                 & The number of items                        \\
$\textbf{S}(\cdot,\cdot)$                & The cosine similarity                      \\
$\textbf{R}$                & The recommendation task                    \\
$\textbf{C}$                 & The preference centers                     \\
$\ell_{CE}(\cdot)$           & The cross-entropy loss                     \\
\hline
\end{tabular}
\label{notation_sum}
\vspace{-10pt}
\end{table}

\section{Methodology}
In this section, we propose a disentangled alignment strategy for collaborative models and LLMs. The overall framework of our method is shown in Fig.\ref{overall}. We first conduct a theoretical analysis of how representation alignment affects downstream tasks, which serves as the rationale behind our approach. Inspired by this analysis, we design two regularization techniques to disentangle the representations in LLMs and collaborative models into two components, i.e., shared and specific representations. Subsequently, in order to facilitate knowledge transfer between LLM and collaborative models without resorting to potentially detrimental perfect alignment, we introduce a structure alignment strategy operating at both local and global scales.
Finally, we define the loss function in our method. We introduce the details in the following sections.

\subsection{Motivation}
Although various alignment strategies between LLM and CM have been explored by several works \cite{RLMRec, ctrl,controlrec}, it is still an open question whether exactly aligning the semantic representations in the latent space is optimal for downstream recommendation tasks. An intuitive idea is to align the semantic representation of collaborative models and LLMs with a small gap. However, it is unclear how the alignment affects the downstream recommendation tasks. To address this problem, we present an illustration in Fig.\ref{motivation}. Due to differences in data organization, training methods, and semantic features, there is a natural gap between the features of LLMs and collaborative models. Inspired by this idea, we conjecture that directly reducing the gap in the latent space does not always lead to better downstream recommendation tasks performance. Nevertheless, it is instructive to theoretically understand how to reduce the gap could be helpful. To this end, we first give a definition of the information gap: $\Delta p = |I(\textbf{D}; Y)-I(\textbf{D'}; Y)|$ to characterize the gap of the two types of model input towards the target label $Y$. It is independent of the encoder network $f_{\textbf{C}(\cdot)}$ and $f_{\textbf{L}(\cdot)}$. Therefore, $\Delta p$ is a constant during the training procedure. In the following, we will provide a theorem. It demonstrates that the information gap will serve as a lower bound of the recommendation tasks error if we attempt to find the representations, which admit a zero gap. Therefore, the information gap is the price for exactly aligning different representations extracted by collaborative models and LLMs. This theorem is presented as follows.

\begin{theorem}\label{motivation_the}
For collaborative models encoder network $f_{\textbf{C}(\cdot)}$ and LLMs encoder network $f_{\textbf{L}(\cdot)}$, if the representations $\textbf{E}^{\textbf{C}}= f_{\textbf{C}}(\textbf{D})$ and $\textbf{E}^{\textbf{L}}= f_{\textbf{L}}(\textbf{D'})$ are exactly aligned in the latent space, i.e., $\textbf{E}^{\textbf{C}} = \textbf{E}^{\textbf{L}}$, we have:
\begin{equation}
{\inf}_{h} \mathbb{E}_{p} [\mathcal{L}_{ce}(h(\textbf{E}^{\textbf{C}}, \textbf{E}^{\textbf{L}}),Y)] - {\inf}_{h'} \mathbb{E}_{p} [\mathcal{L}_{ce}(h'(\textbf{E}^{\textbf{C}}, \textbf{E}^{\textbf{L}}), Y)] \geq \Delta_{p}.\nonumber
\end{equation}
\end{theorem}

Theorem \ref{motivation_the} indicates that the optimal recommendation error with the exactly aligned representations is at least $\Delta p$ larger than we can obtain from the input data if the information gap between collaborative models and LLMs is large. Furthermore, since LLMs and collaborative models have different semantic scenarios and training procedures, there is specific information for each model. Performing exact alignment with all representations will introduce the specific information of collaborative models and LLMs. This specific information may be mutual interference, leading to the downstream recommendation tasks performance decreasing. Therefore, in this paper, we first disentangle the initial representations in both the collaborative model and LLM into specific representation and shared representation. Then, we design a structure alignment strategy at both local and global levels to perform a more slack alignment. We provide the proof in section.\ref{Proof_1}.


\subsection{Representation Disentanglement}\label{split}
Previous alignment strategy for collaborative models and LLMs aims to align the representation directly, e.g., contrastive learning. However, this practice may be suboptimal because collaborative models and LLMs contain different input data types, training manners, and semantic scenarios, thus the direct alignment strategy would introduce the specific information, leading to the unpromising performance of downstream recommendation tasks. Inspired by this intuition, we design a representation disentanglement method to separate the representation into the specific and shared components for collaborative models and LLMs respectively. 


Based on the representation of collaborative models and LLMs, we disentangle the representations into two components, i.e., specific representation and shared representation:

\begin{equation}
\begin{aligned}
\textbf{E}_{sp}^{\textbf{C}} &= f_{sp}^{\textbf{C}}(\textbf{E}^{\textbf{C}}),\textbf{E}_{sh}^{\textbf{C}} = f_{sh}^{\textbf{C}}(\textbf{E}^{\textbf{C}}),\\
\textbf{E}_{sp}^{L} &= f_{sp}^{\textbf{L}}(\textbf{E}^{L}),\textbf{E}_{sh}^{L} = f_{sh}^{\textbf{L}}(\textbf{E}^{L}),
\end{aligned}
\label{speci_shared_encoder}
\end{equation}
where $f_{sh}(\cdot)$ and $f_{sp}(\cdot)$ denote encoder network for the specific representation $\textbf{E}_{sp}$ and shared representation $\textbf{E}_{sh}$, respectively. Here, we adopt MLP as the backbone network for $f_{sh}(\cdot)$ and $f_{sp}(\cdot)$. 

To ensure the specific and the shared representation achieve unique and complementary information, we aim to perform orthogonal constraints on specific and shared representation by minimizing the following equation:


\begin{equation}
\begin{aligned}
\mathcal{L}_{or} = \frac{1}{N}\sum_{i=1}^N(\textbf{S}(\textbf{E}_{{sp}_i}^{L}, \textbf{E}_{{sh}_i}^{L}))^2 + \frac{1}{N}\sum_{i=1}^N(\textbf{S}(\textbf{E}_{{sp}_i}^{C}, \textbf{E}_{{sh}_i}^{C}))^2,
\end{aligned}
\label{loss_or}
\end{equation}
where $\textbf{S}(\cdot,\cdot)$ is the cosine similarity, $N$ is the number of the user and item, i.e., $N = N_U + N_I$.

To avoid the specific representation being non-information noise for the model, we design a strategy to constrain the specific representation for both collaborative models and LLMs. Here, we adopt the uniformity loss \cite{towards} to the specific representation, which maximizes the pairwise Gaussian potential \cite{G_1, G_2}. The uniformity loss can be calculated as:
\begin{equation}
\begin{aligned}
\mathcal{L}_{uni} &= &\text{log} \underset{x,y\sim\textbf{E}_{sp}^C}{\mathbb{E}}{e^{-2||G(x)-G(y)||^2}} \\
&+&\text{log} \underset{x,y\sim\textbf{E}_{sp}^L}{\mathbb{E}}{e^{-2||G(x)-G(y)||^2}},
\end{aligned}
\label{loss_uni}
\end{equation}

\subsection{Structure Alignment}

Inspired by the alignment methods \cite{align1, align2, align3} in other fields, in this paper, we attempt to design the alignment strategy from the structure perspective. The meaningful latent representation structure could preserve potential properties. Therefore, in this subsection, based on Section \ref{split}, we utilize the shared representation for the structure alignment. Specifically, we introduce the method at both global and local levels. Detailed description is as follows.


\subsubsection{Global Structure Alignment.}
Based on the shared representation from collaborative models and LLMs, we design a structure alignment strategy at the global level. To be specific, we first calculate the similarity matrix about the shared representations, which can be expressed as:
\begin{equation}
\begin{aligned}
\textbf{S}_{C}^{G} &= \textbf{E}_{sh}^{C} (\textbf{E}_{sh}^{C})^\top,\\
\textbf{S}_{L}^{G} &= \textbf{E}_{sh}^{L} (\textbf{E}_{sh}^{L})^\top,
\end{aligned}
\label{global_sim}
\end{equation}
where we use matrix multiplication to calculate Eq.\eqref{global_sim}. The shared representation is the concatenation of the user and item representation, which can be considered as the pair-wise instance for the user preference. Through Eq.\eqref{global_sim}, we could obtain the structure of the shared representation with all pair instances at the global level. 

After that, we can align the structure of collaborative models and LLMs' shared representation as follows:
\begin{equation}
\mathcal{L}_{glo} = ||\textbf{S}_{C}^G - \textbf{S}_{L}^G||_F^2.
\label{loss_global}
\end{equation}

\begin{algorithm}[t]
\small
\caption{\textbf{Disentangled Alignment Strategy for collaborative models and LLMs.}}
\label{ALGORITHM}
    \flushleft{\textbf{Input}: LLM representation $\textbf{E}^{{L}}$, Rec representation $\textbf{E}^{{C}}$\\}
\flushleft{\textbf{Output}: The Recommendation Results $\textbf{Res}$.} 
\vspace{-10pt}
\begin{algorithmic}[1]
\FOR{$e=1$ to $step\_size$}
\STATE Disentangle representation into shared and specific with Eq.\eqref{speci_shared_encoder}.
\STATE Calculate orthogonal, uniformity constrains with Eq.\eqref{loss_or} and Eq.\eqref{loss_uni}.
\STATE Obtain global and local structure of shared representation with Eq.\eqref{global_sim} and Eq.\eqref{cluster}.
\STATE Sort the preference centers with Eq.\eqref{sort}.
\STATE Calculate global and local structure alignment with Eq.\eqref{loss_global} and Eq.\eqref{loss_local}.
\STATE Update entire network by minimizing $\mathcal{L}$ in Eq. \eqref{loss_total}.
\ENDFOR
\STATE Inference with recommendation network to obtain the results $\textbf{Res}$. 
\end{algorithmic}
\label{algo}
\end{algorithm}

\subsubsection{Local Structure Alignment.}
 
To comprehensively align the representation structure of the collaborative models and LLMs, we explore the local structure in this subsection. Different from the global structure alignment from the pairwise relationship for all shared representations, the local structure is conducted from a coarse-grained perspective. To be specific, we attempt to use the preference to demonstrate the alignment. Therefore, we first obtain the user's preference in collaborative models and LLMs with shared representation. In this work, we conduct clustering operations in the shared representation as:
\begin{equation}
\begin{aligned}
\textbf{C}_{C} &= f_{C}(\textbf{E}_{sh}^C),\\
\textbf{C}_{L} &= f_{C}(\textbf{E}_{sh}^L),\\
\end{aligned}
\label{cluster}
\end{equation}
where $f_{C}(\cdot)$ is the clustering function, e.g., K-Means \cite{K-means}. $\textbf{C}_C \in \mathbb{R}^{K \times d}$ and $\textbf{C}_L  \in \mathbb{R}^{K \times d}$ indicate the cluster center of collaborative models and LLMs shared representation, respectively. $K$ means the number of the preference centers.

Through Eq.\eqref{cluster}, we could obtain the user preference in both collaborative models and LLMs with different semantic scenarios. Compared with the global structure alignment, the clustering operation could shrink the scale
of the number of users and items. The preference of the user should remain consistent with the collaborative models and LLMs. However, it is a challenge how to align different preference centers rightly since there is no definite target information available. Therefore, we further design an adaptive preference-matching mechanism. The core idea of this mechanism is to seek the most similar preference center adaptively. Specifically, we calculate the Euclidean distance between $i$-th representation in the first preference cluster and $j$-th representation in the second preference cluster for all preference clusters in collaborative models and LLMs:
\begin{equation}
\begin{aligned}
    dis(\textbf{C}_{C}^i, \textbf{C}_{L}^j) = ||\textbf{C}_{C}^i -  \textbf{C}_{L}^j||_2,
\end{aligned}
\end{equation}
where $i, j = 1,2,\dots,K$. Then, we sort $dis$ with a ascending order and adjust $\textbf{C}_C$ and $\textbf{C}_L$, which can be presented as:
\begin{equation}
\begin{aligned}
\text{ind} &= \text{Sort}(dis(\textbf{C}_{C}^i, \textbf{C}_{L}^j)),\\
\textbf{C}_C &= \textbf{C}_C[\text{ind}], 
\textbf{C}_L = \textbf{C}_L[\text{ind}],
\end{aligned}
\label{sort}
\end{equation}
where $\text{Sort}$ is the sort function in ascending order. $\text{ind}$ indicates the index of the sorted preference cluster. Through this operation, the most similar pair-centers could be adjusted into the right position. Then, we mark the sorted centers and select unmarked vectors in $\textbf{C}$ to recalculate the corresponding $dis$ until all preference centers are sorted. In this way, the preference center in collaborative models and LLMs could be roughly corresponding. To perform our local alignment, we calculate the similarity matrix with cosine similarity between different preference centers in collaborative models and LLMs:
\begin{equation}
\begin{aligned}
\textbf{S}_{ij}^{\textbf{C}} = \frac{(\textbf{C}_{C}^i) \cdot (\textbf{C}_{L}^j)}{||\textbf{C}_{C}^i||_2 ||\textbf{C}_{L}^j||_2}.
\end{aligned}
\end{equation}

Then, we minimize the following function to align the different preference centers at the local level:
\begin{equation}
\begin{aligned}
\mathcal{L}_{loc} = \frac{1}{{K}} \sum_{i=1}^{K}(\textbf{S}^\textbf{C}_{ii}-1)^2 + \frac{1}{K^2-K}\sum_{i=1}^K \sum_{i\neq j}{(\textbf{S}^\textbf{C}_{ij})^2},
\end{aligned}
\label{loss_local}
\end{equation}
where $K$ is the number of cluster preference. Through minimizing Eq.\eqref{loss_local}, the same preference centers are forced to agree with each other, and different centers are encouraged to push away.

\subsection{Optimization and Complexity}
In this work, we propose a plug-and-play framework to better align the semantic representation of collaborative models and LLMs. The proposed method is jointly optimized by the following function:
\begin{equation}
\begin{aligned}
\mathcal{L} = \mathcal{L}_{base} + \lambda (\mathcal{L}_{or} + \mathcal{L}_{uni} + \mathcal{L}_{glo} + \mathcal{L}_{loc}),
\end{aligned}
\label{loss_total}
\end{equation}
where $\mathcal{L}_{base}$ is the loss function of the baseline, e.g., classification loss. $\lambda$ indicates the trade-off parameters for the loss function. The detailed learning process of DaRec is shown in Algorithm.\ref{algo}. Here, we analyze the time and space complexity of our proposed loss function in DaRec. We use $N$ and $d$ to denote the number of samples and the dimension of the representation, respectively. For the orthogonal operation in $\mathcal{L}_{or}$, the time complexity is $\mathcal{O}(Nd)$. Moreover, the time complexity of the similarity operation in $\mathcal{L}_{glo}$ is $\mathcal{O}(N^2d)$. Besides, the uniformity loss $\mathcal{L}_{uni}$ exhibits a time complexity of $\mathcal{O}(N^2d)$. Since the dimension of preference center $\textbf{C}$ is $\mathbb{R}^{K \times d}$, the time complexity of $\mathcal{L}_{loc}$ is $\mathcal{O}(K^2d)$. The overall time complexity of the proposed loss function can be approximated as $\mathcal{O}(N^2d+Nd+K^2d)$. Furthermore, the space complexity of the proposed loss function is $\mathcal{O}(N^2+N+K^2)$. In practice, we randomly sample $\hat{N}$ instances for approximation to reduce both computational and space complexity. In Section.\ref{complex_res}, we analyzed the impact of sampling size $\hat{N}$ on model performance. In conclusion, considering that $K << \hat{N}$, the time and space complexity of our proposed loss function are $\mathcal{O}(\hat{N}^2d+\hat{N}d)$ and $\mathcal{O}(\hat{N}^2+\hat{N})$, respectively.

\begin{table}[]
\centering
\caption{Dataset Summary.}
\scalebox{1.1}{
\begin{tabular}{c|cccc}
\hline
\textbf{Dataset}     & \textbf{Users} & \textbf{Items} & \textbf{Interactions} & \textbf{Density} \\ \hline
\textbf{Amazon-book} & 11,000         & 9,332          & 120,464               & 1.2e-3           \\
\textbf{Yelp}        & 11,091         & 11,010         & 166,620               & 1.4e-3           \\
\textbf{Steam}       & 23,310         & 5,237          & 316,190               & 2.6e-3           \\ \hline
\end{tabular}}
\label{data_info}
\end{table}

\begin{table*}[]
\centering
\caption{Recommendation Performance on three datasets with six metrics. The best results are denoted in bold. $\dagger$ denotes results are statistically significant where the p-value is less than {0.05}.}
\scalebox{0.65}{
\begin{tabular}{cc|cccccc|cccccc|cccccc}
\hline
\multicolumn{2}{c|}{{\color[HTML]{333333} Data}}                                                            & \multicolumn{6}{c|}{{\color[HTML]{333333} Amazon-book}}                                                                                                                                                                                             & \multicolumn{6}{c|}{{\color[HTML]{333333} Yelp}}                                                                                                                                                                                                    & \multicolumn{6}{c}{{\color[HTML]{333333} Steam}}                                                                                                                                                                                                    \\ \hline
\multicolumn{1}{c|}{{\color[HTML]{333333} Backbone}}                   & {\color[HTML]{333333} Variants}    & {\color[HTML]{333333} R@5}             & {\color[HTML]{333333} R@10}            & {\color[HTML]{333333} R@20}            & {\color[HTML]{333333} N@5}             & {\color[HTML]{333333} N@10}            & {\color[HTML]{333333} N@20}            & {\color[HTML]{333333} R@5}             & {\color[HTML]{333333} R@10}            & {\color[HTML]{333333} R@20}            & {\color[HTML]{333333} N@5}             & {\color[HTML]{333333} N@10}            & {\color[HTML]{333333} N@20}            & {\color[HTML]{333333} R@5}             & {\color[HTML]{333333} R@10}            & {\color[HTML]{333333} R@20}            & {\color[HTML]{333333} N@5}             & {\color[HTML]{333333} N@10}            & {\color[HTML]{333333} N@20}            \\ \hline
\multicolumn{1}{c|}{{\color[HTML]{333333} }}                           & {\color[HTML]{333333} Baseline}    & {\color[HTML]{333333} 0.0537}          & {\color[HTML]{333333} 0.0872}          & {\color[HTML]{333333} 0.1343}          & {\color[HTML]{333333} 0.0537}          & {\color[HTML]{333333} 0.0653}          & {\color[HTML]{333333} 0.0807}          & {\color[HTML]{333333} 0.039}           & {\color[HTML]{333333} 0.0652}          & {\color[HTML]{333333} 0.01084}         & {\color[HTML]{333333} 0.0451}          & {\color[HTML]{333333} 0.0534}          & {\color[HTML]{333333} 0.068}           & {\color[HTML]{333333} 0.05}            & {\color[HTML]{333333} 0.0826}          & {\color[HTML]{333333} 0.1313}          & {\color[HTML]{333333} 0.0556}          & {\color[HTML]{333333} 0.0665}          & {\color[HTML]{333333} 0.083}           \\
\multicolumn{1}{c|}{{\color[HTML]{333333} }}                           & {\color[HTML]{333333} RLMRec-Con}  & {\color[HTML]{333333} 0.0561}          & {\color[HTML]{333333} 0.0899}          & {\color[HTML]{333333} 0.1395}          & {\color[HTML]{333333} 0.0562}          & {\color[HTML]{333333} 0.0679}          & {\color[HTML]{333333} 0.0842}          & {\color[HTML]{333333} 0.0409}          & {\color[HTML]{333333} 0.0685}          & {\color[HTML]{333333} 0.1144}          & {\color[HTML]{333333} 0.0474}          & {\color[HTML]{333333} 0.0562}          & {\color[HTML]{333333} 0.0719}          & {\color[HTML]{333333} 0.0538}          & {\color[HTML]{333333} 0.0883}          & {\color[HTML]{333333} 0.1398}          & {\color[HTML]{333333} 0.0597}          & {\color[HTML]{333333} 0.0713}          & {\color[HTML]{333333} 0.0888}          \\
\multicolumn{1}{c|}{{\color[HTML]{333333} }}                           & {\color[HTML]{333333} RLMRec-Gen}  & {\color[HTML]{333333} 0.0551}          & {\color[HTML]{333333} 0.0891}          & {\color[HTML]{333333} 0.1372}          & {\color[HTML]{333333} 0.0559}          & {\color[HTML]{333333} 0.0675}          & {\color[HTML]{333333} 0.0832}          & {\color[HTML]{333333} 0.0393}          & {\color[HTML]{333333} 0.0654}          & {\color[HTML]{333333} 0.1074}          & {\color[HTML]{333333} 0.0454}          & {\color[HTML]{333333} 0.0535}          & {\color[HTML]{333333} 0.0678}          & {\color[HTML]{333333} 0.0532}          & {\color[HTML]{333333} 0.0874}          & {\color[HTML]{333333} 0.1385}          & {\color[HTML]{333333} 0.0588}          & {\color[HTML]{333333} 0.0702}          & {\color[HTML]{333333} 0.0875}          \\
\multicolumn{1}{c|}{{\color[HTML]{333333} }}                           & {\color[HTML]{333333} Ours}        & {\color[HTML]{333333} $\textbf{0.0562}$$^{\dagger}$} & {\color[HTML]{333333} \textbf{0.0906}$^{\dagger}$} & {\color[HTML]{333333} \textbf{0.1413}$^{\dagger}$} & {\color[HTML]{333333} \textbf{0.0563}$^{\dagger}$} & {\color[HTML]{333333} \textbf{0.0684}$^{\dagger}$} & {\color[HTML]{333333} \textbf{0.085}$^{\dagger}$}  & {\color[HTML]{333333} \textbf{0.0422}$^{\dagger}$} & {\color[HTML]{333333} \textbf{0.0713}$^{\dagger}$} & {\color[HTML]{333333} \textbf{0.1205}$^{\dagger}$} & {\color[HTML]{333333} \textbf{0.048}$^{\dagger}$}  & {\color[HTML]{333333} \textbf{0.0574}$^{\dagger}$} & {\color[HTML]{333333} \textbf{0.0742}$^{\dagger}$} & {\color[HTML]{333333} \textbf{0.0547}$^{\dagger}$}           & {\color[HTML]{333333} \textbf{0.0900}$^{\dagger}$}            & {\color[HTML]{333333} \textbf{0.1415}$^{\dagger}$}          & {\color[HTML]{333333} \textbf{0.0603}$^{\dagger}$}          & {\color[HTML]{333333} \textbf{0.0721}$^{\dagger}$}          & {\color[HTML]{333333} \textbf{0.0896}$^{\dagger}$}          \\
\multicolumn{1}{c|}{\multirow{-5}{*}{{\color[HTML]{333333} GCCF}}}     & {\color[HTML]{333333} Improvement} & {\color[HTML]{333333} 0.18\%}          & {\color[HTML]{333333} 0.78\%}          & {\color[HTML]{333333} 1.29\%}          & {\color[HTML]{333333} 0.18\%}          & {\color[HTML]{333333} 0.74\%}          & {\color[HTML]{333333} 0.95\%}          & {\color[HTML]{333333} 3.18\%}          & {\color[HTML]{333333} 4.09\%}          & {\color[HTML]{333333} 5.33\%}          & {\color[HTML]{333333} 1.27\%}          & {\color[HTML]{333333} 2.14\%}          & {\color[HTML]{333333} 3.20\%}          & {\color[HTML]{333333} 1.67\%}              & {\color[HTML]{333333} 1.93\%}              & {\color[HTML]{333333} 1.22\%}              & {\color[HTML]{333333} 1.01\%}              & {\color[HTML]{333333} 1.12\%}              & {\color[HTML]{333333} 0.90\%}              \\ \hline
\multicolumn{1}{c|}{{\color[HTML]{333333} }}                           & {\color[HTML]{333333} Baseline}    & {\color[HTML]{333333} 0.057}           & {\color[HTML]{333333} 0.0915}          & {\color[HTML]{333333} 0.1411}          & {\color[HTML]{333333} 0.0574}          & {\color[HTML]{333333} 0.0694}          & {\color[HTML]{333333} 0.0856}          & {\color[HTML]{333333} 0.0421}          & {\color[HTML]{333333} 0.0706}          & {\color[HTML]{333333} 0.1157}          & {\color[HTML]{333333} 0.0491}          & {\color[HTML]{333333} 0.058}           & {\color[HTML]{333333} 0.0733}          & {\color[HTML]{333333} 0.0518}          & {\color[HTML]{333333} 0.0852}          & {\color[HTML]{333333} 0.1348}          & {\color[HTML]{333333} 0.0575}          & {\color[HTML]{333333} 0.0687}          & {\color[HTML]{333333} 0.0855}          \\
\multicolumn{1}{c|}{{\color[HTML]{333333} }}                           & {\color[HTML]{333333} RLMRec-Con}  & {\color[HTML]{333333} 0.0608}          & {\color[HTML]{333333} 0.0969}          & {\color[HTML]{333333} 0.1483}          & {\color[HTML]{333333} 0.0606}          & {\color[HTML]{333333} 0.0734}          & {\color[HTML]{333333} 0.0903}          & {\color[HTML]{333333} 0.0445}          & {\color[HTML]{333333} 0.0754}          & {\color[HTML]{333333} 0.123}           & {\color[HTML]{333333} 0.0518}          & {\color[HTML]{333333} 0.0614}          & {\color[HTML]{333333} 0.0776}          & {\color[HTML]{333333} 0.0548}          & {\color[HTML]{333333} 0.0895}          & {\color[HTML]{333333} 0.01421}         & {\color[HTML]{333333} 0.0608}          & {\color[HTML]{333333} 0.0724}          & {\color[HTML]{333333} 0.0902}          \\
\multicolumn{1}{c|}{{\color[HTML]{333333} }}                           & {\color[HTML]{333333} RLMRec-Gen}  & {\color[HTML]{333333} 0.0596}          & {\color[HTML]{333333} 0.0948}          & {\color[HTML]{333333} 0.1446}          & {\color[HTML]{333333} 0.0605}          & {\color[HTML]{333333} 0.0724}          & {\color[HTML]{333333} 0.0887}          & {\color[HTML]{333333} 0.0435}          & {\color[HTML]{333333} 0.0734}          & {\color[HTML]{333333} 0.1209}          & {\color[HTML]{333333} 0.0505}          & {\color[HTML]{333333} 0.06}            & {\color[HTML]{333333} 0.0761}          & {\color[HTML]{333333} 0.055}           & {\color[HTML]{333333} 0.0907}          & {\color[HTML]{333333} 0.1433}          & {\color[HTML]{333333} 0.0607}          & {\color[HTML]{333333} 0.0729}          & {\color[HTML]{333333} 0.0907}          \\
\multicolumn{1}{c|}{{\color[HTML]{333333} }}                           & {\color[HTML]{333333} Ours}        & {\color[HTML]{333333} \textbf{0.0628}$^{\dagger}$} & {\color[HTML]{333333} \textbf{0.0976}$^{\dagger}$} & {\color[HTML]{333333} \textbf{0.1495}$^{\dagger}$} & {\color[HTML]{333333} \textbf{0.0621}$^{\dagger}$} & {\color[HTML]{333333} \textbf{0.0742}$^{\dagger}$} & {\color[HTML]{333333} \textbf{0.091}$^{\dagger}$}  & {\color[HTML]{333333} \textbf{0.0461}$^{\dagger}$} & {\color[HTML]{333333} \textbf{0.0759}$^{\dagger}$} & {\color[HTML]{333333} \textbf{0.1246}$^{\dagger}$} & {\color[HTML]{333333} \textbf{0.0537}$^{\dagger}$} & {\color[HTML]{333333} \textbf{0.0625}$^{\dagger}$} & {\color[HTML]{333333} \textbf{0.0789}$^{\dagger}$} & {\color[HTML]{333333} \textbf{0.0558}$^{\dagger}$} & {\color[HTML]{333333} \textbf{0.0917}$^{\dagger}$} & {\color[HTML]{333333} \textbf{0.1456}$^{\dagger}$} & {\color[HTML]{333333} \textbf{0.0609}$^{\dagger}$} & {\color[HTML]{333333} \textbf{0.073}$^{\dagger}$}  & {\color[HTML]{333333} \textbf{0.0914}$^{\dagger}$} \\
\multicolumn{1}{c|}{\multirow{-5}{*}{{\color[HTML]{333333} LightGCN}}} & {\color[HTML]{333333} Improvement} & {\color[HTML]{333333} 3.29\%}          & {\color[HTML]{333333} 0.72\%}          & {\color[HTML]{333333} 0.81\%}          & {\color[HTML]{333333} 2.48\%}          & {\color[HTML]{333333} 1.09\%}          & {\color[HTML]{333333} 0.78\%}          & {\color[HTML]{333333} 3.60\%}          & {\color[HTML]{333333} 0.66\%}          & {\color[HTML]{333333} 1.30\%}          & {\color[HTML]{333333} 3.67\%}          & {\color[HTML]{333333} 1.79\%}          & {\color[HTML]{333333} 1.68\%}          & {\color[HTML]{333333} 1.45\%}          & {\color[HTML]{333333} 1.10\%}          & {\color[HTML]{333333} 1.61\%}          & {\color[HTML]{333333} 0.33\%}          & {\color[HTML]{333333} 0.14\%}          & {\color[HTML]{333333} 0.77\%}          \\ \hline
\multicolumn{1}{c|}{{\color[HTML]{333333} }}                           & {\color[HTML]{333333} Baseline}    & {\color[HTML]{333333} 0.0637}          & {\color[HTML]{333333} 0.0994}          & {\color[HTML]{333333} 0.1473}          & {\color[HTML]{333333} 0.0632}          & {\color[HTML]{333333} 0.0756}          & {\color[HTML]{333333} 0.0913}          & {\color[HTML]{333333} 0.0432}          & {\color[HTML]{333333} 0.0722}          & {\color[HTML]{333333} 0.1197}          & {\color[HTML]{333333} 0.0501}          & {\color[HTML]{333333} 0.0592}          & {\color[HTML]{333333} 0.0753}          & {\color[HTML]{333333} 0.0565}          & {\color[HTML]{333333} 0.0919}          & {\color[HTML]{333333} 0.1444}          & {\color[HTML]{333333} 0.0618}          & {\color[HTML]{333333} 0.0738}          & {\color[HTML]{333333} 0.0917}          \\
\multicolumn{1}{c|}{{\color[HTML]{333333} }}                           & {\color[HTML]{333333} RLMRec-Con}  & {\color[HTML]{333333} 0.0655}          & {\color[HTML]{333333} 0.1017}          & {\color[HTML]{333333} 0.1528}          & {\color[HTML]{333333} 0.0652}          & {\color[HTML]{333333} 0.0778}          & {\color[HTML]{333333} 0.0945}          & {\color[HTML]{333333} 0.0452}          & {\color[HTML]{333333} 0.0763}          & {\color[HTML]{333333} 0.1248}          & {\color[HTML]{333333} 0.053}           & {\color[HTML]{333333} 0.0626}          & {\color[HTML]{333333} 0.079}           & {\color[HTML]{333333} 0.0589}          & {\color[HTML]{333333} 0.0956}          & {\color[HTML]{333333} 0.1489}          & {\color[HTML]{333333} 0.0645}          & {\color[HTML]{333333} 0.0768}          & {\color[HTML]{333333} 0.095}           \\
\multicolumn{1}{c|}{{\color[HTML]{333333} }}                           & {\color[HTML]{333333} RLMRec-Gen}  & {\color[HTML]{333333} 0.0644}          & {\color[HTML]{333333} 0.1015}          & {\color[HTML]{333333} 0.1537}          & {\color[HTML]{333333} 0.0648}          & {\color[HTML]{333333} 0.0777}          & {\color[HTML]{333333} 0.0947}          & {\color[HTML]{333333} 0.0467}          & {\color[HTML]{333333} 0.0771}          & {\color[HTML]{333333} 0.1263}          & {\color[HTML]{333333} 0.0537}          & {\color[HTML]{333333} 0.0631}          & {\color[HTML]{333333} 0.0798}          & {\color[HTML]{333333} 0.0574}          & {\color[HTML]{333333} 0.094}           & {\color[HTML]{333333} 0.1476}          & {\color[HTML]{333333} 0.0629}          & {\color[HTML]{333333} 0.0752}          & {\color[HTML]{333333} 0.0934}          \\
\multicolumn{1}{c|}{{\color[HTML]{333333} }}                           & {\color[HTML]{333333} Ours}        & {\color[HTML]{333333} \textbf{0.0667}$^{\dagger}$} & {\color[HTML]{333333} \textbf{0.102}$^{\dagger}$}  & {\color[HTML]{333333} \textbf{0.1536}$^{\dagger}$} & {\color[HTML]{333333} \textbf{0.0662}$^{\dagger}$} & {\color[HTML]{333333} \textbf{0.0785}$^{\dagger}$} & {\color[HTML]{333333} \textbf{0.0952}$^{\dagger}$} & {\color[HTML]{333333} \textbf{0.0471}$^{\dagger}$} & {\color[HTML]{333333} \textbf{0.0785}$^{\dagger}$} & {\color[HTML]{333333} \textbf{0.1284}$^{\dagger}$} & {\color[HTML]{333333} \textbf{0.0545}$^{\dagger}$} & {\color[HTML]{333333} \textbf{0.064}$^{\dagger}$}  & {\color[HTML]{333333} \textbf{0.081}$^{\dagger}$}  & {\color[HTML]{333333} \textbf{0.0599}$^{\dagger}$} & {\color[HTML]{333333} \textbf{0.0968}$^{\dagger}$} & {\color[HTML]{333333} \textbf{0.15}$^{\dagger}$}   & {\color[HTML]{333333} \textbf{0.0655}$^{\dagger}$} & {\color[HTML]{333333} \textbf{0.0778}$^{\dagger}$} & {\color[HTML]{333333} \textbf{0.0958}$^{\dagger}$} \\
\multicolumn{1}{c|}{\multirow{-5}{*}{{\color[HTML]{333333} SGL}}}      & {\color[HTML]{333333} Improvement} & {\color[HTML]{333333} 1.83\%}          & {\color[HTML]{333333} 0.29\%}          & {\color[HTML]{333333} 0.52\%}          & {\color[HTML]{333333} 1.53\%}          & {\color[HTML]{333333} 0.90\%}          & {\color[HTML]{333333} 0.74\%}          & {\color[HTML]{333333} 1.06\%}          & {\color[HTML]{333333} 1.82\%}          & {\color[HTML]{333333} 1.66\%}          & {\color[HTML]{333333} 1.49\%}          & {\color[HTML]{333333} 1.43\%}          & {\color[HTML]{333333} 1.50\%}          & {\color[HTML]{333333} 1.70\%}          & {\color[HTML]{333333} 1.26\%}          & {\color[HTML]{333333} 0.74\%}          & {\color[HTML]{333333} 1.55\%}          & {\color[HTML]{333333} 1.30\%}          & {\color[HTML]{333333} 0.84\%}          \\ \hline
\multicolumn{1}{c|}{{\color[HTML]{333333} }}                           & {\color[HTML]{333333} Baseline}    & {\color[HTML]{333333} 0.0618}          & {\color[HTML]{333333} 0.0992}          & {\color[HTML]{333333} 0.1512}          & {\color[HTML]{333333} 0.0619}          & {\color[HTML]{333333} 0.0749}          & {\color[HTML]{333333} 0.0919}          & {\color[HTML]{333333} 0.0467}          & {\color[HTML]{333333} 0.0772}          & {\color[HTML]{333333} 0.1254}          & {\color[HTML]{333333} 0.0546}          & {\color[HTML]{333333} 0.0638}          & {\color[HTML]{333333} 0.0801}          & {\color[HTML]{333333} 0.0564}          & {\color[HTML]{333333} 0.0918}          & {\color[HTML]{333333} 0.1436}          & {\color[HTML]{333333} 0.0618}          & {\color[HTML]{333333} 0.0738}          & {\color[HTML]{333333} 0.0915}          \\
\multicolumn{1}{c|}{{\color[HTML]{333333} }}                           & {\color[HTML]{333333} RLMRec-Con}  & {\color[HTML]{333333} 0.0633}          & {\color[HTML]{333333} 0.1011}          & {\color[HTML]{333333} 0.1552}          & {\color[HTML]{333333} 0.0633}          & {\color[HTML]{333333} 0.0765}          & {\color[HTML]{333333} 0.0942}          & {\color[HTML]{333333} 0.047}           & {\color[HTML]{333333} 0.0784}          & {\color[HTML]{333333} 0.1292}          & {\color[HTML]{333333} 0.0546}          & {\color[HTML]{333333} 0.0642}          & {\color[HTML]{333333} 0.0814}          & {\color[HTML]{333333} 0.0582}          & {\color[HTML]{333333} 0.0945}          & {\color[HTML]{333333} 0.1482}          & {\color[HTML]{333333} 0.0638}          & {\color[HTML]{333333} 0.076}           & {\color[HTML]{333333} 0.0942}          \\
\multicolumn{1}{c|}{{\color[HTML]{333333} }}                           & {\color[HTML]{333333} RLMRec-Gen}  & {\color[HTML]{333333} 0.0617}          & {\color[HTML]{333333} 0.0991}          & {\color[HTML]{333333} 0.1524}          & {\color[HTML]{333333} 0.0622}          & {\color[HTML]{333333} 0.0752}          & {\color[HTML]{333333} 0.0925}          & {\color[HTML]{333333} 0.0464}          & {\color[HTML]{333333} 0.0767}          & {\color[HTML]{333333} 0.1267}          & {\color[HTML]{333333} 0.0541}          & {\color[HTML]{333333} 0.0634}          & {\color[HTML]{333333} 0.0803}          & {\color[HTML]{333333} 0.0572}          & {\color[HTML]{333333} 0.0929}          & {\color[HTML]{333333} 0.1456}          & {\color[HTML]{333333} 0.0627}          & {\color[HTML]{333333} 0.0747}          & {\color[HTML]{333333} 0.0926}          \\
\multicolumn{1}{c|}{{\color[HTML]{333333} }}                           & {\color[HTML]{333333} Ours}        & {\color[HTML]{333333} \textbf{0.0648}$^{\dagger}$} & {\color[HTML]{333333} \textbf{0.103}$^{\dagger}$}  & {\color[HTML]{333333} \textbf{0.1563}$^{\dagger}$} & {\color[HTML]{333333} \textbf{0.0651}$^{\dagger}$} & {\color[HTML]{333333} \textbf{0.0781}$^{\dagger}$} & {\color[HTML]{333333} \textbf{0.0954}$^{\dagger}$} & {\color[HTML]{333333} \textbf{0.0479}$^{\dagger}$} & {\color[HTML]{333333} \textbf{0.0804}$^{\dagger}$} & {\color[HTML]{333333} \textbf{0.1317}$^{\dagger}$} & {\color[HTML]{333333} \textbf{0.0553}$^{\dagger}$} & {\color[HTML]{333333} \textbf{0.0656}$^{\dagger}$} & {\color[HTML]{333333} \textbf{0.0831}$^{\dagger}$} & {\color[HTML]{333333} \textbf{0.0588}$^{\dagger}$} & {\color[HTML]{333333} \textbf{0.095}$^{\dagger}$}  & {\color[HTML]{333333} \textbf{0.1497}$^{\dagger}$} & {\color[HTML]{333333} \textbf{0.0642}$^{\dagger}$} & {\color[HTML]{333333} \textbf{0.0762}$^{\dagger}$} & {\color[HTML]{333333} \textbf{0.0947}$^{\dagger}$} \\
\multicolumn{1}{c|}{\multirow{-5}{*}{{\color[HTML]{333333} SimGCL}}}   & {\color[HTML]{333333} Improvement} & {\color[HTML]{333333} 2.37\%}          & {\color[HTML]{333333} 1.88\%}          & {\color[HTML]{333333} 0.71\%}          & {\color[HTML]{333333} 2.84\%}          & {\color[HTML]{333333} 2.09\%}          & {\color[HTML]{333333} 1.27\%}          & {\color[HTML]{333333} 1.91\%}          & {\color[HTML]{333333} 2.55\%}          & {\color[HTML]{333333} 1.93\%}          & {\color[HTML]{333333} 1.28\%}          & {\color[HTML]{333333} 2.18\%}          & {\color[HTML]{333333} 2.09\%}          & {\color[HTML]{333333} 1.03\%}          & {\color[HTML]{333333} 0.53\%}          & {\color[HTML]{333333} 1.01\%}          & {\color[HTML]{333333} 0.63\%}          & {\color[HTML]{333333} 0.26\%}          & {\color[HTML]{333333} 0.53\%}          \\ \hline
\multicolumn{1}{c|}{{\color[HTML]{333333} }}                           & {\color[HTML]{333333} Baseline}    & {\color[HTML]{333333} 0.0662}          & {\color[HTML]{333333} 0.1019}          & {\color[HTML]{333333} 0.1517}          & {\color[HTML]{333333} 0.0658}          & {\color[HTML]{333333} 0.078}           & {\color[HTML]{333333} 0.0943}          & {\color[HTML]{333333} 0.0468}          & {\color[HTML]{333333} 0.0778}          & {\color[HTML]{333333} 0.1249}          & {\color[HTML]{333333} 0.0543}          & {\color[HTML]{333333} 0.064}           & {\color[HTML]{333333} 0.08}            & {\color[HTML]{333333} 0.0561}          & {\color[HTML]{333333} 0.0915}          & {\color[HTML]{333333} 0.1437}          & {\color[HTML]{333333} 0.0618}          & {\color[HTML]{333333} 0.0736}          & {\color[HTML]{333333} 0.0914}          \\
\multicolumn{1}{c|}{{\color[HTML]{333333} }}                           & {\color[HTML]{333333} RLMRec-Con}  & {\color[HTML]{333333} 0.0665}          & {\color[HTML]{333333} 0.104}           & {\color[HTML]{333333} 0.1563}          & {\color[HTML]{333333} 0.0668}          & {\color[HTML]{333333} 0.0798}          & {\color[HTML]{333333} 0.0968}          & {\color[HTML]{333333} 0.0486}          & {\color[HTML]{333333} 0.0813}          & {\color[HTML]{333333} 0.1321}          & {\color[HTML]{333333} 0.0561}          & {\color[HTML]{333333} 0.0663}          & {\color[HTML]{333333} 0.0836}          & {\color[HTML]{333333} 0.0572}          & {\color[HTML]{333333} 0.0929}          & {\color[HTML]{333333} 0.1459}          & {\color[HTML]{333333} 0.0627}          & {\color[HTML]{333333} 0.0747}          & {\color[HTML]{333333} 0.0927}          \\
\multicolumn{1}{c|}{{\color[HTML]{333333} }}                           & {\color[HTML]{333333} RLMRec-Gen}  & {\color[HTML]{333333} 0.0666}          & {\color[HTML]{333333} 0.1046}          & {\color[HTML]{333333} 0.1559}          & {\color[HTML]{333333} 0.067}           & {\color[HTML]{333333} 0.0801}          & {\color[HTML]{333333} 0.0969}          & {\color[HTML]{333333} 0.0475}          & {\color[HTML]{333333} 0.0785}          & {\color[HTML]{333333} 0.1281}          & {\color[HTML]{333333} 0.0549}          & {\color[HTML]{333333} 0.0646}          & {\color[HTML]{333333} 0.0815}          & {\color[HTML]{333333} 0.057}           & {\color[HTML]{333333} 0.0918}          & {\color[HTML]{333333} 0.143}           & {\color[HTML]{333333} 0.0625}          & {\color[HTML]{333333} 0.0741}          & {\color[HTML]{333333} 0.0915}          \\
\multicolumn{1}{c|}{{\color[HTML]{333333} }}                           & {\color[HTML]{333333} Ours}        & {\color[HTML]{333333} \textbf{0.0677}$^{\dagger}$} & {\color[HTML]{333333} 0.1045}          & {\color[HTML]{333333} \textbf{0.1582}$^{\dagger}$} & {\color[HTML]{333333} \textbf{0.0674}$^{\dagger}$} & {\color[HTML]{333333} \textbf{0.0807}$^{\dagger}$} & {\color[HTML]{333333} \textbf{0.0981}$^{\dagger}$} & {\color[HTML]{333333} \textbf{0.0495}$^{\dagger}$} & {\color[HTML]{333333} \textbf{0.0826}$^{\dagger}$} & {\color[HTML]{333333} \textbf{0.1352}$^{\dagger}$} & {\color[HTML]{333333} \textbf{0.0569}$^{\dagger}$} & {\color[HTML]{333333} \textbf{0.0673}$^{\dagger}$} & {\color[HTML]{333333} \textbf{0.0850}$^{\dagger}$}  & {\color[HTML]{333333} \textbf{0.0586}$^{\dagger}$} & {\color[HTML]{333333} \textbf{0.0938}$^{\dagger}$} & {\color[HTML]{333333} \textbf{0.1479}$^{\dagger}$} & {\color[HTML]{333333} \textbf{0.0638}$^{\dagger}$} & {\color[HTML]{333333} \textbf{0.0751}$^{\dagger}$} & {\color[HTML]{333333} \textbf{0.0937}$^{\dagger}$} \\
\multicolumn{1}{c|}{\multirow{-5}{*}{{\color[HTML]{333333} DCCF}}}     & {\color[HTML]{333333} Improvement} & {\color[HTML]{333333} 1.65\%}          & {\color[HTML]{333333} -0.10\%}         & {\color[HTML]{333333} 1.48\%}          & {\color[HTML]{333333} 0.60\%}          & {\color[HTML]{333333} 0.75\%}          & {\color[HTML]{333333} 1.24\%}          & {\color[HTML]{333333} 1.85\%}          & {\color[HTML]{333333} 1.60\%}          & {\color[HTML]{333333} 2.35\%}          & {\color[HTML]{333333} 1.43\%}          & {\color[HTML]{333333} 1.51\%}          & {\color[HTML]{333333} 1.67\%}          & {\color[HTML]{333333} 2.45\%}          & {\color[HTML]{333333} 0.97\%}          & {\color[HTML]{333333} 1.37\%}          & {\color[HTML]{333333} 1.75\%}          & {\color[HTML]{333333} 0.54\%}          & {\color[HTML]{333333} 1.08\%}          \\ \hline
\multicolumn{1}{c|}{{\color[HTML]{333333} }}                           & {\color[HTML]{333333} Baseline}    & {\color[HTML]{333333} 0.0689}          & {\color[HTML]{333333} 0.1055}          & {\color[HTML]{333333} 0.1536}          & {\color[HTML]{333333} 0.0705}          & {\color[HTML]{333333} 0.0828}          & {\color[HTML]{333333} 0.0984}          & {\color[HTML]{333333} 0.0469}          & {\color[HTML]{333333} 0.0789}          & {\color[HTML]{333333} 0.128}           & {\color[HTML]{333333} 0.0547}          & {\color[HTML]{333333} 0.0647}          & {\color[HTML]{333333} 0.0813}          & {\color[HTML]{333333} 0.0519}          & {\color[HTML]{333333} 0.0853}          & {\color[HTML]{333333} 0.1358}          & {\color[HTML]{333333} 0.0572}          & {\color[HTML]{333333} 0.0684}          & {\color[HTML]{333333} 0.0855}          \\
\multicolumn{1}{c|}{{\color[HTML]{333333} }}                           & {\color[HTML]{333333} RLMRec-Con}  & {\color[HTML]{333333} 0.0695}          & {\color[HTML]{333333} 0.1083}          & {\color[HTML]{333333} 0.1586}          & {\color[HTML]{333333} 0.0704}          & {\color[HTML]{333333} 0.0837}          & {\color[HTML]{333333} 0.1001}          & {\color[HTML]{333333} 0.0488}          & {\color[HTML]{333333} 0.0814}          & {\color[HTML]{333333} 0.1319}          & {\color[HTML]{333333} 0.0562}          & {\color[HTML]{333333} 0.0663}          & {\color[HTML]{333333} 0.0835}          & {\color[HTML]{333333} 0.054}           & {\color[HTML]{333333} 0.0876}          & {\color[HTML]{333333} 0.1372}          & {\color[HTML]{333333} 0.0593}          & {\color[HTML]{333333} 0.0704}          & {\color[HTML]{333333} 0.0872}          \\
\multicolumn{1}{c|}{{\color[HTML]{333333} }}                           & {\color[HTML]{333333} RLMRec-Gen}  & {\color[HTML]{333333} 0.0693}          & {\color[HTML]{333333} 0.1069}          & {\color[HTML]{333333} 0.1581}          & {\color[HTML]{333333} 0.0701}          & {\color[HTML]{333333} 0.083}           & {\color[HTML]{333333} 0.0996}          & {\color[HTML]{333333} 0.0493}          & {\color[HTML]{333333} 0.0828}          & {\color[HTML]{333333} 0.133}           & {\color[HTML]{333333} 0.0572}          & {\color[HTML]{333333} 0.0677}          & {\color[HTML]{333333} 0.0848}          & {\color[HTML]{333333} 0.0539}          & {\color[HTML]{333333} 0.0888}          & {\color[HTML]{333333} 0.1410}           & {\color[HTML]{333333} 0.0593}          & {\color[HTML]{333333} 0.071}           & {\color[HTML]{333333} 0.0886}          \\
\multicolumn{1}{c|}{{\color[HTML]{333333} }}                           & {\color[HTML]{333333} Ours}        & {\color[HTML]{333333} \textbf{0.0714}$^{\dagger}$} & {\color[HTML]{333333} \textbf{0.1102}$^{\dagger}$} & {\color[HTML]{333333} \textbf{0.159}$^{\dagger}$}  & {\color[HTML]{333333} \textbf{0.0725}$^{\dagger}$} & {\color[HTML]{333333} \textbf{0.0856}$^{\dagger}$} & {\color[HTML]{333333} \textbf{0.1016}$^{\dagger}$} & {\color[HTML]{333333} \textbf{0.0512}$^{\dagger}$} & {\color[HTML]{333333} \textbf{0.0841}$^{\dagger}$} & {\color[HTML]{333333} \textbf{0.1344}$^{\dagger}$} & {\color[HTML]{333333} \textbf{0.059}$^{\dagger}$}  & {\color[HTML]{333333} \textbf{0.0691}$^{\dagger}$} & {\color[HTML]{333333} \textbf{0.0861}$^{\dagger}$} & {\color[HTML]{333333} \textbf{0.0554}$^{\dagger}$} & {\color[HTML]{333333} \textbf{0.0900}$^{\dagger}$}   & {\color[HTML]{333333} \textbf{0.1422}$^{\dagger}$} & {\color[HTML]{333333} \textbf{0.0604}$^{\dagger}$} & {\color[HTML]{333333} \textbf{0.0719}$^{\dagger}$} & {\color[HTML]{333333} \textbf{0.0895}$^{\dagger}$} \\
\multicolumn{1}{c|}{\multirow{-5}{*}{{\color[HTML]{333333} AutoCF}}}   & {\color[HTML]{333333} Improvement} & {\color[HTML]{333333} 2.73\%}          & {\color[HTML]{333333} 1.75\%}          & {\color[HTML]{333333} 0.25\%}          & {\color[HTML]{333333} 2.98\%}          & {\color[HTML]{333333} 2.27\%}          & {\color[HTML]{333333} 1.50\%}          & {\color[HTML]{333333} 3.85\%}          & {\color[HTML]{333333} 1.57\%}          & {\color[HTML]{333333} 1.05\%}          & {\color[HTML]{333333} 3.15\%}          & {\color[HTML]{333333} 2.07\%}          & {\color[HTML]{333333} 1.53\%}          & {\color[HTML]{333333} 2.59\%}          & {\color[HTML]{333333} 1.35\%}          & {\color[HTML]{333333} 0.85\%}          & {\color[HTML]{333333} 1.85\%}          & {\color[HTML]{333333} 1.27\%}          & {\color[HTML]{333333} 1.02\%}          \\ \hline
\end{tabular}}
\label{com_res}
\end{table*}



\section{Theoretical analysis}
In this section, we explore the rationality of our proposed disentangled alignment framework from the theoretical perspective. We give the following notation for the sake of convenience. Let $\widehat{\textbf{E}}$ denote the concatenated shared and specific representations of our method, and use $\widetilde{\textbf{E}}$ to denote the representations extracted by the previous undisentangled methods. We have:

\begin{theorem}
For the recommendation downstream task $\textbf{R}$, the representations $\widehat{\textbf{E}}$ contain more relevant information and less irrelevant information than $\widetilde{\textbf{E}}$ extracted by previous methods, which can be presented as:
\begin{equation}
\begin{aligned}
I(\widehat{\textbf{E}}^{\textbf{D}}, \textbf{R}) &\ge I(\widetilde{\textbf{E}}^{\textbf{D}},\textbf{R}), \\
H(\widehat{\textbf{E}}^{\textbf{D}}|\textbf{R}) & \le H(\widetilde{\textbf{E}}^{\textbf{D}}|\textbf{R}),
\end{aligned}
\label{theorem_better}
\end{equation} 
where $I(\textbf{E}^{\textbf{D}}, \textbf{R})$ means the mutual information between the representations and recommendation tasks, $H(\textbf{E}^{\textbf{D}}|\textbf{R})$ denotes the entropy of the representation conditioned on recommendation tasks.
\end{theorem} 

We provide the proof in section~\ref{proof_2}.

\section{Experiment}
In this section, we conduct experiments to evaluate the effectiveness of our proposed method. The specific effectiveness can be illustrated by answering the following questions. 


\begin{itemize}
    \item \textbf{RQ1}: How does our proposed disentangled alignment framework improve the performance of existing state-of-the-art recommender methods?
    \item \textbf{RQ2}: How do the proposed modules influence the recommendation performance?
    \item \textbf{RQ3}: How do the hyper-parameters impact the performance of DaRec?
    \item \textbf{RQ4}: What is the preference center revealed by DaRec?
\end{itemize}

\subsection{Experimental Settings}

\textbf{Benchmark Datasets.}
The experimental results are evaluated in three widely used benchmark datasets, including Amazon Book, Yelp, and Steam.

A detailed description of the dataset is shown in Table.\ref{data_info}. Following previous works \cite{NGCF,xia2022hypergraph}, we filter out the interactions with the ratings below 3 in all datasets for data preprocessing. Moreover, we adopt the sparse splitting with a 3:1:1 ratio for all datasets.

\textbf{Compared Methods}
In this paper, we compare our proposed alignment framework DaRec into six baselines, i.e., GCCF \cite{GCCF}, LightGCN \cite{LightGCN}, SGL \cite{SGL}, SimGCL \cite{SimGCL}, DCCF \cite{DCCF}, and AutoCF \cite{autocf}, RLMRec \cite{RLMRec}, and KAR \cite{KAR}. The details of baselines are described as follows.

\begin{table}[]
\centering
\caption{Recommendation Performance with LLMs-enhanced Methods on two datasets.}
\begin{tabular}{cc|cc|cc}
\hline
\multicolumn{2}{c|}{{\color[HTML]{333333} Data}}                                                           & \multicolumn{2}{c|}{{\color[HTML]{333333} Amazon-book}}                         & \multicolumn{2}{c}{{\color[HTML]{333333} Yelp}}                                 \\ \hline
\multicolumn{1}{c|}{{\color[HTML]{333333} Backbone}}                   & {\color[HTML]{333333} Variants}   & {\color[HTML]{333333} R@20}            & {\color[HTML]{333333} N@20}            & {\color[HTML]{333333} R@20}            & {\color[HTML]{333333} N@20}            \\ \hline
\multicolumn{1}{c|}{{\color[HTML]{333333} }}                           & {\color[HTML]{333333} Baseline}   & {\color[HTML]{333333} 0.1411}          & {\color[HTML]{333333} 0.0856}          & {\color[HTML]{333333} 0.1157}          & {\color[HTML]{333333} 0.0733}          \\
\multicolumn{1}{c|}{{\color[HTML]{333333} }}                           & {\color[HTML]{333333} RLMRec-Con} & {\color[HTML]{333333} 0.1483}          & {\color[HTML]{333333} 0.0903}          & {\color[HTML]{333333} 0.123}           & {\color[HTML]{333333} 0.0776}          \\
\multicolumn{1}{c|}{{\color[HTML]{333333} }}                           & {\color[HTML]{333333} RLMRec-Gen} & {\color[HTML]{333333} 0.1446}          & {\color[HTML]{333333} 0.0887}          & {\color[HTML]{333333} 0.1209}          & {\color[HTML]{333333} 0.0761}          \\
\multicolumn{1}{c|}{{\color[HTML]{333333} }}                           & {\color[HTML]{333333} KAR}        & {\color[HTML]{333333} 0.1416}          & {\color[HTML]{333333} 0.0863}          & {\color[HTML]{333333} 0.1194}          & {\color[HTML]{333333} 0.0756}          \\
\multicolumn{1}{c|}{\multirow{-5}{*}{{\color[HTML]{333333} LightGCN}}} & {\color[HTML]{333333} Ours}       & {\color[HTML]{333333} \textbf{0.1495}} & {\color[HTML]{333333} \textbf{0.091}}  & {\color[HTML]{333333} \textbf{0.1246}} & {\color[HTML]{333333} \textbf{0.0789}} \\ \hline
\multicolumn{1}{c|}{{\color[HTML]{333333} }}                           & {\color[HTML]{333333} Baseline}   & {\color[HTML]{333333} 0.1473}          & {\color[HTML]{333333} 0.0913}          & {\color[HTML]{333333} 0.1197}          & {\color[HTML]{333333} 0.0753}          \\
\multicolumn{1}{c|}{{\color[HTML]{333333} }}                           & {\color[HTML]{333333} RLMRec-Con} & {\color[HTML]{333333} 0.1528}          & {\color[HTML]{333333} 0.0945}          & {\color[HTML]{333333} 0.1248}          & {\color[HTML]{333333} 0.0790}           \\
\multicolumn{1}{c|}{{\color[HTML]{333333} }}                           & {\color[HTML]{333333} RLMRec-Gen} & {\color[HTML]{333333} 0.1537}          & {\color[HTML]{333333} 0.0947}          & {\color[HTML]{333333} 0.1263}          & {\color[HTML]{333333} 0.0798}          \\
\multicolumn{1}{c|}{{\color[HTML]{333333} }}                           & {\color[HTML]{333333} KAR}        & {\color[HTML]{333333} 0.1436}          & {\color[HTML]{333333} 0.0875}          & {\color[HTML]{333333} 0.1208}          & {\color[HTML]{333333} 0.0761}          \\
\multicolumn{1}{c|}{\multirow{-5}{*}{{\color[HTML]{333333} SGL}}}      & {\color[HTML]{333333} Ours}       & {\color[HTML]{333333} \textbf{0.1536}} & {\color[HTML]{333333} \textbf{0.0952}} & {\color[HTML]{333333} \textbf{0.1284}} & {\color[HTML]{333333} \textbf{0.081}}  \\ \hline
\end{tabular}
\label{com_res_l}
\end{table}

\begin{figure*}[h]
\centering
\small
\begin{minipage}{0.24\linewidth}
\centerline{\includegraphics[width=1.0\textwidth]{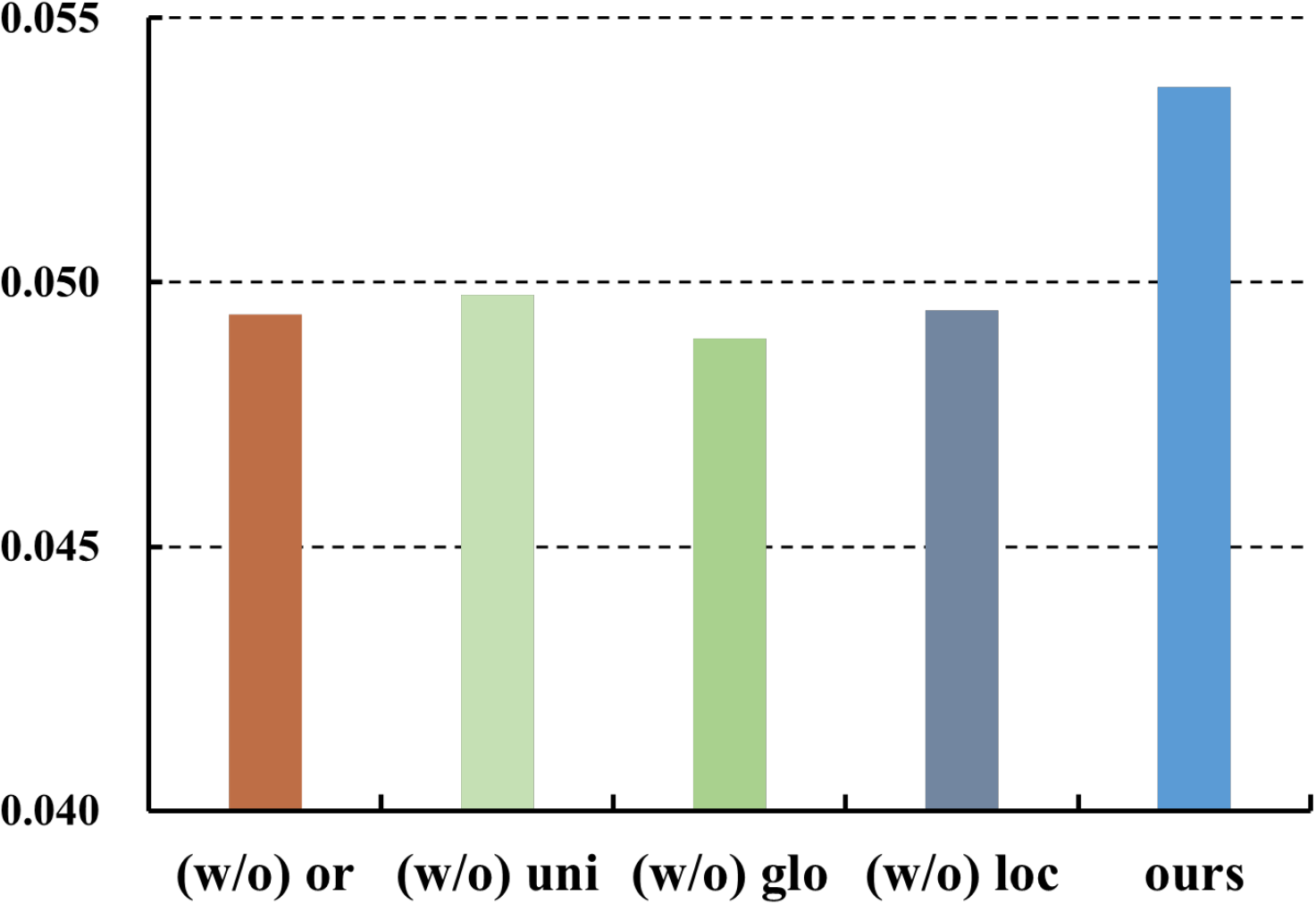}}
\centerline{\textbf{LightGCN-Yelp}}
\vspace{5pt}
\centerline{\includegraphics[width=1.0\textwidth]{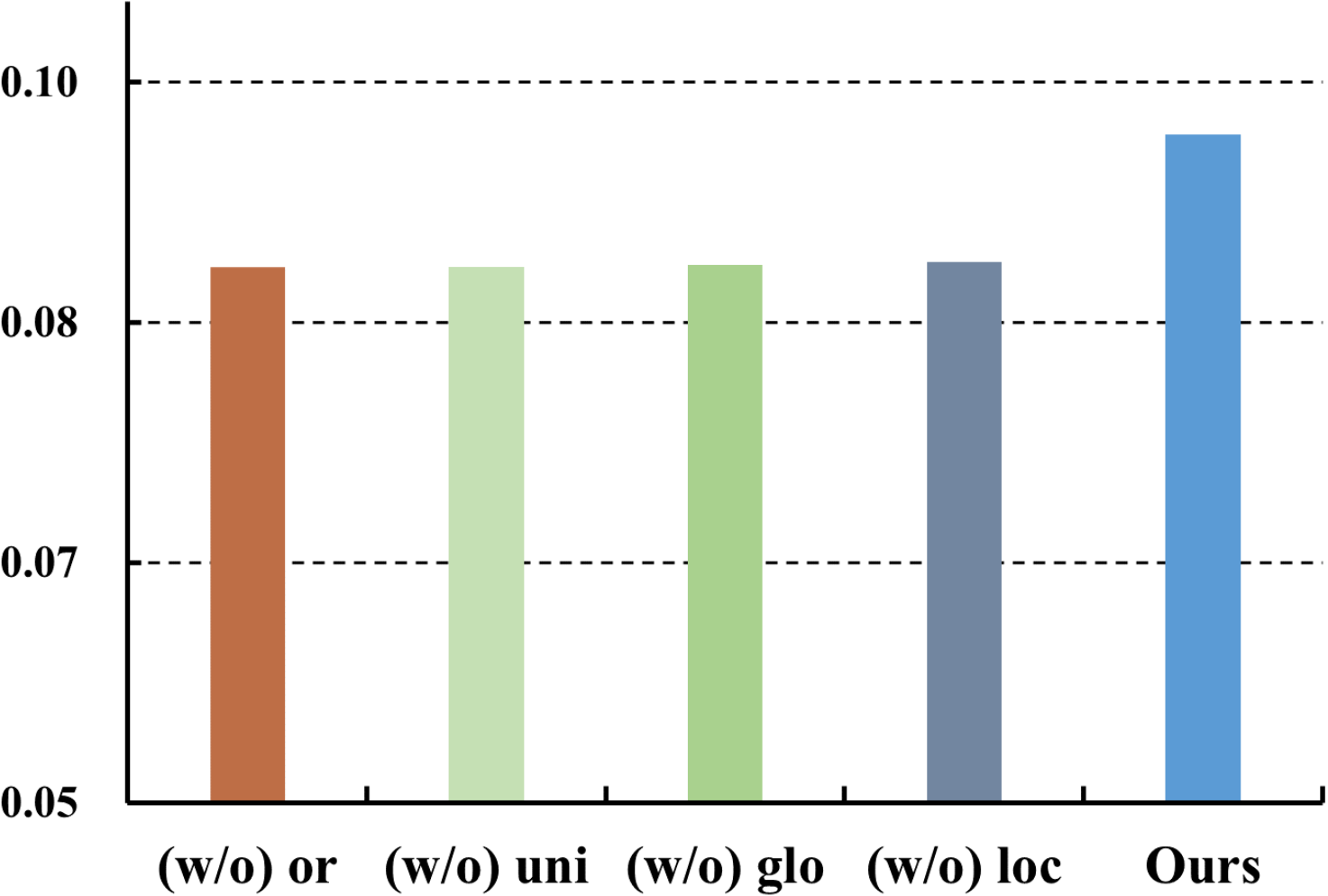}}
\centerline{\textbf{LightGCN-Steam}}
\vspace{5pt}
\centerline{\includegraphics[width=1.0\textwidth]{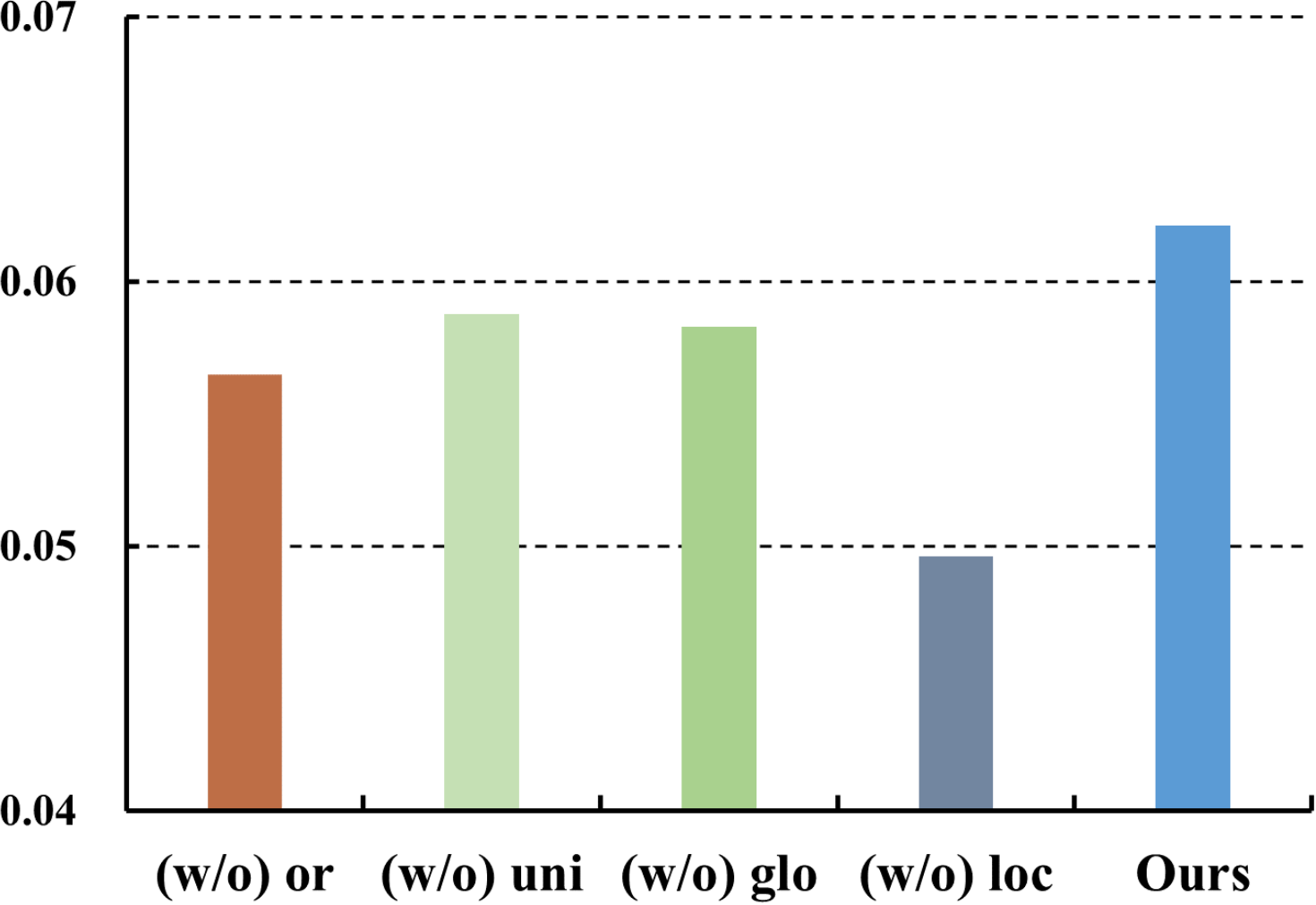}}
\centerline{\textbf{LightGCN-Amazon}}
\vspace{5pt}
\centerline{\includegraphics[width=1.0\textwidth]{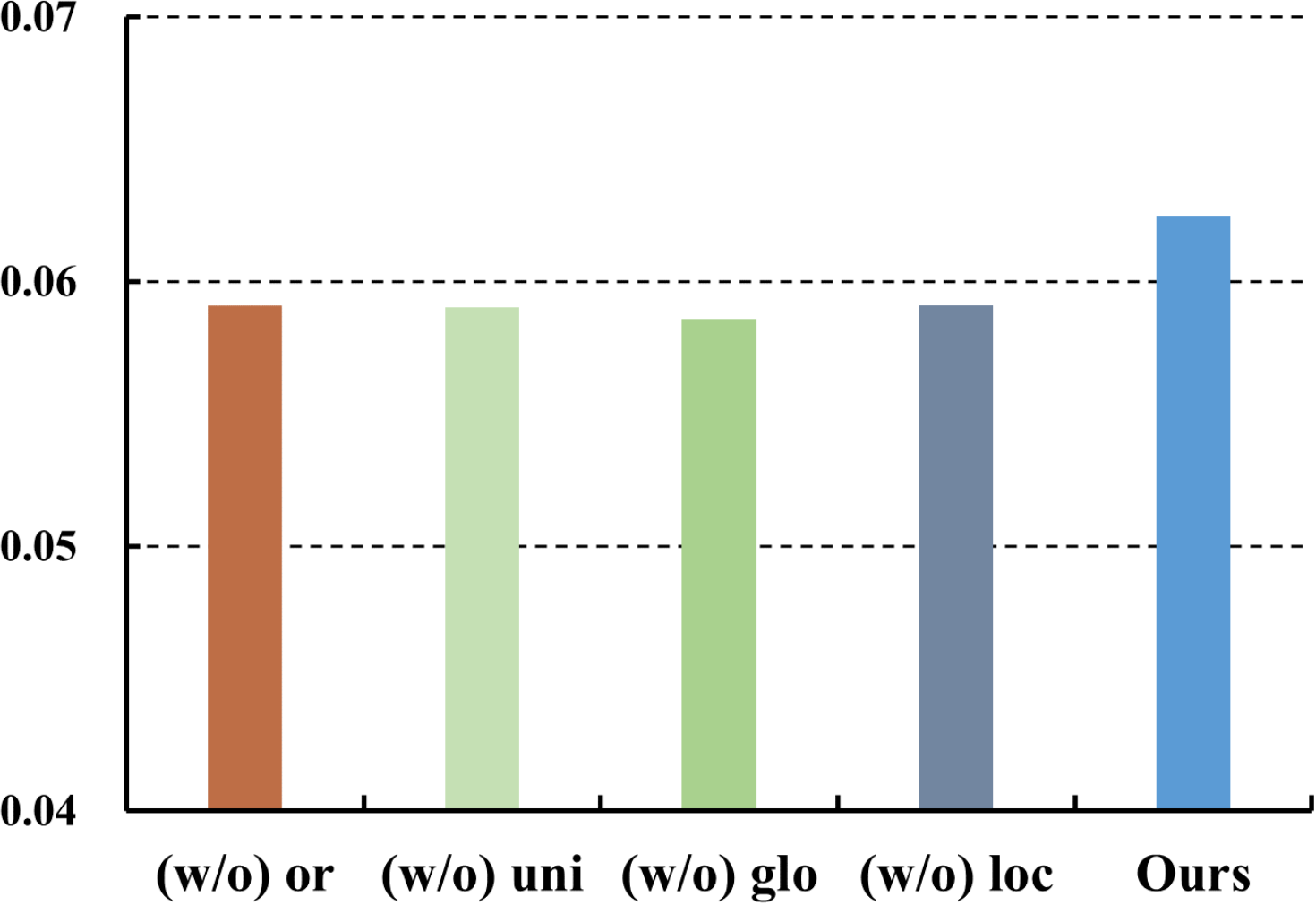}}
\centerline{\textbf{LightGCN-Yelp}}
\vspace{5pt}
\end{minipage}
\begin{minipage}{0.24\linewidth}
\centerline{\includegraphics[width=1.0\textwidth]{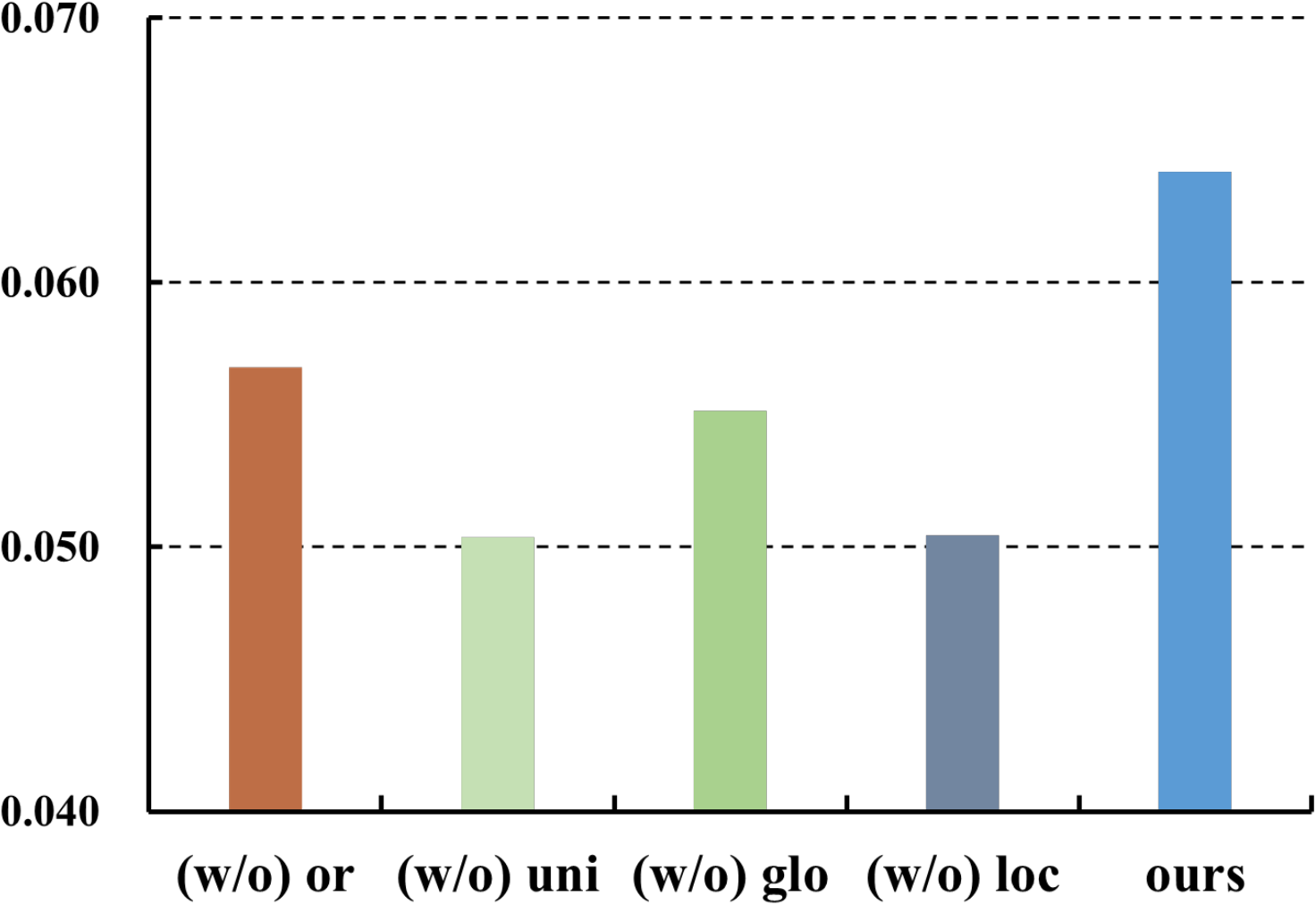}}
\centerline{\textbf{SimGCL-Steam}}
\vspace{5pt}
\centerline{\includegraphics[width=1.0\textwidth]{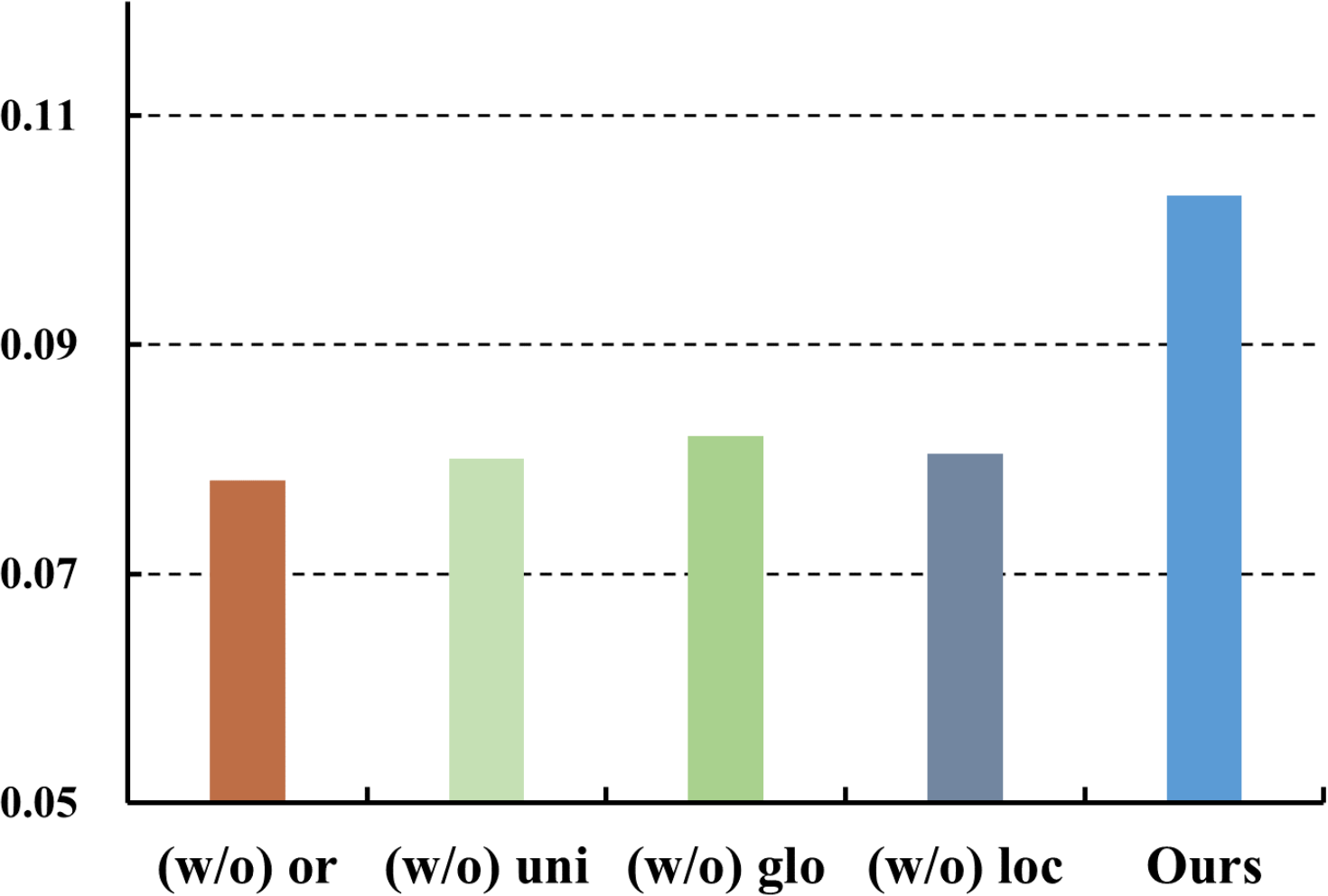}}
\centerline{\textbf{SimGCL-Amazon}}
\vspace{5pt}
\centerline{\includegraphics[width=1.0\textwidth]{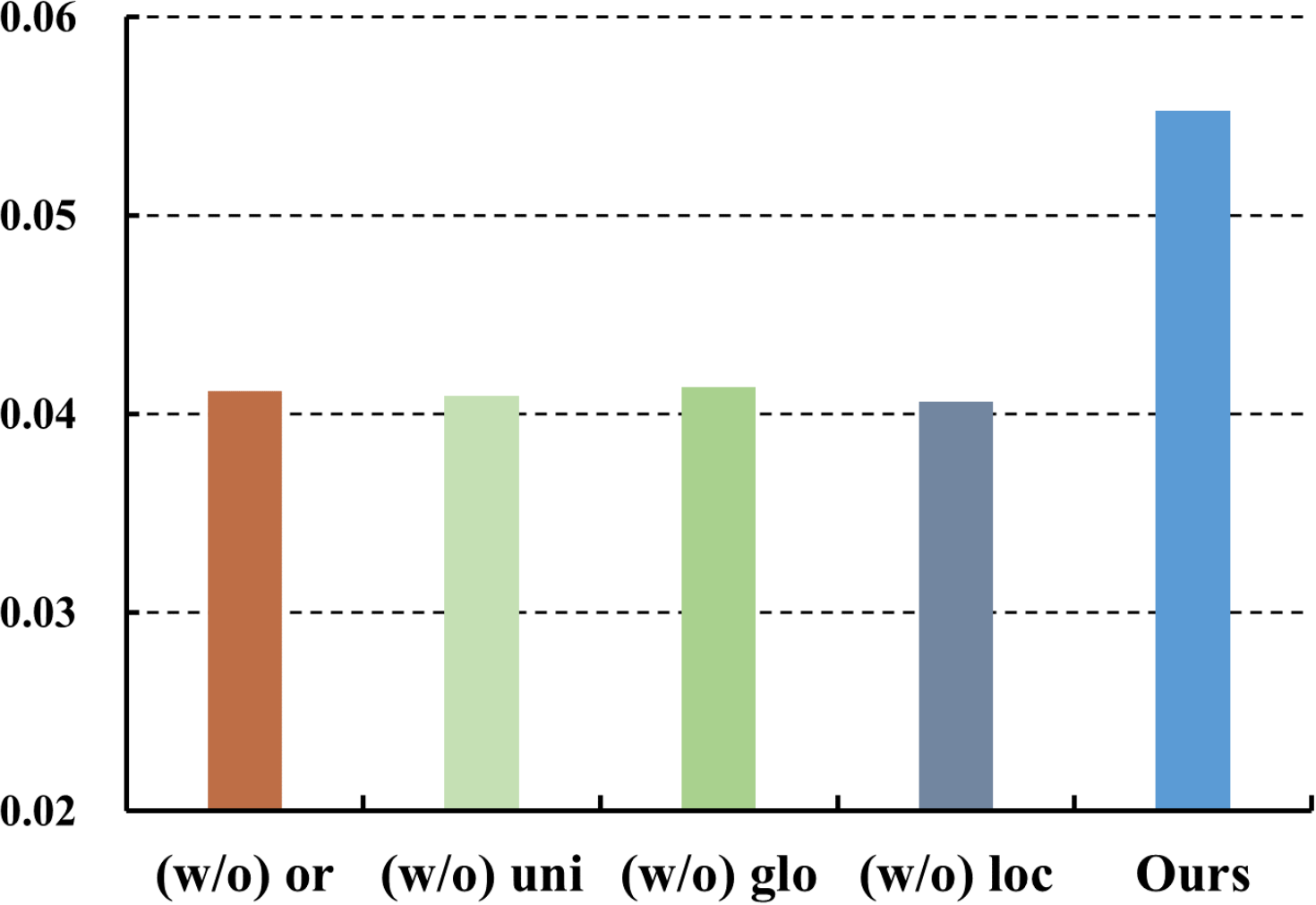}}
\centerline{\textbf{SimGCL-Yelp}}
\vspace{5pt}
\centerline{\includegraphics[width=1.0\textwidth]{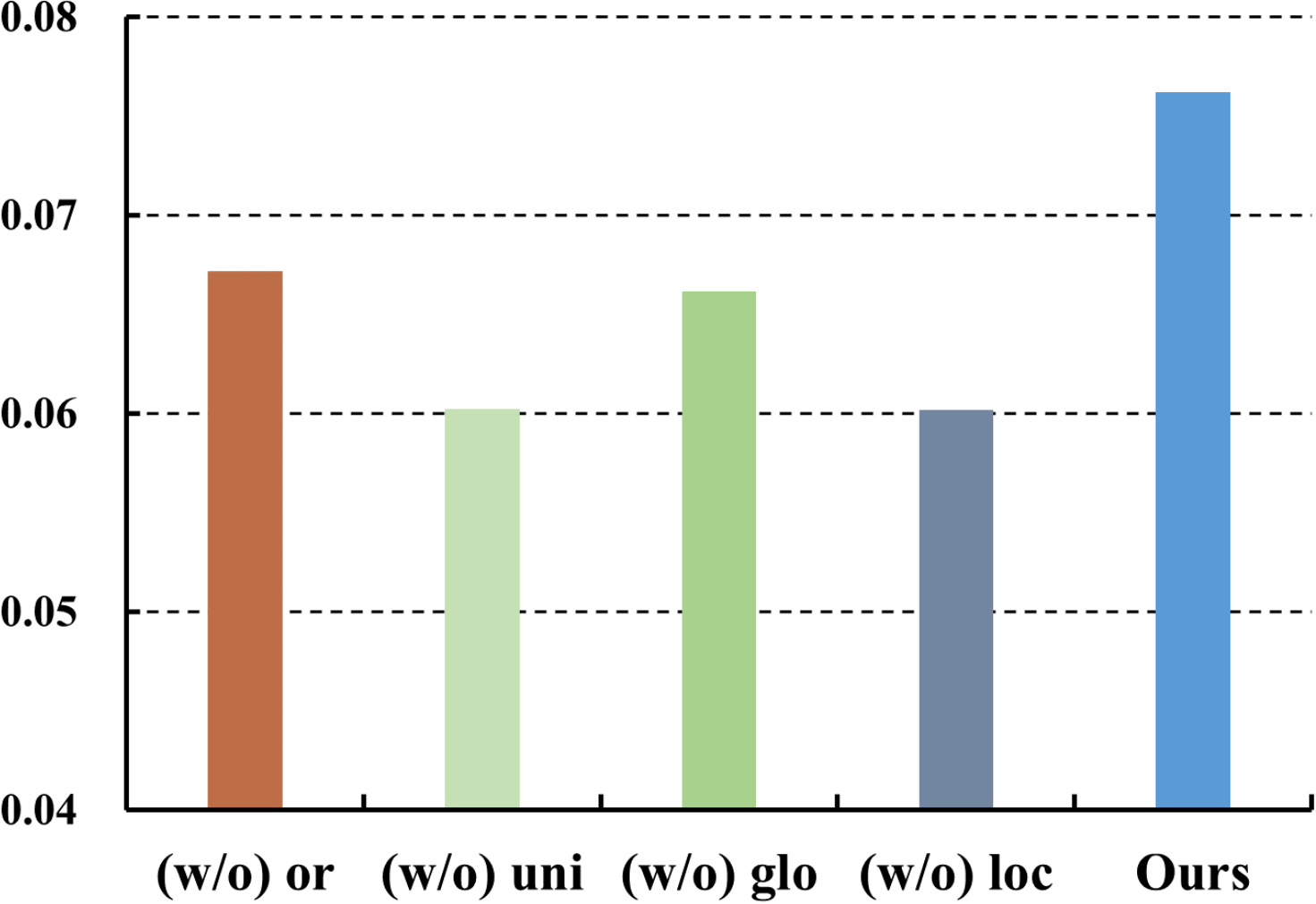}}
\centerline{\textbf{SimGCL-Steam}}
\vspace{5pt}
\end{minipage}
\begin{minipage}{0.24\linewidth}
\centerline{\includegraphics[width=1.0\textwidth]{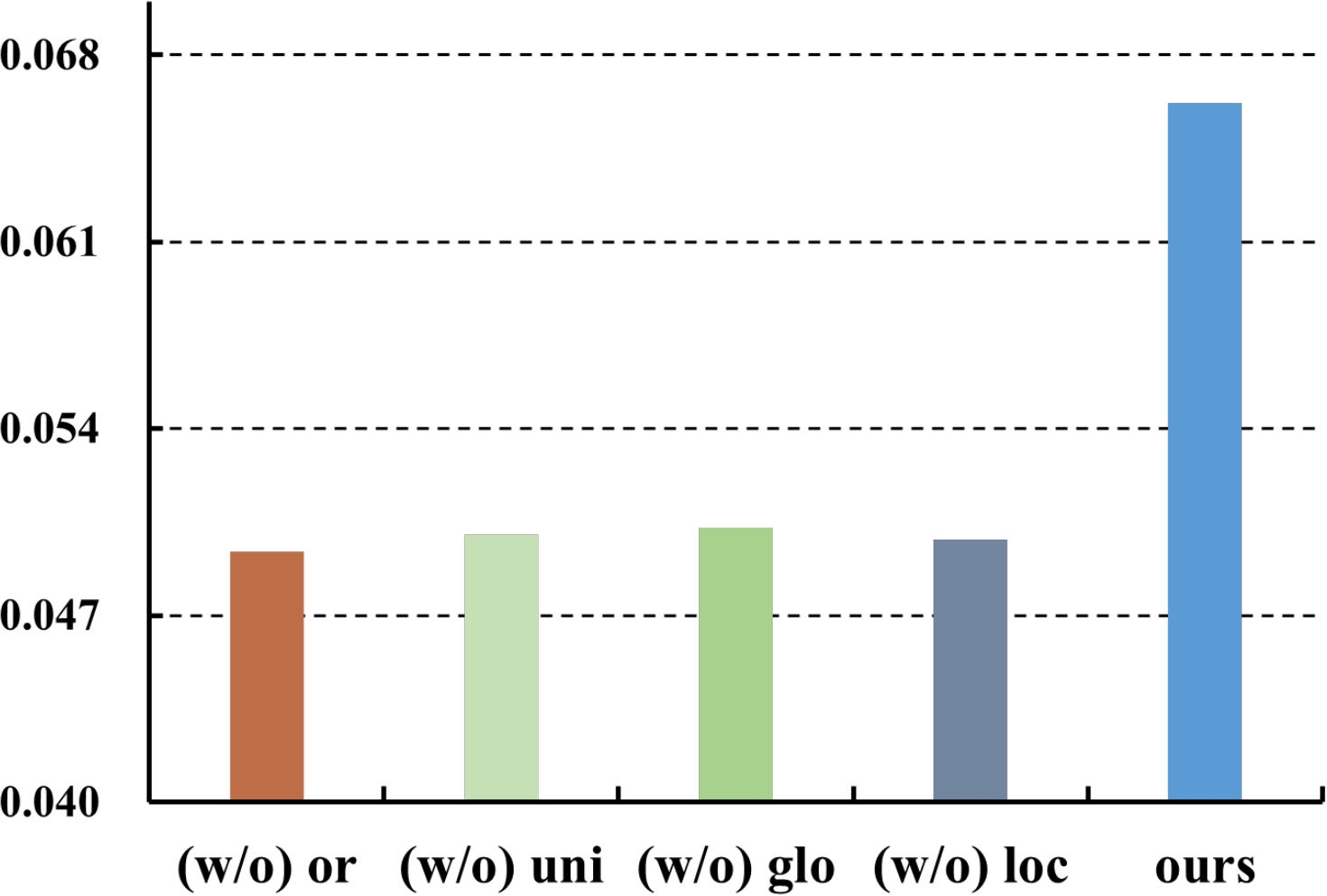}}
\centerline{\textbf{SGL-Amazon}}
\vspace{5pt}
\centerline{\includegraphics[width=1.0\textwidth]{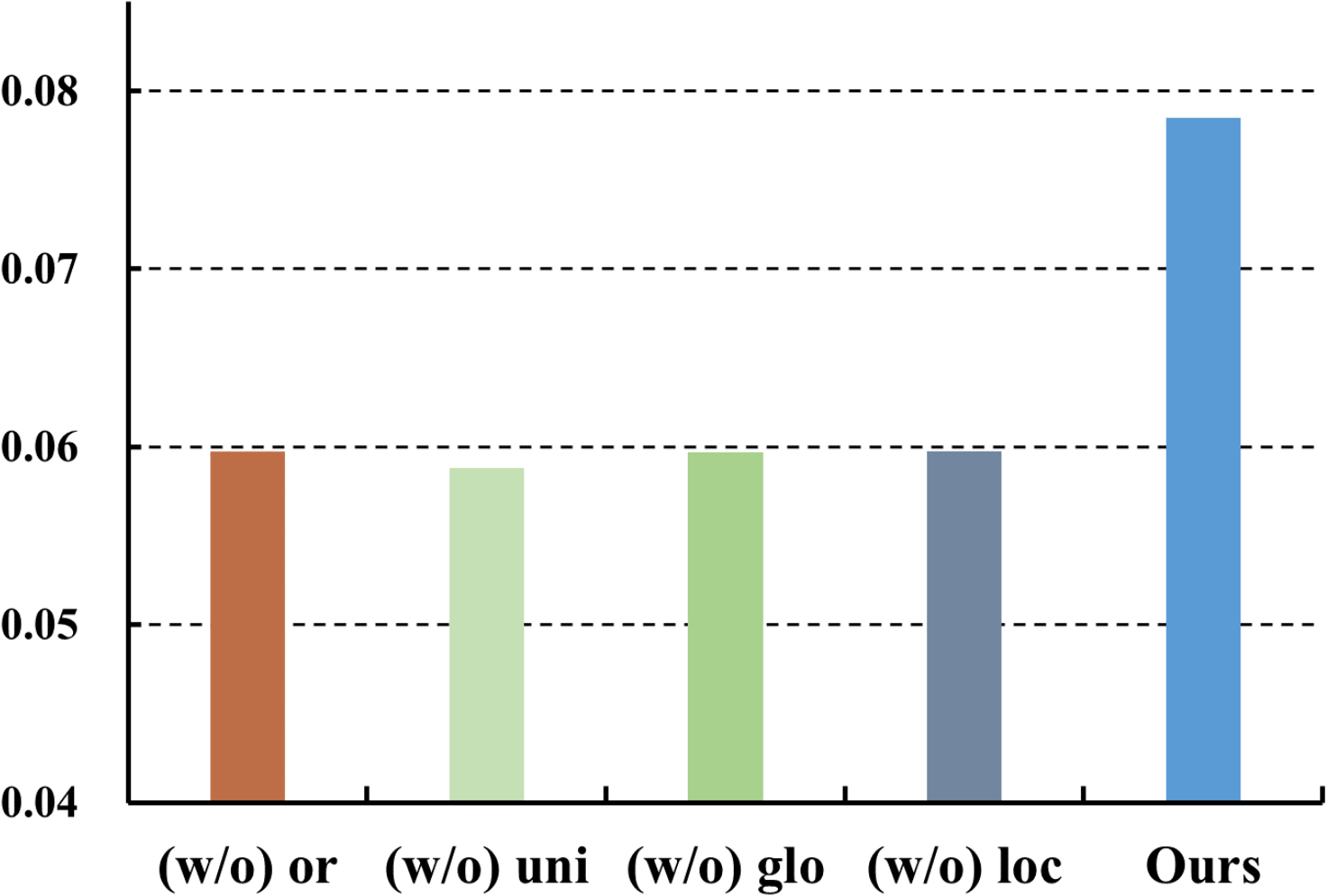}}
\centerline{\textbf{SGL-Yelp}}
\vspace{5pt}
\centerline{\includegraphics[width=1.0\textwidth]{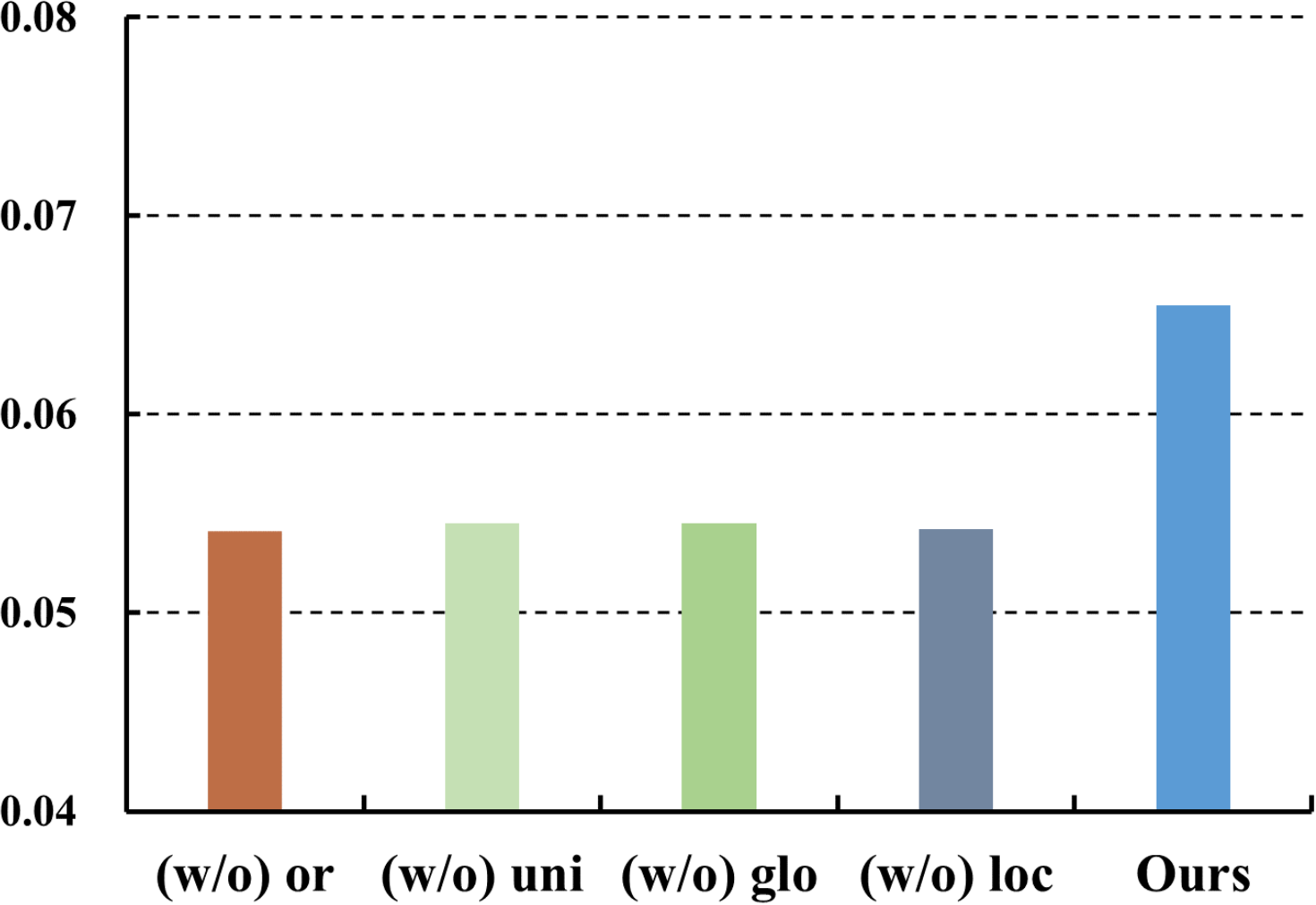}}
\centerline{\textbf{SGL-Steam}}
\vspace{5pt}
\centerline{\includegraphics[width=1.0\textwidth]{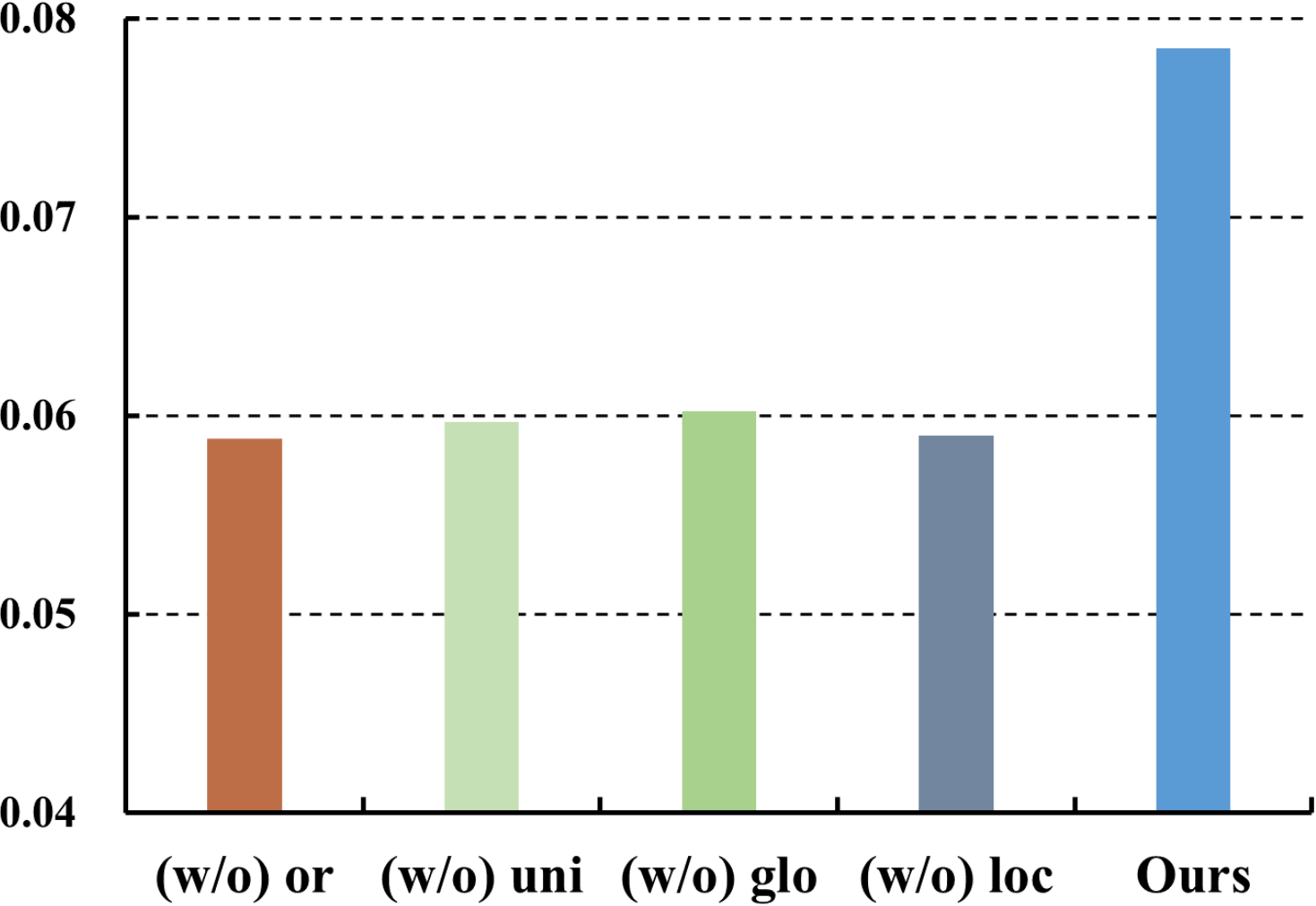}}
\centerline{\textbf{SGL-Amazon}}
\vspace{5pt}
\end{minipage}
\begin{minipage}{0.24\linewidth}
\centerline{\includegraphics[width=1.0\textwidth]{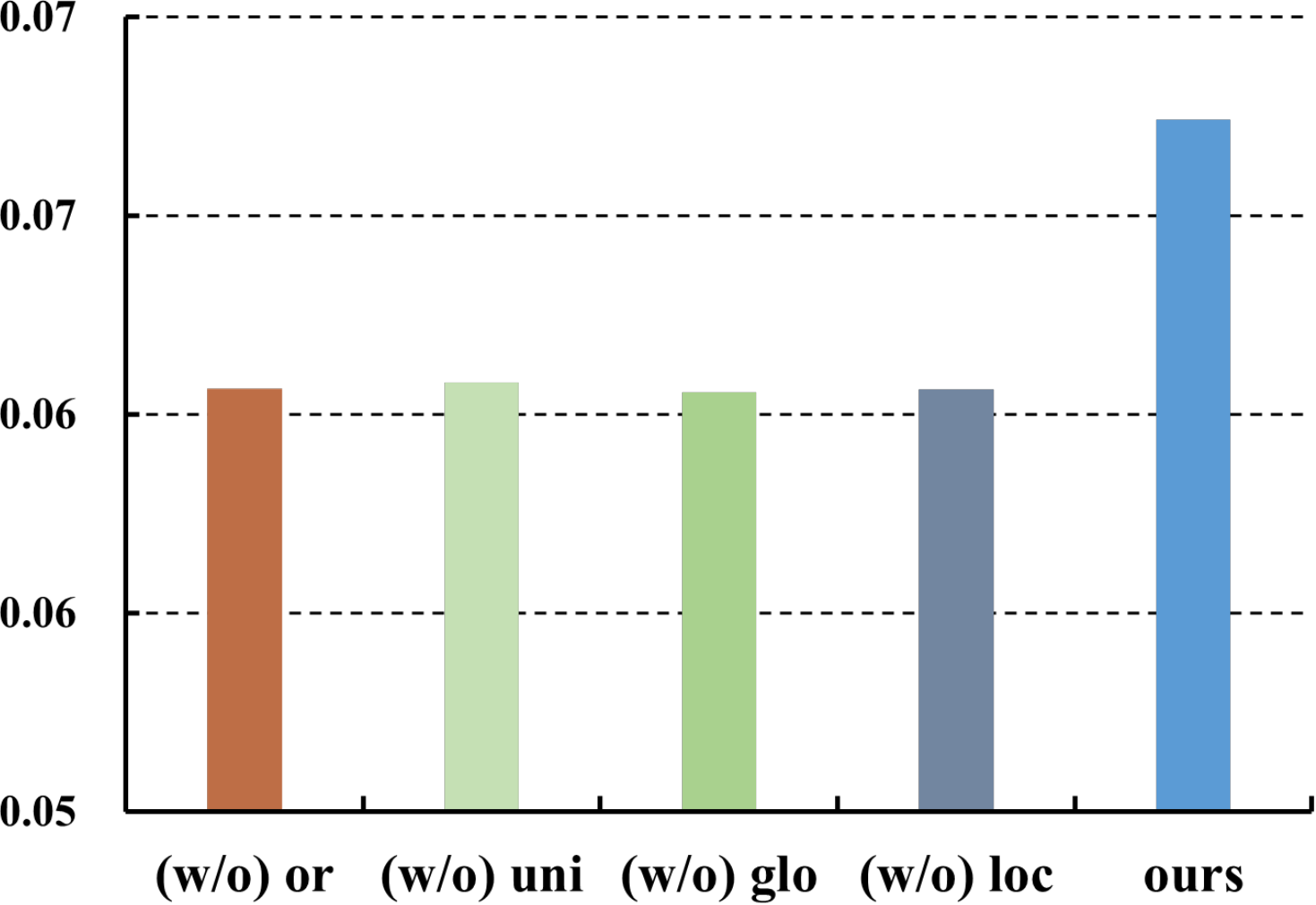}}
\centerline{\textbf{DCCF-Amazon}}
\vspace{5pt}
\centerline{\includegraphics[width=1.0\textwidth]{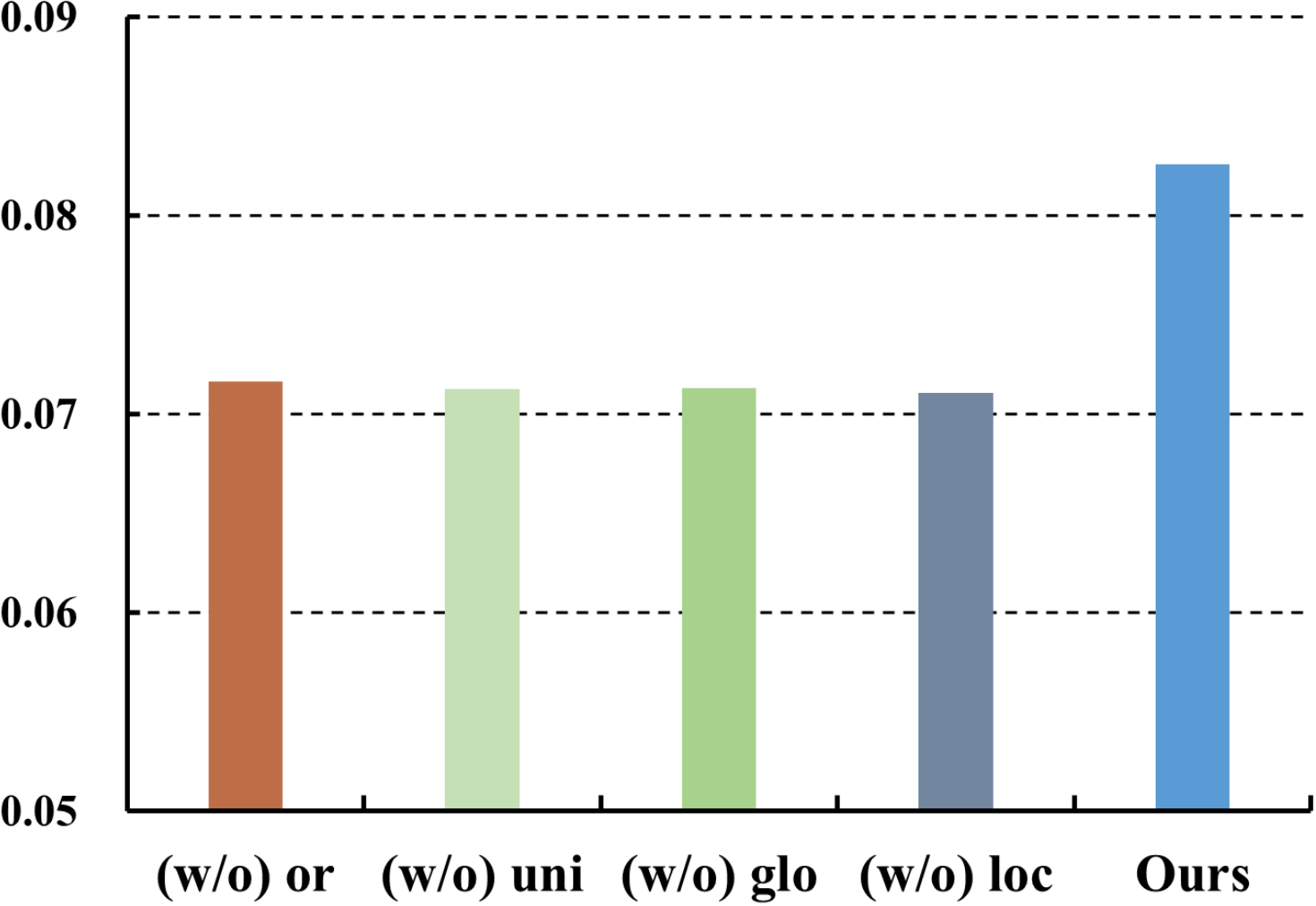}}
\centerline{\textbf{DCCF-Yelp}}
\vspace{5pt}
\centerline{\includegraphics[width=1.0\textwidth]{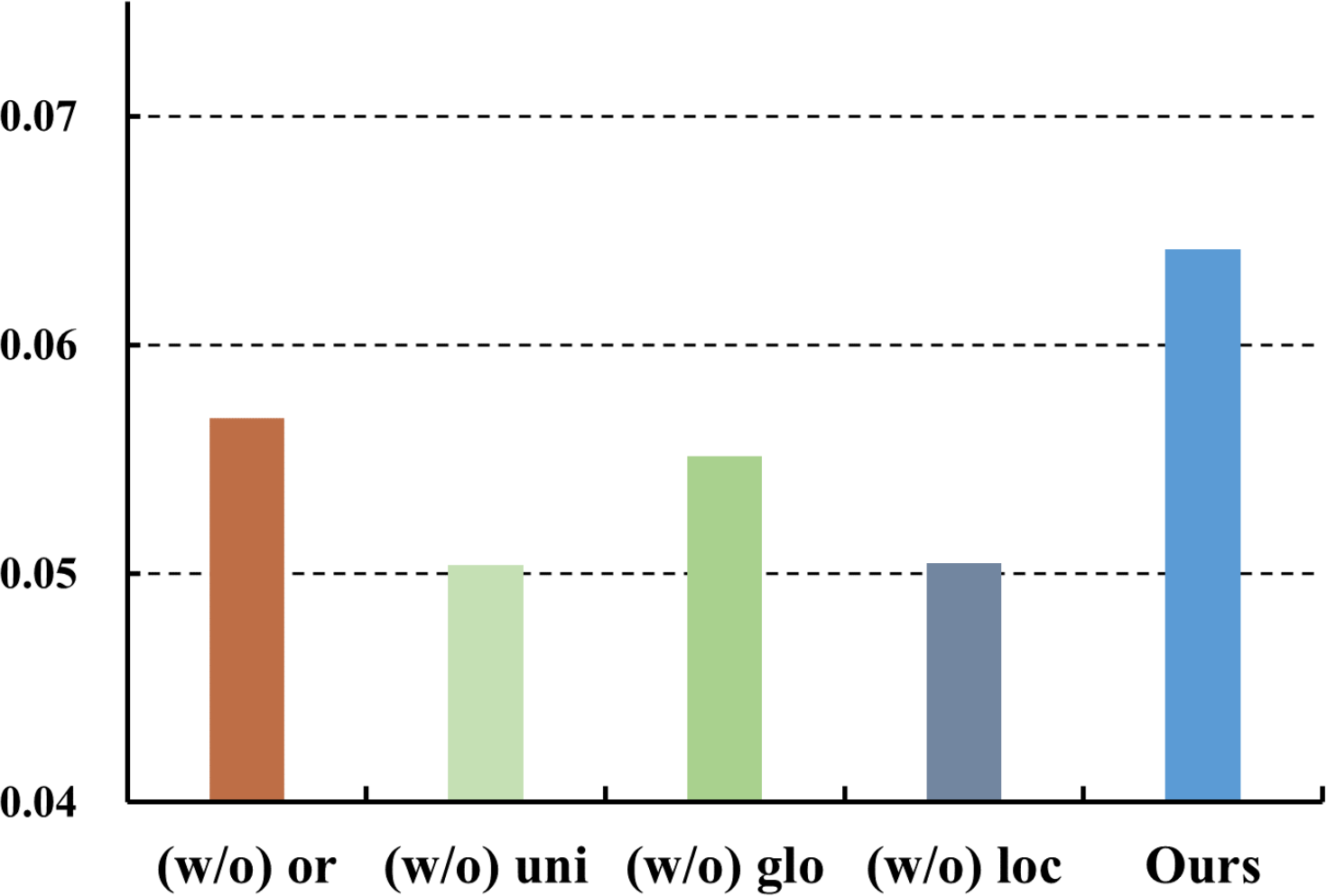}}
\centerline{\textbf{DCCF-Steam}}
\vspace{5pt}
\centerline{\includegraphics[width=1.0\textwidth]{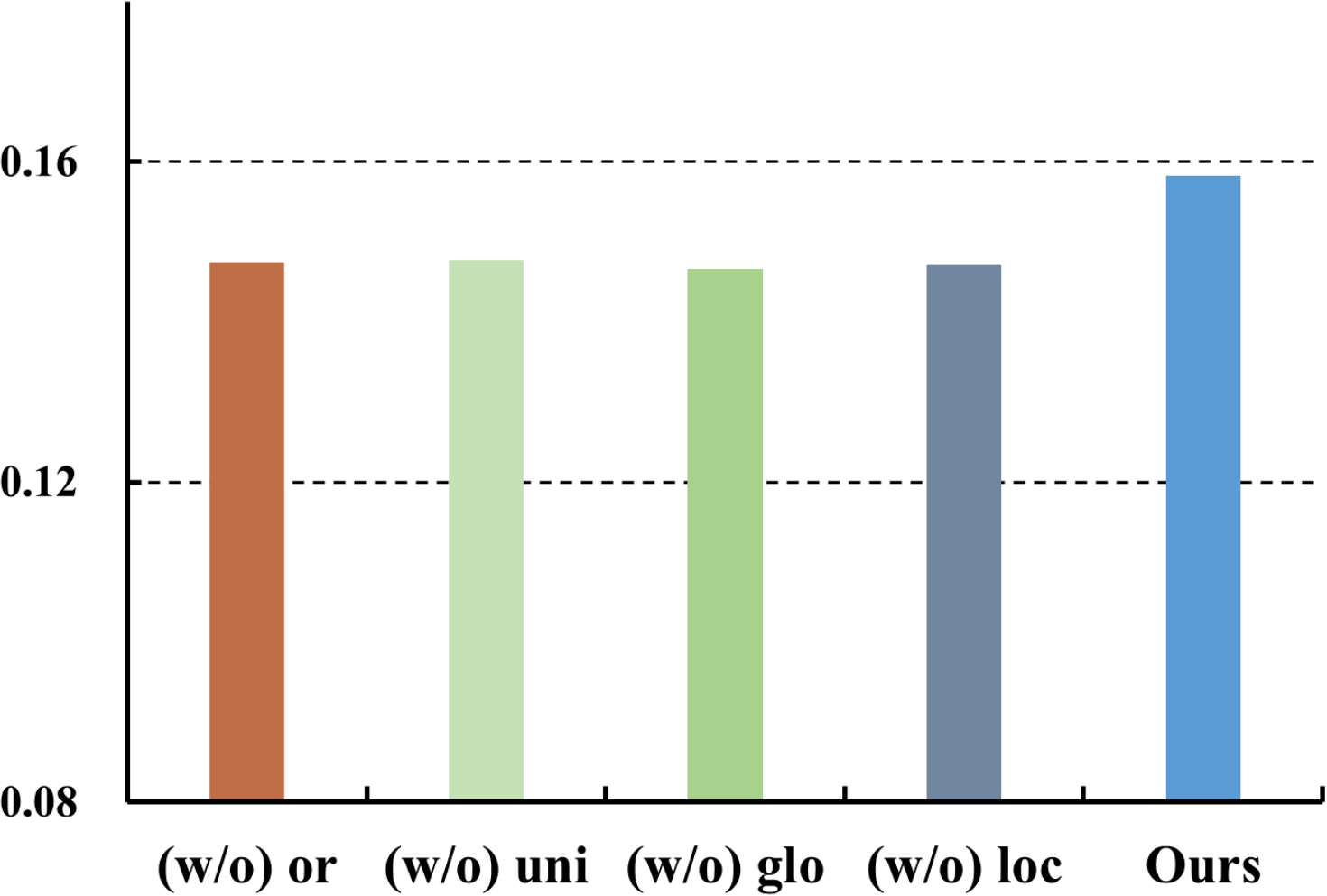}}
\centerline{\textbf{DCCF-Amazon}}
\vspace{5pt}
\end{minipage}
\caption{Ablation studies of our proposed method with four baselines in three datasets. The first row, the second row, the third row and the fouth row correspond with Recall@5, Recall@10, NDCG@5, NDCG@10 Metric, respectively.}
\label{ab_res}
\end{figure*}


\begin{itemize}
    \item GCCF empirically demonstrates that removing non-linearities improves recommendation performance. The authors design a residual network structure for collaborative filtering with user-item interaction modeling.
    \item LightGCN simplifies the design of Graph Convolutional Networks (GCNs) for recommendation tasks. It learns user and item embeddings through linear propagation operations on the user-item interaction graph. This simplification makes the model easier to implement and train.
    \item SGL explores self-supervised learning with a user-item graph. It generates augmented views through node dropout, edge dropout, and random walk. Theoretical analyses indicate that SGL can effectively mine hard negatives.
    \item SimGCL reveals that graph augmentation is important for recommendation performance. Instead of using complex data augmentations to the embeddings, SimGCL generates views in a simpler way.
    \item DCCF addresses two questions in graph contrastive recommendation: the oversight of user-item interaction behaviors and the presence of noisy information in data augmentation. It implements disentanglement for self-supervised learning in an adaptive manner.
    \item AutoCF designs a unified recommendation framework that automatically conducts data augmentation. It enhances the model’s discriminative capacity by employing contrastive learning strategies.
    \item RLMRec proposes a paradigm integrating Large Language Models (LLMs) with recommendation models. It aligns auxiliary textual information in the semantic space through cross-view alignment.
    \item KAR leverages comprehensive world knowledge by introducing factorization prompting.
\end{itemize}

\textbf{Evaluation Metrics.}
The recommendation performance is evaluated using two widely used metrics: Recall@K and NDCG@K. These metrics are applied under the all-ranking protocol \cite{DGCF}, which evaluates the top-K items selected from the entire set of items that were not interacted with by the users.

\textbf{Training Details.}
The experiments are conducted on the PyTorch deep learning platform with the 32G V100. For the baselines, we adopt their source with original settings. In our model, the learning rate is set to 1e-3 for all datasets and baselines with Adam optimizer. Following RLMRec\cite{RLMRec}, we combine the system prompt and the user/item profile to generate the prompt. Moreover, we utilize the GPT-3.5-turbo and text-embedding-ada-002 \cite{text-code} to generate the representations $\textbf{E}^{\textbf{L}}$. Moreover, we set the trade-off hyper-parameter $\lambda$ as $0.1$ for all datasets and baselines. The sampling number $\hat{N}$ is set to 4096 for all experiments. 


\begin{figure*}[h]
\centering
\small
\begin{minipage}{0.24\linewidth}
\centerline{\includegraphics[width=1.0\textwidth]{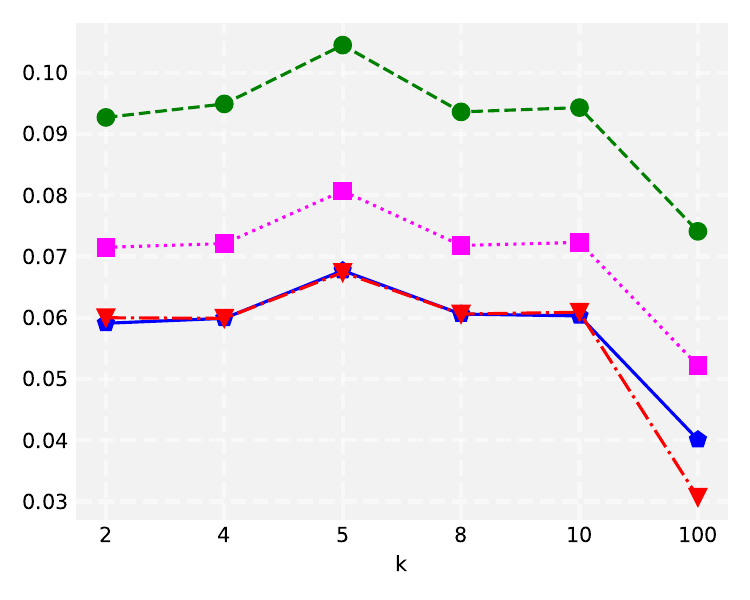}}
\centerline{\textbf{DCCF-Amazon-K}}
\vspace{5pt}
\centerline{\includegraphics[width=1.0\textwidth]{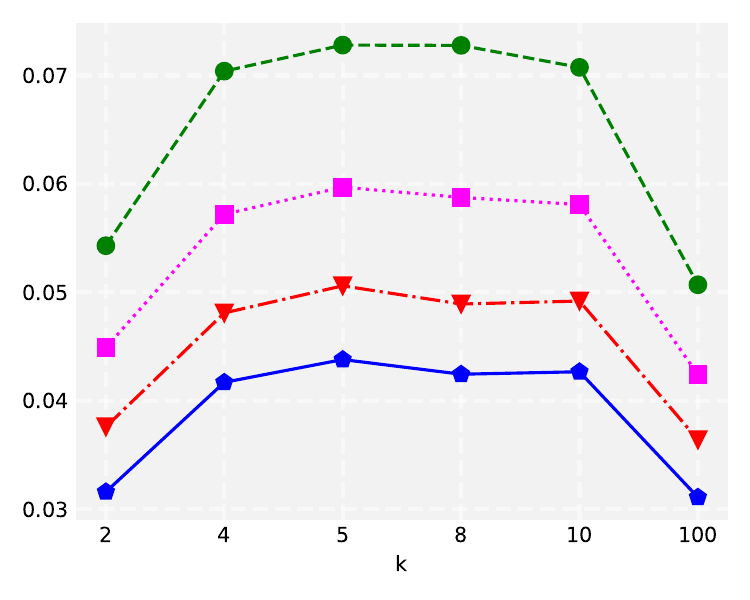}}
\centerline{\textbf{DCCF-Yelp-K}}
\vspace{5pt}
\centerline{\includegraphics[width=1.0\textwidth]{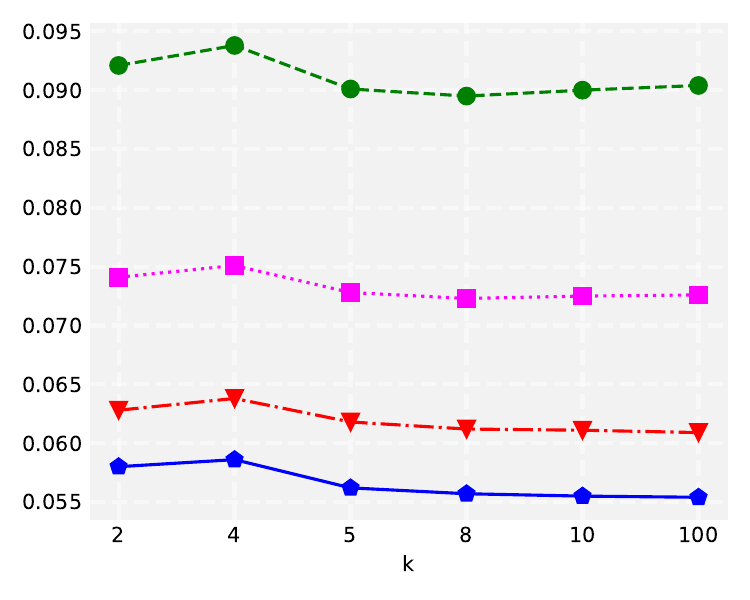}}
\centerline{\textbf{DCCF-Steam-K}}
\vspace{5pt}
\end{minipage}
\begin{minipage}{0.24\linewidth}
\centerline{\includegraphics[width=1.0\textwidth]{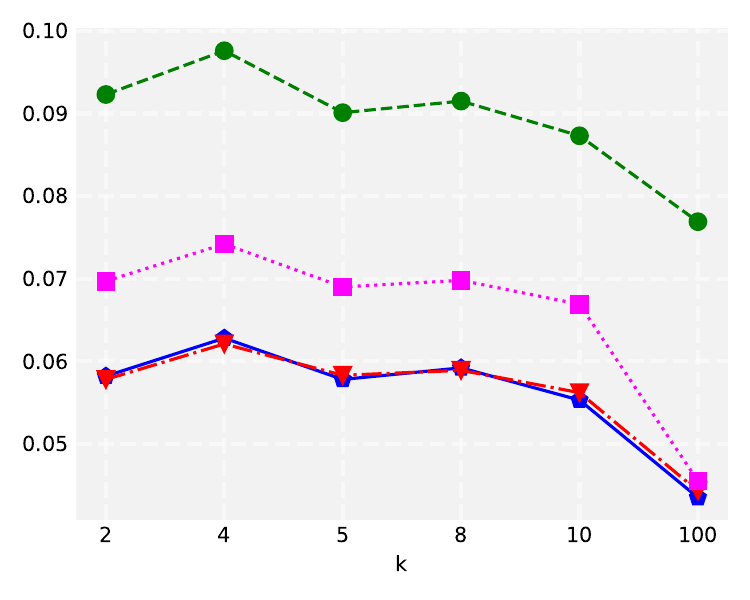}}
\centerline{\textbf{LightGCN-Amazon-K}}
\vspace{5pt}
\centerline{\includegraphics[width=1.0\textwidth]{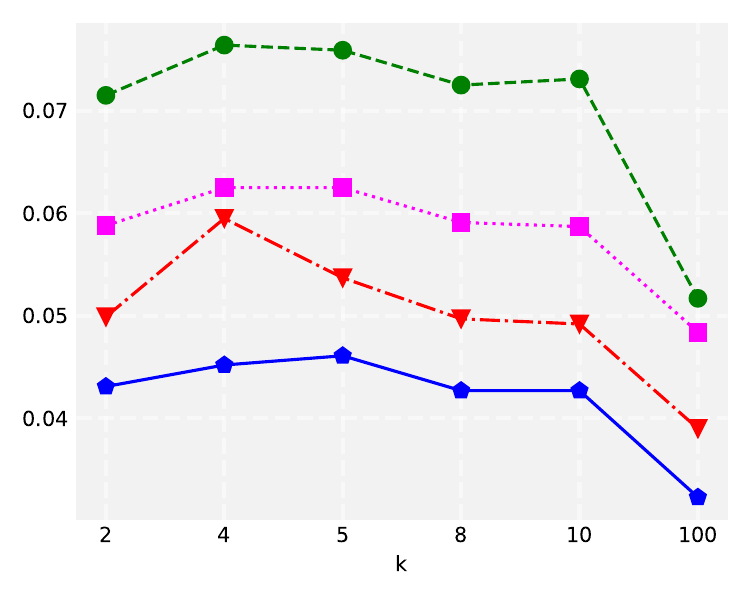}}
\centerline{\textbf{LightGCN-Yelp-K}}
\vspace{5pt}
\centerline{\includegraphics[width=1.0\textwidth]{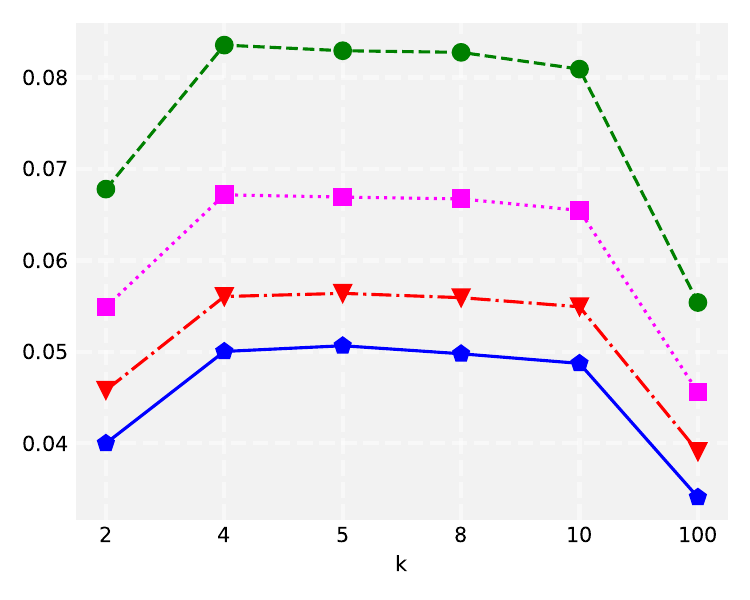}}
\centerline{\textbf{LightGCN-Steam-K}}
\vspace{5pt}
\end{minipage}
\begin{minipage}{0.24\linewidth}
\centerline{\includegraphics[width=1.0\textwidth]{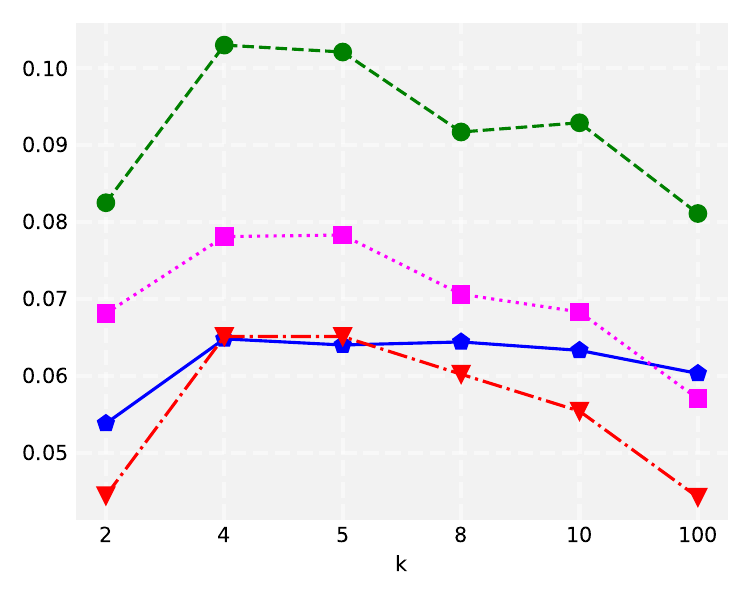}}
\centerline{\textbf{SimGCL-Amazon-K}}
\vspace{5pt}
\centerline{\includegraphics[width=1.0\textwidth]{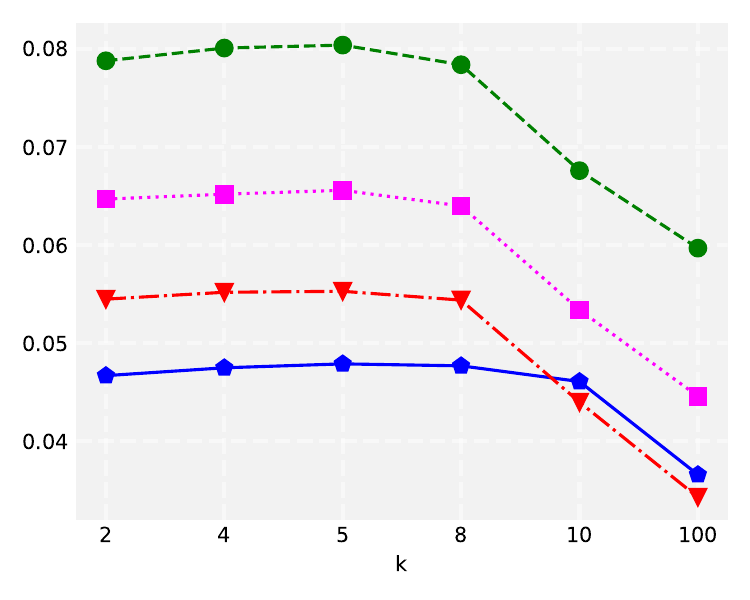}}
\centerline{\textbf{SimGCL-Yelp-K}}
\vspace{5pt}
\centerline{\includegraphics[width=1.0\textwidth]{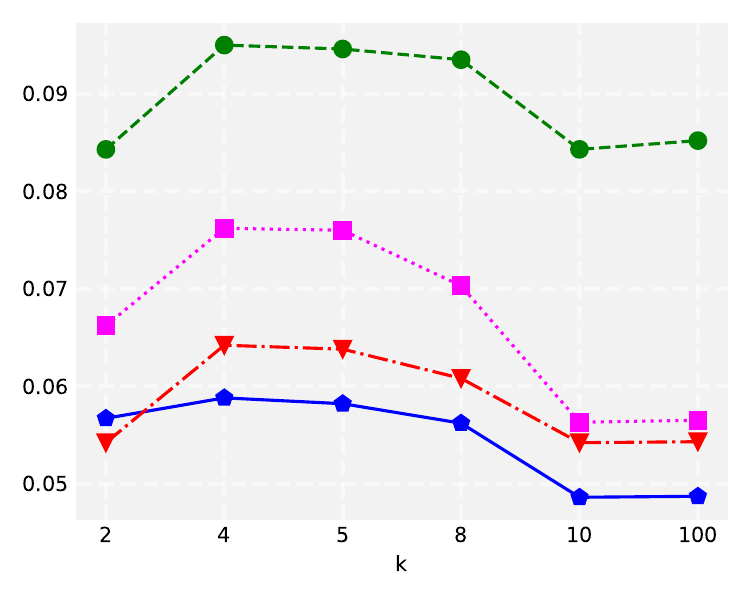}}
\centerline{\textbf{SimGCL-Steam-K}}
\vspace{5pt}
\end{minipage}
\begin{minipage}{0.24\linewidth}
\centerline{\includegraphics[width=1.0\textwidth]{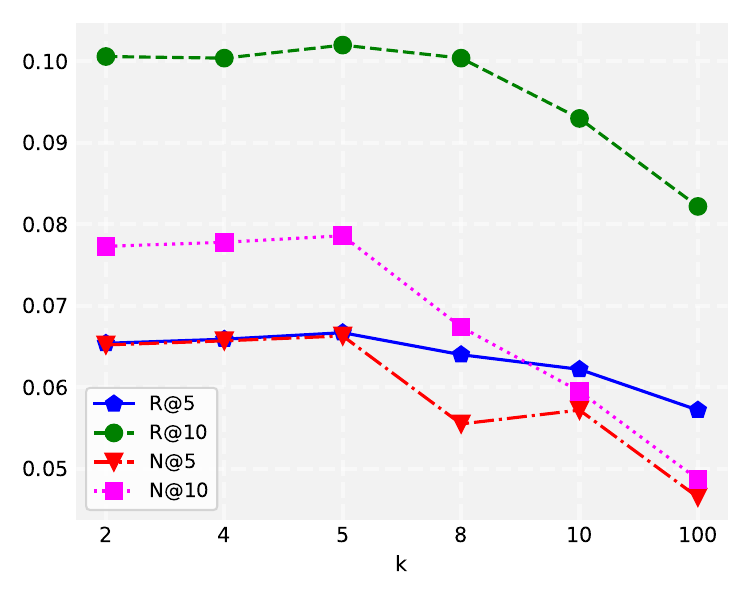}}
\centerline{\textbf{SGL-Amazon-K}}
\vspace{5pt}
\centerline{\includegraphics[width=1.0\textwidth]{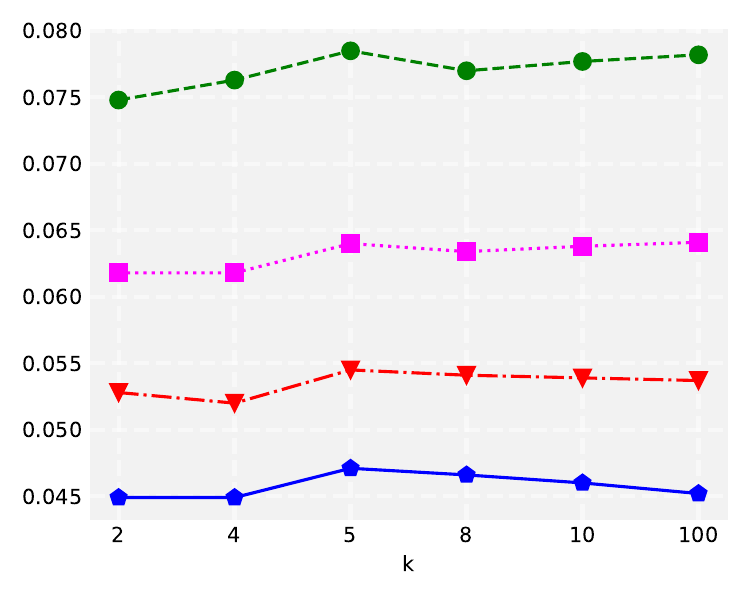}}
\centerline{\textbf{SGL-Yelp-K}}
\vspace{5pt}
\centerline{\includegraphics[width=1.0\textwidth]{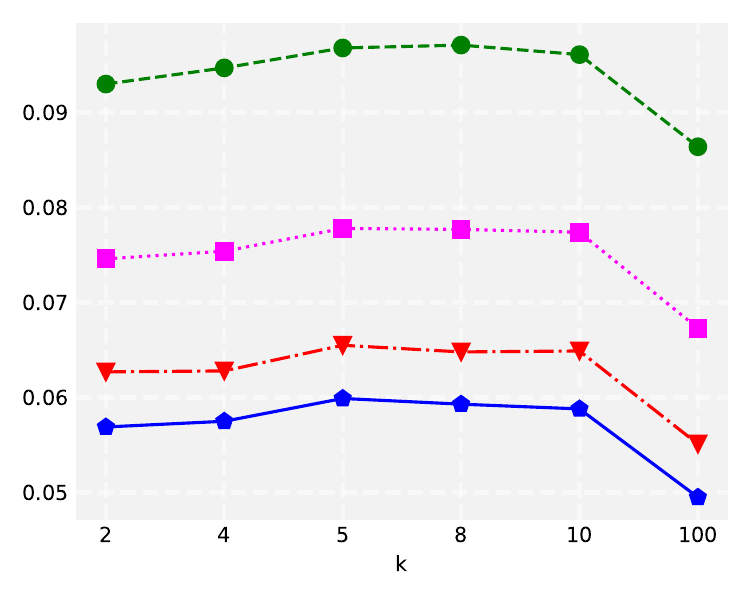}}
\centerline{\textbf{SGL-Steam-K}}
\vspace{5pt}
\end{minipage}
\vspace{-2pt}
\caption{Sensitive analysis with four baselines in three datasets for hyper-parameter $K$.}
\label{sen_k}
\end{figure*}

\subsection{Performance Comparison (\textbf{RQ1})}
To demonstrate the effectiveness and superiority of our proposed DaRec, in this subsection, we conduct experiments with nine state-of-the-art baselines on three datasets with six metrics. The compared algorithms can be roughly divided into two categories, i.e., traditional collaborative filtering methods (GCCF\cite{GCCF}, LightGCN \cite{LightGCN}, SGL\cite{SGL}, SimGCL\cite{SimGCL}, DCCF\cite{DCCF}, AutoCF\cite{autocf}), and LLMs-enhanced recommendation methods (RLMRec-Con\cite{RLMRec}, RLMRec-Gene \cite{RLMRec}, KAR\cite{KAR}). Here, RLMRec-Con and RLMRec-Gene denote two methods in RLMRec\cite{RLMRec}.

In this work, we design a plug-and-play disentangled framework for better aligning the collaborative models and LLMs. The results are shown in Table.\ref{com_res} and Table.\ref{com_res_l}. From the results, we could observe as follows. 

\begin{itemize}
    \item Compared with the traditional collaborative filtering methods (GCCF \cite{GCCF}, LightGCN \cite{LightGCN}, SGL \cite{SGL}, SimGCL \cite{SimGCL}, DCCF \cite{DCCF}, AutoCF \cite{autocf}), our proposed DaRec could achieve better recommendation performance. The reason we analyze this is that the representations are enhanced by the LLMs, leading to more semantic information for the representations.
    \item LLMs-enhanced recommendation methods (RLMRec \cite{RLMRec} and KAR \cite{KAR}) achieve sub-optimal recommendation performance compared with our proposed method. We conjecture that we could perform a better alignment for collaborative models and LLMs with our disentangled alignment strategy. 
    \item Our proposed DaRec outperforms other recommendation methods in three datasets with six metrics. Taking the results of AutoCF on the Yelp dataset for example, with our plug-and-play framework, DaRec improves the AutoCF to exceed the second-best recommendation method by margins of 3.85\%, 1.57\%, 3.15\%, 2.07\% in R@5, R@10, N@5, and N@10, respectively.
\end{itemize}


\begin{figure*}[h]
\centering
\small
\begin{minipage}{0.24\linewidth}
\centerline{\includegraphics[width=1.0\textwidth]{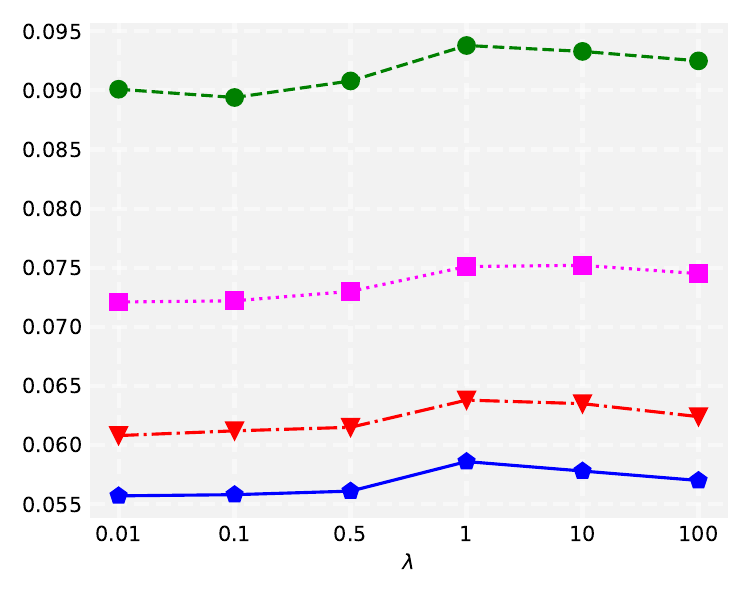}}
\centerline{\textbf{DCCF-Steam-Trade}}
\vspace{5pt}
\centerline{\includegraphics[width=1.0\textwidth]{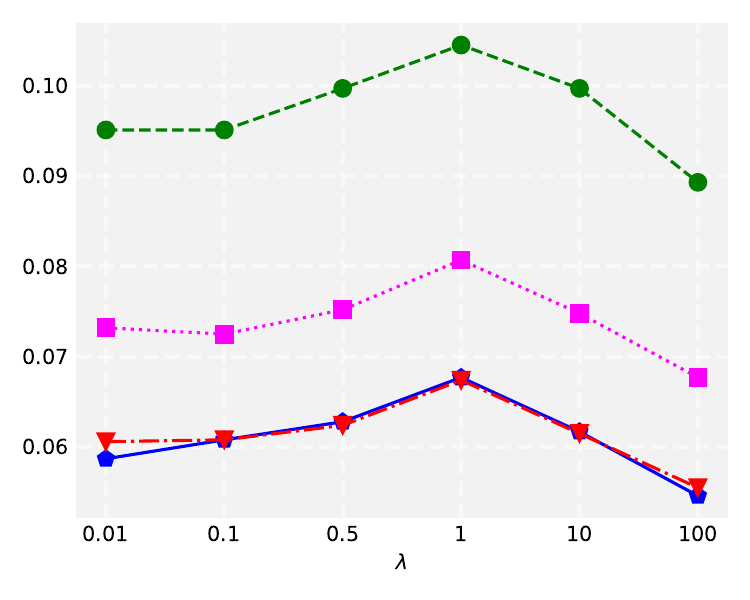}}
\centerline{\textbf{DCCF-Amazon-Trade}}
\vspace{5pt}
\centerline{\includegraphics[width=1.0\textwidth]{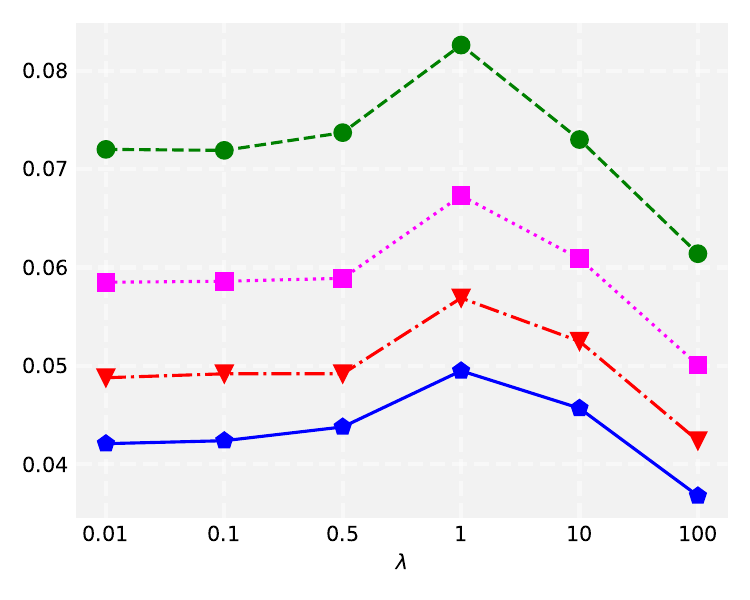}}
\centerline{\textbf{DCCF-Yelp-Trade}}
\vspace{5pt}
\end{minipage}
\begin{minipage}{0.24\linewidth}
\centerline{\includegraphics[width=1.0\textwidth]{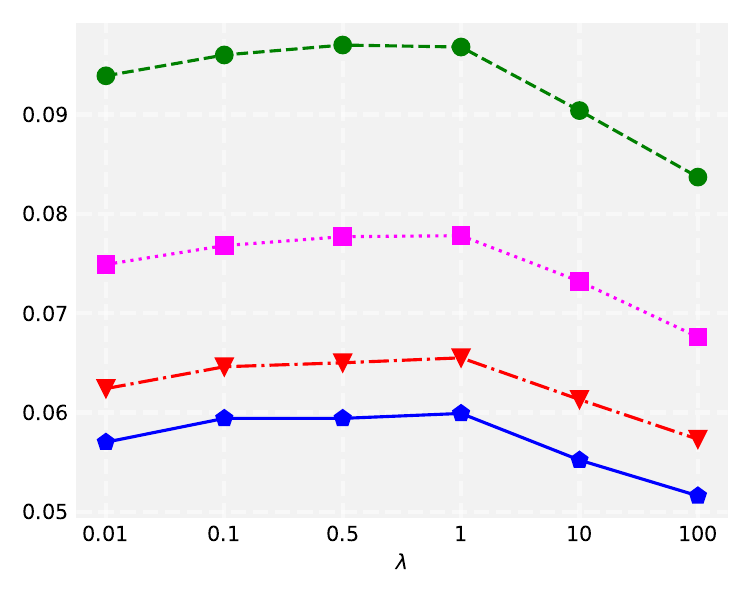}}
\centerline{\textbf{SGL-Steam-Trade}}
\vspace{5pt}
\centerline{\includegraphics[width=1.0\textwidth]{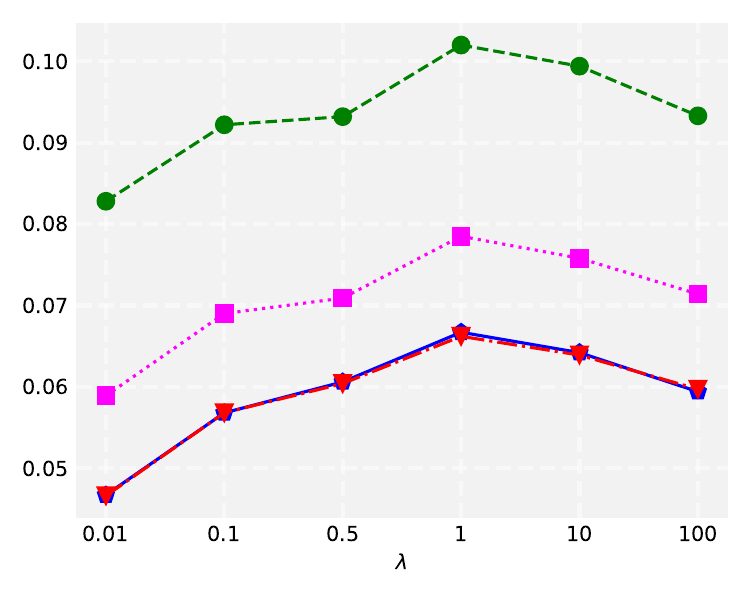}}
\centerline{\textbf{SGL-Amazon-Trade}}
\vspace{5pt}
\centerline{\includegraphics[width=1.0\textwidth]{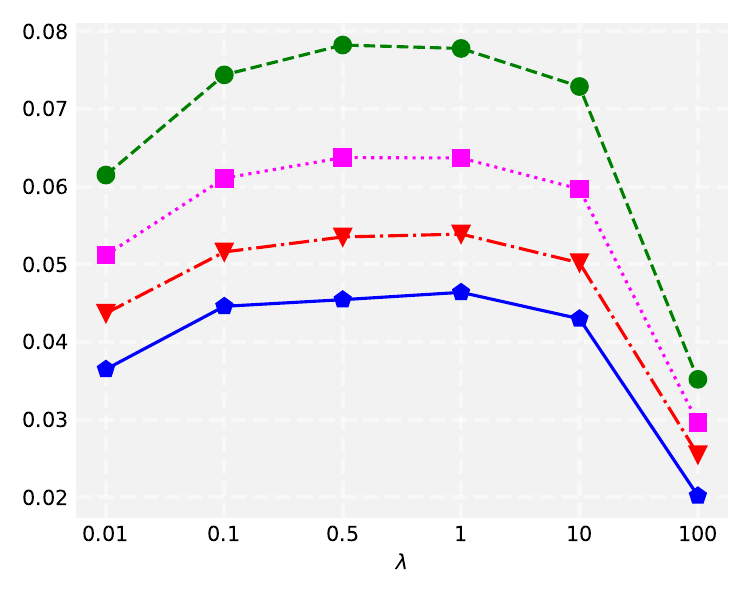}}
\centerline{\textbf{SGL-Yelp-Trade}}
\vspace{5pt}
\end{minipage}
\begin{minipage}{0.24\linewidth}
\centerline{\includegraphics[width=1.0\textwidth]{Figure/sen/new/dccf_steam_trade.pdf}}
\centerline{\textbf{DCCF-Steam-Trade}}
\vspace{5pt}
\centerline{\includegraphics[width=1.0\textwidth]{Figure/sen/new/dccf_amazon_trade.pdf}}
\centerline{\textbf{DCCF-Amazon-Trade}}
\vspace{5pt}
\centerline{\includegraphics[width=1.0\textwidth]{Figure/sen/new/dccf_yelp_trade.pdf}}
\centerline{\textbf{DCCF-Yelp-Trade}}
\vspace{5pt}
\end{minipage}
\begin{minipage}{0.24\linewidth}
\centerline{\includegraphics[width=1.0\textwidth]{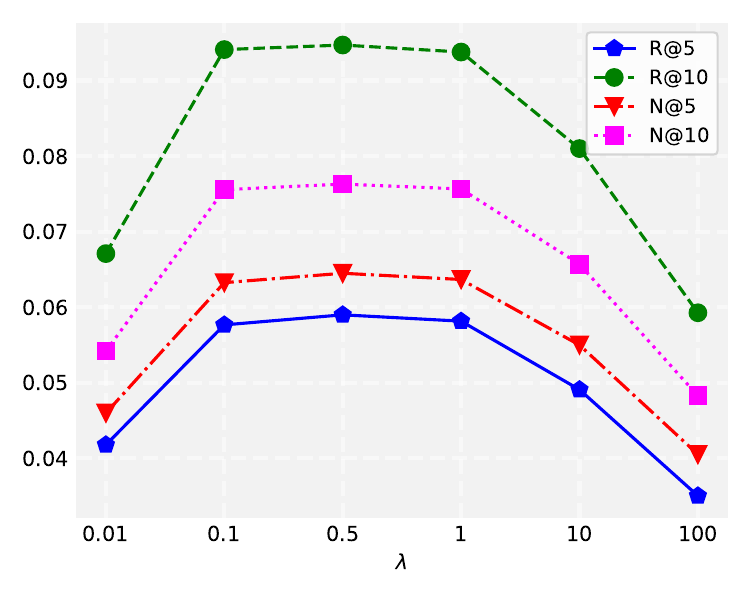}}
\centerline{\textbf{SimGCL-Steam-Trade}}
\vspace{5pt}
\centerline{\includegraphics[width=1.0\textwidth]{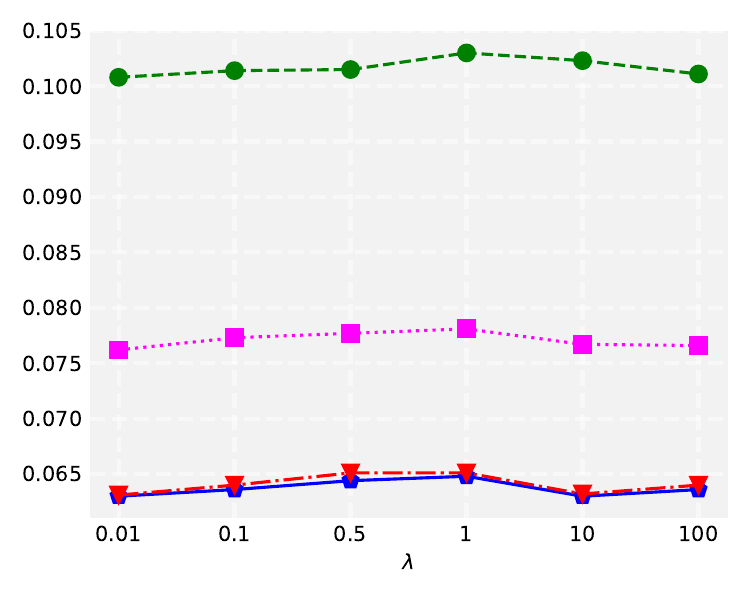}}
\centerline{\textbf{SimGCL-Amazon-Trade}}
\vspace{5pt}
\centerline{\includegraphics[width=1.0\textwidth]{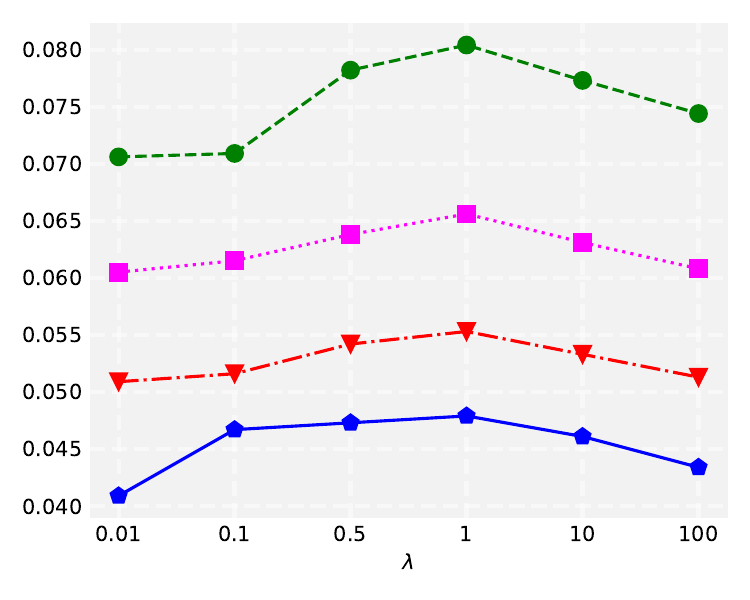}}
\centerline{\textbf{SimGCL-Yelp-Trade}}
\vspace{5pt}
\end{minipage}
\vspace{-5pt}
\caption{Sensitive analysis with four baselines in three datasets for hyper-parameter trade-off parameter $\lambda$, respectively.}
\label{sen_trade}
\end{figure*}

\subsection{Ablation Study (\textbf{RQ2})}
Our proposed method contains the orthogonal loss, the uniformity loss, the global loss, and the local loss. In this subsection, we conduct ablation studies to verify the effectiveness of our designed modules. To be specific, we utilize ``(w/o) or'', ``(w/o) uni'', ``(w/o) glo'', and ``(w/o) loc'' to denote reduced models by individually removing the orthogonal loss, the uniformity loss, the global loss, and the local loss. The results are shown in Fig.\ref{ab_res}. From the results, we could observe that the removal of any of the designed losses leads to a noticeable decline in recommendation performance, indicating that each loss contributes to the overall performance. We further analyze the reasons as follows.

\begin{figure}[]
\centering
\small
\begin{minipage}{0.45\linewidth}
\centerline{\includegraphics[width=1\textwidth]{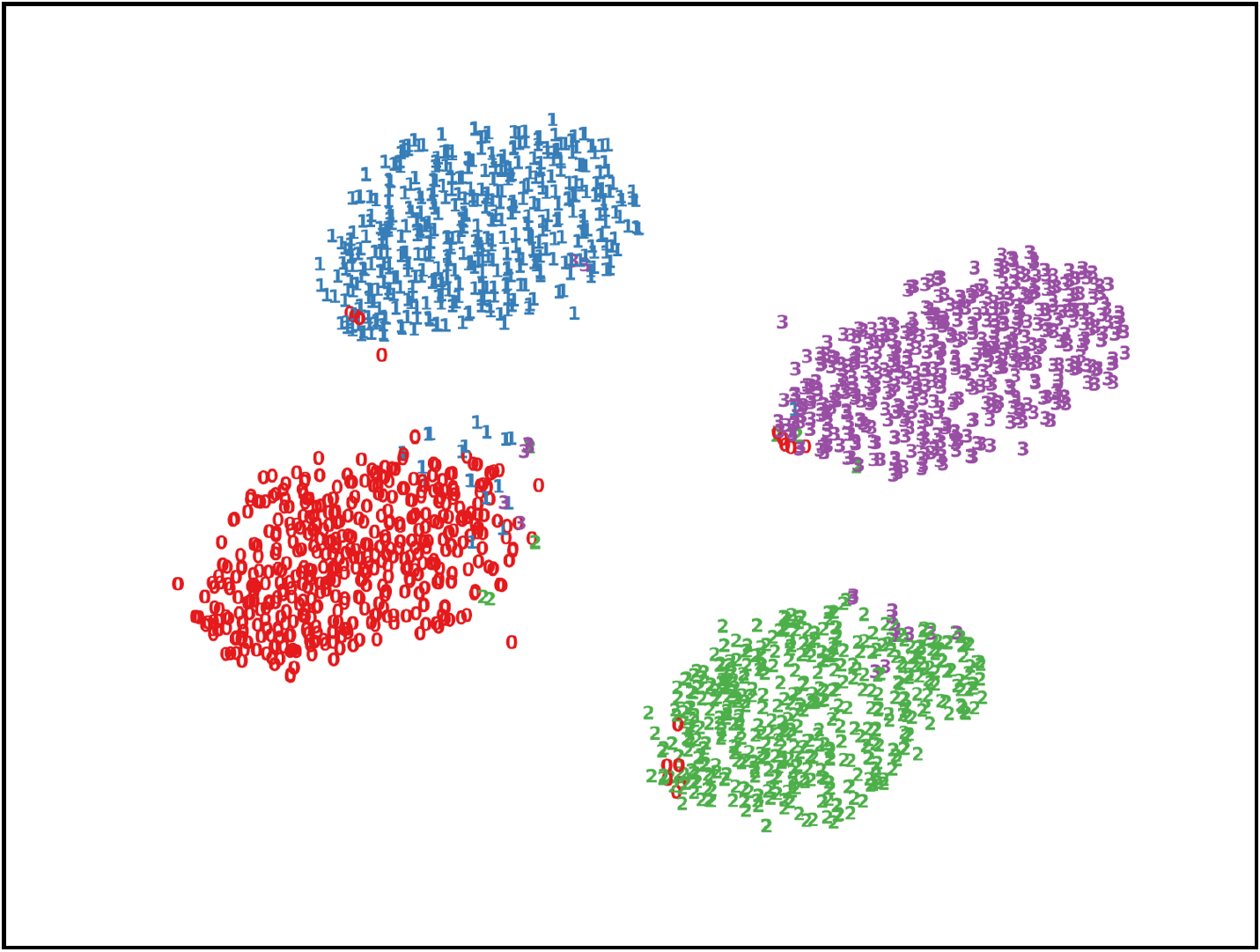}}
\vspace{3pt}
\textbf{\centerline{LLMs-Steam}}
\end{minipage}\hspace{4mm}
\begin{minipage}{0.45\linewidth}
\centerline{\includegraphics[width=\textwidth]{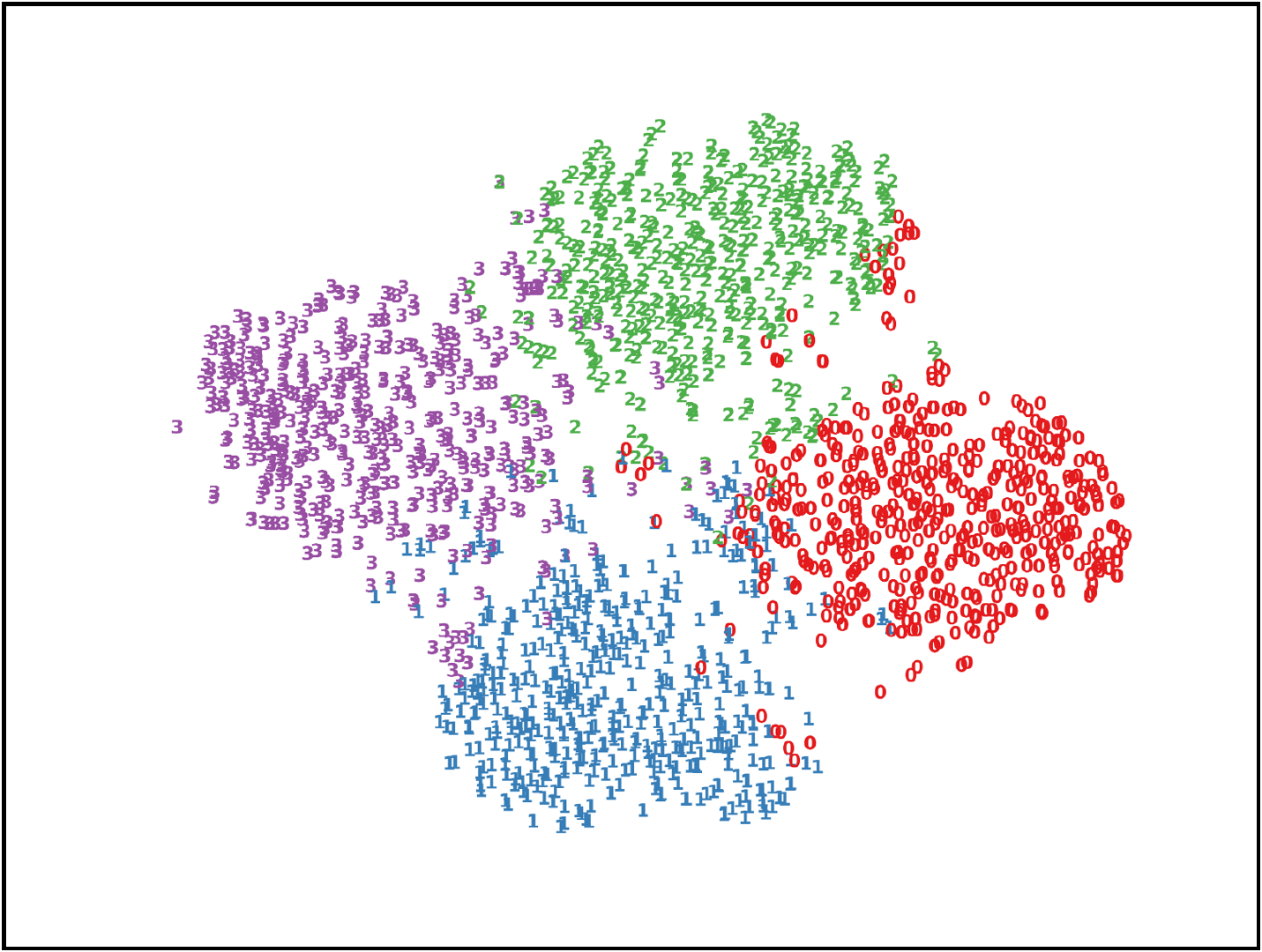}}
\vspace{3pt}
\textbf{\centerline{LightGCN-Steam}}
\end{minipage}
\caption{2D $t$-SNE visualization of the shared representation on Steam dataset from LLMs and LightGCN\cite{LightGCN}.}
\label{vis_res}
\end{figure}

\begin{itemize}
    \item Instead of exactly aligning all representations from collaborative models and LLMs, we disentangle the representation into two components, i.e., specific and shared representation. The orthogonal loss and the uniformity loss could effectively keep informative. 
    \item The global and local structure alignment strategies could better transfer the semantic knowledge from LLMs to collaborative models. Compared with the previous alignment strategy, our designed structure methods could benefit the model to obtain better performance by modeling the structure of the representations.
\end{itemize}


\subsection{Hyper-parameter Analysis (\textbf{RQ3})}

\subsubsection{Sensitivity Analysis of Cluster Number $K$}
In this subsection, we conduct experiments to evaluate the influence of the parameter $K$, which represents the number of preference centers. We varied the value of $K$ within the range of $\{2, 4, 5, 8, 10, 100\}$. The results are shown in Fig. \ref{sen_k}. Based on the results, we have the following observations. 

\begin{itemize}
    \item The model achieve best recommendation performance when $K$ is in $[4, 8]$. When $K$ takes extreme values, e.g., $K=100$, the performance will decrease dramatically. We speculate that this is because the interest centers become too scattered, making it difficult to accurately reflect the true preferences of users. 
    \item A similar situation occurs when $K=2$, where having too few interest centers fails to effectively capture the diverse preferences of users.
\end{itemize}

\subsubsection{Sensitivity Analysis of trade-off hyper-parameters}
Furthermore, we conduct experiments to evaluate the robustness of our proposed DaRec for the trade-off parameter $\lambda$. Here, we investigated the values of trade-off parameters in the range of $\{0.01, 0.1, 0.5, 1.0, 10, 100\}$. The experimental results are shown in Fig. \ref{sen_trade}. We could obtain the following observations.

\begin{itemize}
    \item When the value of trade-off is set to extreme values, e.g., $0.01$ or $100$, the recommendation performance tends to decrease. Extreme values can disrupt the balance between different loss components.
    \item The collaborative models achieve promising performance when the trade-off values in $[0.1, 1.0]$.
\end{itemize}

\subsubsection{Sensitivity Analysis of sampling size $\hat{N}$}\label{complex_res}

\begin{figure}[]
\centering
\small
\begin{minipage}{0.45\linewidth}
\centerline{\includegraphics[width=1\textwidth]{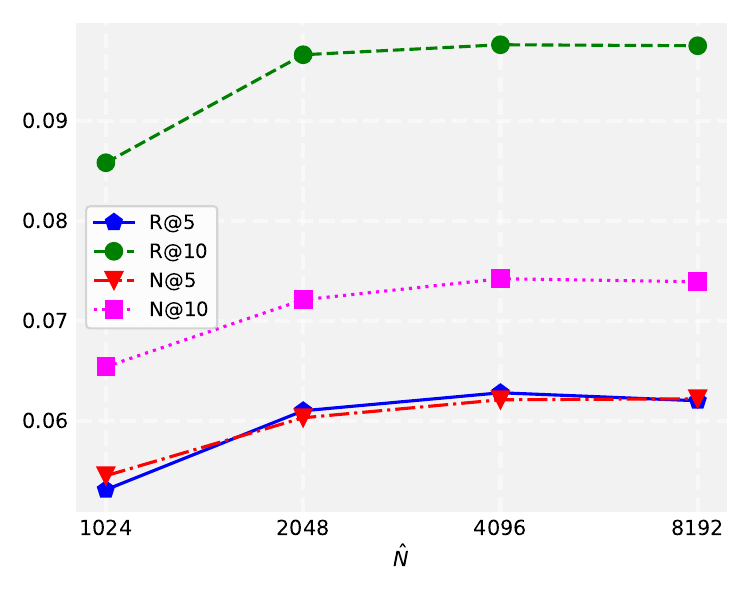}}
\vspace{3pt}
\textbf{\centerline{LightGCN-Amazon}}
\end{minipage}\hspace{4mm}
\begin{minipage}{0.45\linewidth}
\centerline{\includegraphics[width=\textwidth]{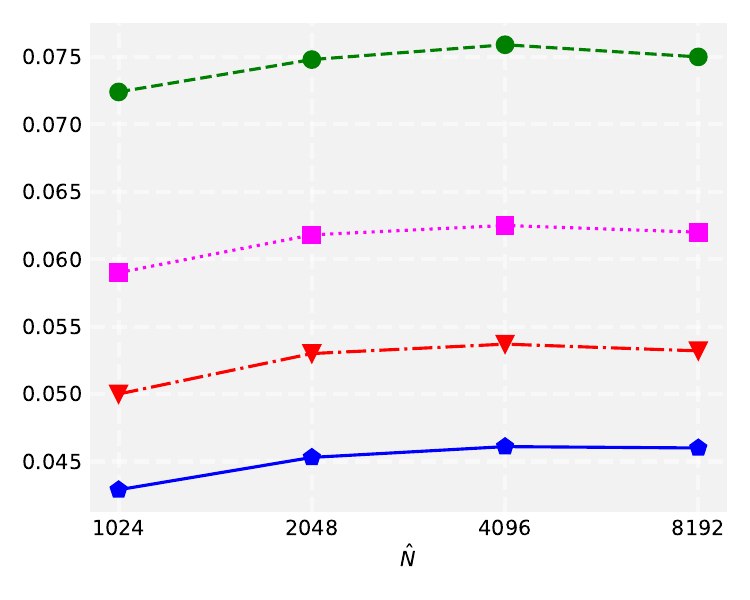}}
\vspace{3pt}
\textbf{\centerline{LightGCN-Yelp}}
\end{minipage}
\caption{Sensitive analysis for the sampling number $\hat{N}$.}
\label{sen_n}
\end{figure}

Moreover, in this subsection, we implement experiments to verify the influence of the sampling number $\hat{N}$ on recommendation performance. The experimental results are shown in Fig.\ref{sen_n}. For our experimental setup, we employ LightGCN \cite{LightGCN} as the backbone and utilized datasets from Amazon and Yelp to implement the experiments. We explore the values of sampling number within the range of $\{1024, 2048, 4096, 8192\}$. From the results, we could observe as follows. 


\begin{itemize}
    \item When the sampling number is set to a lower value, such as $\hat{N} = 1024$, the recommendation performance is suboptimal. We attribute this to the fact that a small sample size fails to accurately approximate the distribution of the entire dataset.
    \item The recommendation performance stabilizes when the sampling number $\hat{N}$ is within the range of $[4096, 8192]$. To balance performance and computational efficiency, we have opted to set the sampling number to 4096 for all subsequent experiments.
\end{itemize}

\subsection{Visualization Analysis (\textbf{RQ4})}
In this subsection, we conduct visualization analysis to demonstrate the user preference, i.e., the inherent interest clustering structure. To be specific, we utilize the $t$-SNE algorithm \cite{T_SNE} to show the clustering results. We perform $t$-SNE on the representation $\textbf{E}^\textbf{C}$ and $\textbf{E}^\textbf{L}$ from collaborative models and LLMs, repectively. Here, we use the LightGCN \cite{LightGCN} as the collaborative model to obtain the $\textbf{E}^\textbf{C}$. The visualization results are shown in Fig.\ref{vis_res}, we can observe that our proposed DaRec approach successfully captures and represents the underlying interest clusters.


\section{Case Study}
\begin{figure}[]
\centering
\scalebox{0.3}{
\includegraphics{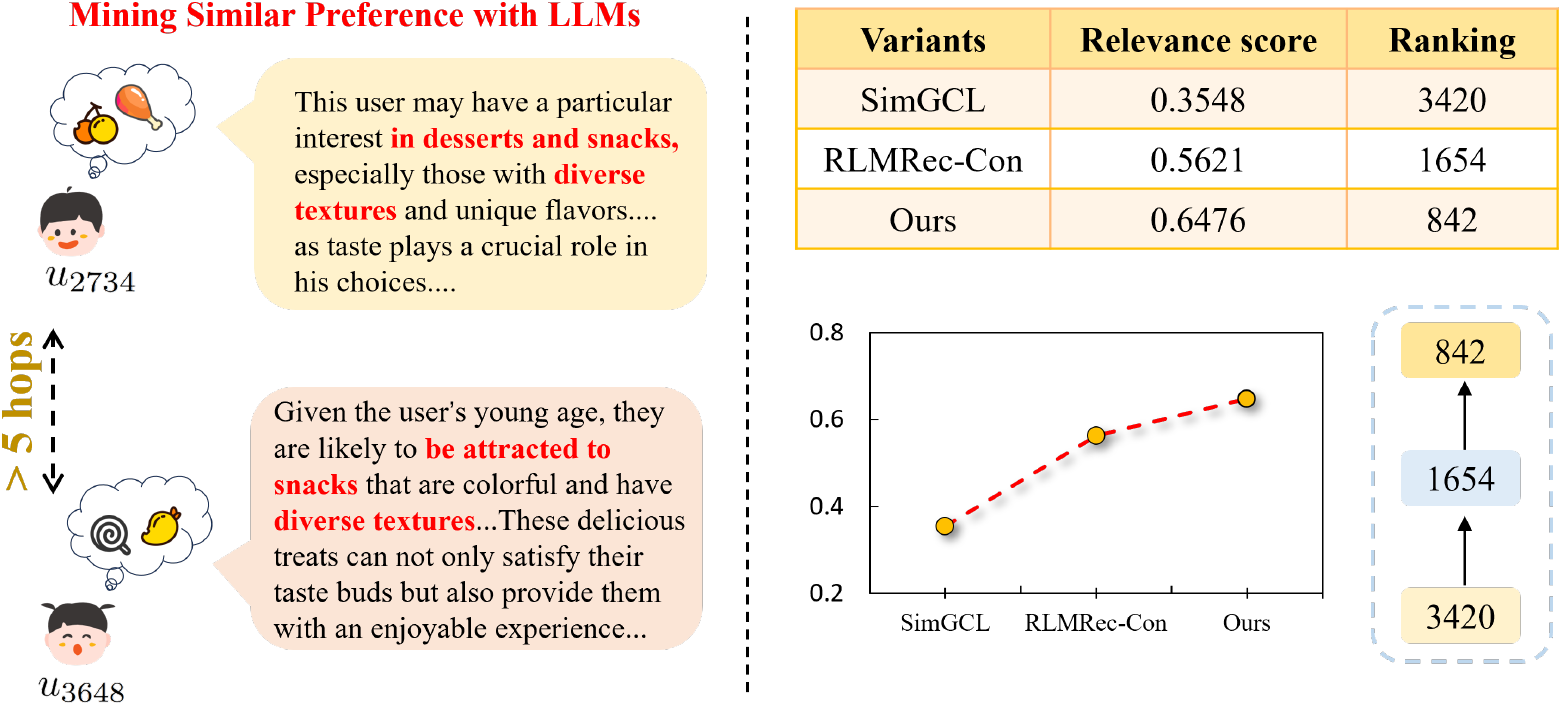}}
\caption{{Case study to demonstrate the ability on capturing global user dependencies.}}
\label{case_study}
\vspace{-10pt}
\end{figure}

In this section, we conduct a case study to demonstrate the effectiveness of our DaRec framework. We explore how LLMs enhance the semantic features of collaborative models through our designed alignment framework. Specifically, we leverage the model’s ability to capture global user dependencies. We focus on users who are separated by multiple hops ($>$ 5 hops) in the network. To evaluate the model’s ability to capture these global relationships, we calculate the similarity of user representations. For this purpose, we adopt SimGCL \cite{SimGCL}, RLMRec-Con \cite{RLMRec}, and our DaRec as baselines, all employing the same backbone. The dataset used for this study is Yelp. The relationships are evaluated using two metrics: relevance score and the ranking of long-distance neighbors based on this score. The relevance score is determined using the cosine similarity function. The case study is presented in Fig.~\ref{case_study}. In this scenario, we focus on user $u_{2734}$ and user $u_{3648}$. From the results, we observe that with our designed alignment framework DaRec, the semantic information are better aligned between $u_{2734}$ and $u_{3648}$, i.e., "snacks" and  "diverse textures". The relevance score and the ranking are increasing. This demonstrates that the learned representations from our DaRec capture global collaborative relationships beyond other recommendation methods.

\section{Related Work}

\subsection{GNN-based Recommendation}
Within the realm of recommender systems, collaborative filtering stands as a cornerstone technology, exerting a significant influence on the operation of these systems. Existing methods always utilize Graph Neural Networks (GNNs), such as LightGCN \cite{LightGCN}, NGCF \cite{NGCF} and GCCF \cite{GCCF}, to model the historical user-item interactions, thereby facilitating the capture of more complex relationships. Nonetheless, the implicit feedback data from users frequently contains considerable noise, which can compromise the performance of these Graph Neural Network (GNN)-based methods \cite{SCGC,CCGC,Graphlearner,CONVERT,MGCN,Dealmvc,asymmetric}. In response to the aforementioned challenges, a self-supervised learning method, commonly referred to as contrastive learning, takes precedence. Representative approaches, such as SGL \cite{SGL}, LightGCL \cite{LightGCL}, and NCL \cite{NCL}, employ the contrastive augmented data to boost the robustness of the whole recommendations and take out more promising performance.

\subsection{Large Language Models}
As the adoption of LLMs \cite{GPT, GPT4} becomes more widespread, the challenge of how to efficiently adapt these models for recommender systems has emerged as a pivotal research focus within the recommendation community \cite{liao2024llara,du2024enhancing,wang2024llm}. Several researchers \cite{calrec, RLMRec, controlrec, ctrl} take a step forward to study how to integrate the powerful representation ability of large language models into the recommendation system by using the contrastive learning mentioned above. For example, RLMRec \cite{RLMRec} utilizes contrastive and generative alignment techniques to align CF-side relational embeddings with LLMs-side semantic representations, such strategic integration effectively combines the advantages of general recommenders with those of Language Models, creating a robust system that leverages the strengths of both. ControlRec \cite{controlrec} narrows the semantic gap between language models and general recommenders via two auxiliary contrastive objectives, enhancing the performance of the proposed model by improving the ability to integrate the two types of data sources. CTRL \cite{ctrl} handles tabular data and transformed textual data as two separate modalities, harnessing the power of contrastive learning for a more precise alignment and integration of knowledge. While the aforementioned methods have made noteworthy advancements, we have theoretically demonstrated that such methods, which depend solely on direct alignment, may produce unsatisfactory results. To address this issue, our approach employs a disentangled alignment strategy for both the collaborative models and LLMs. This implementation will lead to substantial enhancements in the performance of LLMs-based recommender systems.

\section{Conclusion}
In this work, we present a novel plug-and-play structure framework for aligning collaborative models and LLMs. We first theoretically analyze that reducing the gap to zero may not always lead to promising performance. Therefore, we disentangle the representation into two components, i.e., shared and specific parts. Moreover, we design a structure alignment strategy at both local and global levels to explore the structure of the shared representation. We further provide proof that the shared and specific representations obtained by our method contain more relevant and less irrelevant information with downstream recommendation tasks. Extensive experimental results on benchmark datasets show the effectiveness of our method.

\section*{Acknowledgment}

This work was supported by the National Key R\&D Program of China 2020AAA0107100, the Natural Science Foundation of China (project no. 62325604, 62276271, 62476281). 

\section{Proof of Theorem 1}\label{Proof_1}
\begin{proof}
Consider the joint mutual information, $I(\textbf{E}^{\textbf{C}}, \textbf{E}^{\textbf{L}}; Y)$. By the chain rule, we have the following decompositions:

\begin{equation}
\begin{aligned}
 I(\textbf{E}^{\textbf{C}}, \textbf{E}^{\textbf{L}}; Y) &= I(\textbf{E}^{\textbf{C}}; Y) + I(\textbf{E}^{\textbf{L}}; Y | \textbf{E}^{\textbf{C}})\\
 &= I(\textbf{E}^{\textbf{L}}; Y) + I(\textbf{E}^{\textbf{C}}; Y | \textbf{E}^{\textbf{L}}).     
\end{aligned}
\end{equation}

Since the collaborative model's representation $\textbf{E}^{\textbf{C}}$ and LLMs representation $\textbf{E}^{\textbf{L}}$ are exactly aligned by various strategies, e.g., contrastive learning, we have:

\begin{equation}
I(\textbf{E}^{\textbf{L}}; Y | \textbf{E}^{\textbf{C}}) = I(\textbf{E}^{\textbf{C}}; Y | \textbf{E}^{\textbf{L}}) = 0, 
\end{equation}

Therefore, 

\begin{equation}
I(\textbf{E}^{\textbf{C}}, \textbf{E}^{\textbf{L}}; Y) = I(\textbf{E}^{\textbf{L}}; Y) = I(\textbf{E}^{\textbf{C}}; Y). 
\end{equation}

On the other hand, by the celebrated data-processing inequality, we have:

\begin{equation}
\begin{aligned}
&I(\textbf{E}^{\textbf{C}}; Y) \leq I(\textbf{D}; Y),\\
&I(\textbf{E}^{\textbf{L}}; Y) \leq I(\textbf{D'}; Y).   
\end{aligned}
\end{equation}

Thus, we have the chain of inequalities:
\begin{equation}
\begin{aligned}
I(\textbf{E}^\textbf{C}, \textbf{E}^{\textbf{L}}; Y)&=\min\{I(\textbf{E}^\textbf{C}; Y), I(\textbf{E}^{\textbf{L}}; Y)\} \\
&\leq \min\{I(\textbf{D}; Y), I(\textbf{D'}; Y)\} \\
&\leq \max\{I(\textbf{D}; Y), I(\textbf{D'}; Y)\} \\
&\leq I(\textbf{D}, \textbf{D'}; Y),
\end{aligned}
\end{equation}
where the last inequality follows from the fact that joint mutual information $I(\textbf{D}, \textbf{D'}; Y)$ is at least as larger as any one of $I(\textbf{D}; Y)$ and $I(\textbf{D'}; Y)$. Thus, with the variational form of the conditional entropy, we have:
\begin{equation}
\begin{aligned}
&{\inf}_{h} {\mathbb{E}_{p}}[\ell_{CE}(h(\textbf{E}^\textbf{C}, \textbf{E}^{\textbf{L}}),Y)] - {\inf}_{h'}{\mathbb{E}_{p}}[\ell_{CE}(h'(\textbf{E}^\textbf{C}, \textbf{E}^{\textbf{L}}),Y)]
\\
&= H(Y | \textbf{E}^\textbf{C}, \textbf{E}^{\textbf{L}}) - H(Y | \textbf{D}, \textbf{D'}) \\
&= I(\textbf{D}, \textbf{D'}; Y) - I(\textbf{E}^\textbf{C}, \textbf{E}^{\textbf{L}}; Y)  \\
&\geq \max\{I(\textbf{D}; Y), I(\textbf{D'}; Y)\} - \min\{I(\textbf{D}; Y), I(\textbf{D'}; Y)\} \\
&= H(Y | \textbf{E}^\textbf{C}, \textbf{E}^{\textbf{L}}) - H(Y | \textbf{D}, \textbf{D'}) \\
&= \Delta_p.
\nonumber
\end{aligned}
\end{equation}

\end{proof}

\section{Proof of Theorem 2}\label{proof_2}
To prove Theorem \ref{theorem_better}, we define some notations, Let $\textbf{D}$ be the model input and $\textbf{E}_{sh}^*$ be the optimal shared representation in both collaborative models and LLMs. We first introduce the following lemmas:

\begin{lemma}\label{le_1}

For the input $\textbf{D}$, we have $\textbf{E}_{sh} = f_{sh}^{\textbf{D}}(\textbf{D}) = \rho(\textbf{E}_{sh}^*)$, where $\rho(\cdot)$ is an invertible function.
\end{lemma}

\begin{lemma}\label{le_2}
With the representations $\widehat{\textbf{E}}$ extracted by our DaRec and $\widetilde{\textbf{E}}$ extracted by previous methods in recommendation tasks $\textbf{R}$, we have:
\begin{equation}
\begin{aligned}
&I(\widehat{\textbf{E}}, \textbf{D'}, \textbf{R}) = I(\widetilde{\textbf{E}},\textbf{D'},\textbf{R}) = I(\textbf{D}, \textbf{D'}, \textbf{R}),\\
&H(\widehat{\textbf{E}}) - H(\widetilde{\textbf{E}}) = H(\widehat{\textbf{E}}|\textbf{D'}) - H(\widetilde{\textbf{E}}|\textbf{D'}),
\end{aligned}
\label{lemma}
\end{equation}
where $\textbf{D}$ and $\textbf{D'}$ are the two types for the collaborative models and LLMs, respectively. 
\end{lemma}

\textbf{Remark:} Through Lemma.\ref{le_1}, the optimal shared representation and the shared representation learned by our model can be transformed from each other with the invertibility function $\rho(\cdot)$. Therefore, we could extract the complete shared representation. Here we give the following proof for Lemma.\ref{le_1}.
\begin{proof}
In our method, we split the representation into specific and shared components, which denotes that shared representations from LLMs and collaborative models are exactly aligned, i.e., $\textbf{E}_{sh}^{L} = \textbf{E}_{sh}^{C}$, we have:
\begin{equation}
\begin{aligned}
\textbf{E}_{sh}^{L} &= \textbf{E}_{sh}^{C},\\
f_{sh}^{\textbf{D}}(\textbf{D}) &= f_{sh}^{\textbf{D'}}(\textbf{D'}) ,
\end{aligned}
\label{sh=sh}
\end{equation}
where $\textbf{D}$ and $\textbf{D'}$ are the input for collaborative models and LLMs. $f_{sh}^{\textbf{D}}(\cdot)$ and $f_{sh}^{\textbf{D}}(\cdot)$ indicate the encoder network to obtain the shared specific representation for collaborative models and LLMs. Here, we adopt the MLP as the backbone network for the encoder network. According Eq.\ref{loss_or}, the specific representation $\textbf{E}_{sp}$ and the shared representation $\textbf{E}_{sh}$ are expected to be independent. We assume that $f_{sh}^{\textbf{D}}(\cdot), f_{sh}^{\textbf{D'}}(\cdot)$ are invertible, and we utilize $g_{sh}^{D}$ to denote ${f_{sh}^{\textbf{D}}}^{(-1)}$. Besides, let $\textbf{E}_{sh}^*$ and $\textbf{E}_{sp}^{\textbf{D}*}, \textbf{E}_{sp}^{\textbf{D}'*}$ indicate the optimal shared and specific representations, which are also independent. With the encoder network $f_{sh}^{\textbf{D}}(\cdot)$ and $f_{sh}^{\textbf{D'}}(\cdot)$, we can transform Eq.\eqref{sh=sh} into:
\begin{equation}
\begin{aligned}
f_{sh}^{\textbf{D}}\left ( \left [ \begin{matrix}
\textbf{E}_{sh}^* \\ 
\\ \textbf{E}_{sp}^{\textbf{D}*}
\end{matrix} \right ]  \right ) 
=f_{sh}^{\textbf{D'}}\left ( \left [ \begin{matrix}
\textbf{E}_{sh}^* \\ 
\\ \textbf{E}_{sp}^{\textbf{D'}*}
\end{matrix} \right ]  \right ).
\end{aligned}
\end{equation}
%
Therefore, to prove the shared representation extracted function $f_{sh}^{\textbf{D}}(\cdot)$ can extract the complete shared information, we only have to demonstrate $f_{sh}(\cdot)$ is the function of only $\textbf{E}_{sh}^*$ but the not the function of $\textbf{E}_{sp}^*$. To this end, we calculate the Jacobian of $f_{sh}(\cdot)$ to analyze the first-order partial derivatives of $f_{sh}(\cdot)$ and $f_{sp}(\cdot)$ w.r.t. $\textbf{E}_{sh}^*$ and $\textbf{E}_{sp}^*$. Let $\theta^{\textbf{D}}$ as the $[\textbf{E}_{sh}^{*\mathsf{T}}, (\textbf{E}_{sp}^{\textbf{D}*})\mathsf{T}]\mathsf{T}$. The Jacobian matrices of $f_{sh}^{\textbf{D}}(\cdot)$ can be calculate as:
\begin{equation}
\begin{aligned}
\textbf{J}^{\textbf{D}} &= \begin{bmatrix}
 \textbf{J}_{11}^{\textbf{D}} &\textbf{J}_{12}^{\textbf{D}}
 \\
 \\
 \textbf{J}_{21}^{\textbf{D}}&\textbf{J}_{22}^{\textbf{D}}
\end{bmatrix},
\end{aligned}
\end{equation}
where the elements can be presented as:
\begin{equation}
\begin{aligned}
[\textbf{J}_{11}^{\textbf{D}}]_{i,j} = \frac{\partial[f_{sh}^{\textbf{D}}(\theta)^{\textbf{D}}]_i}{\partial\textbf{E}_{sh_{j}}^{\textbf{D}*}} &, [\textbf{J}_{12}^{\textbf{D}}]_{i,k} = \frac{\partial[f_{sh}^{\textbf{D}}(\theta)^{\textbf{D}}]_i}{\partial\textbf{E}_{sp_{k}}^{\textbf{D}*}},\\
[\textbf{J}_{21}^{\textbf{D}}]_{k,i} = \frac{\partial[f_{sp}^{\textbf{D}}(\theta)^{\textbf{D}}]_k}{\partial\textbf{E}_{sh_{i}}^{*\textbf{D}}}&, [\textbf{J}_{22}^{\textbf{D}}]_{k,l} = \frac{\partial[f_{sp}^{\textbf{D}}(\theta)^{\textbf{D}}]_k}{\partial\textbf{E}_{sp_{l}}^{\textbf{D}*}},
\end{aligned}
\end{equation}
where $\textbf{J}_{11}^{\textbf{D}} \in \mathbb{R}^{N \times N}$, $\textbf{J}_{12}^{\textbf{D}} \mathbb{R}^{N \times N}$, $\textbf{J}_{21}^{\textbf{D}} \in \mathbb{R}^{n\times N}$ and $\textbf{J}_{22}^{\textbf{D}} \in \mathbb{R}^{n\times n}$. $i, j \in [1,N]$ and $k,l \in [1,n]$. After that, we only have to proof $\textbf{J}_{12}$ is an all-zero matrix while the determinant of $\textbf{J}_{11}^{\textbf{D}}$ is non-zero to show that the matrix consisting of all the partial derivatives of $f_{sh}^{\textbf{D}}(\cdot)$ w.r.t. $\textbf{E}_{sh}^*$ is full rank while any partial derivatives of $f_{sh}^{\textbf{D}}(\cdot)$ w.r.t. $\textbf{E}_{sp}^{\textbf{D}*}$ is zero. With any fixed $\bar{\textbf{E}}_{sh}^*$ and $\bar{\textbf{E}}_{sp}^{\textbf{D'}*}$, for all $\textbf{E}_{sh}^{\textbf{D}*}$, we have:
\begin{equation}
\begin{aligned}
f_{sh}^{\textbf{D}}\left ( \left [ \begin{matrix}
\bar{\textbf{E}}_{sh}^* \\ 
\\ \textbf{E}_{sp}^{\textbf{D}*}
\end{matrix} \right ]  \right ) 
=f_{sh}^{\textbf{D'}}\left ( \left [ \begin{matrix}
\bar{\textbf{E}}_{sh}^* \\ 
\\ \bar{\textbf{E}}_{sp}^{\textbf{D'}*}
\end{matrix} \right ]  \right ).
\end{aligned}
\label{bar_E}
\end{equation}
After that, we take the partial derivatives of Eq.\eqref{bar_E} with $\textbf{E}_{sp}^{\textbf{D}}$ for $j \in [1,n]$. Besides, we have $\textbf{J}_{12}^{\textbf{D}}|_{\bar{\textbf{E}}_{sh},\textbf{E}_{sp}^{\textbf{D}}}$ = $\textbf{J}_{12}^{\textbf{D'}}|_{\bar{\textbf{E}}_{sh},\bar{\textbf{E}}_{sp}^{\textbf{D'}}}$. According to the chain rules and taking derivatives of constants, we can obtain:
\begin{equation}
\begin{aligned}
\textbf{J}_{12}^{\textbf{D'}}|_{\bar{\textbf{E}}_sh, \bar{\textbf{E}}_{sp}^{\textbf{D'}}} = \left ( \textbf{J}_{f_{sh}^{\textbf{D'}}|_{\bar{\textbf{E}}_{sh}, \bar{\textbf{E}}_{sp}^{\textbf{D'}}}} \right ) \begin{bmatrix}
\textbf{0}_{N \times n} \\
 \\\textbf{0}_{N \times n}
\end{bmatrix} = \textbf{0}_{N \times n},
\end{aligned}
\end{equation}
where $\textbf{J}_{f_{sh}^{\textbf{D'}}} \in \mathbb{R}^{N \times (N+n)}$ is the Jacobian of $f_{sh}^{\textbf{D'}}$. The above proof is based on any fixed $\bar{\textbf{E}}_{sh}^*$ and $\bar{\textbf{E}}_{sp}^{\textbf{D}*}$. So, the same derivation holds for all $\textbf{E}_{sh}^*$ and $\textbf{E}_{sp}^{\textbf{D}*}$. Therefore, $\textbf{J}_{12}^{\textbf{D}}$ is an all-zero matrix and the learned $f_{sh}^{\textbf{D}}\theta^{\textbf{D}}$.
\end{proof}
\vspace{-10pt}

Based on the proof of Lemma.\ref{le_1}, we give the proof of Lemma.\ref{le_2} as follows.
\vspace{-7pt}
\begin{proof}
According to the proof of Lemma.\ref{le_1}, our proposed method could obtain the complete shared information for two types input $\textbf{D}$ and $\textbf{D'}$. Therefore, we have:
\begin{equation}
I(\widehat{\textbf{E}}^{\textbf{D}}, \textbf{D'}) = I(\textbf{D}, \textbf{D'}).
\label{le2_1}
\end{equation}

Most alignment strategies adopt contrastive learning, which maximizes the mutual information for collaborative models and LLMs. Assume previous contrastive learning methods could obtain complete information, thus we have:

\begin{equation}
I(\widetilde{\textbf{E}}^{\textbf{D}}, \textbf{D'}) = I(\textbf{D}, \textbf{D'}).
\label{le2_2}
\end{equation}

Following the previous works \cite{
wang2022rethinking}, if the random variable $c$ is observed, the random variable $a$ is conditionally independent of any other variable $b$, we assume that $I(a,b|c)=0,\forall b$. Thus, we have:
\begin{equation}
\begin{aligned}
&\quad I(\textbf{D}, \textbf{D'}, \textbf{R}) - I(\widehat{\textbf{E}}^\textbf{D}, \textbf{D'}, \textbf{R})\\ 
&= [I(\textbf{D}, \textbf{D'} - I(\textbf{D}, \textbf{D'} \mid \textbf{R})] - [I(\widehat{\textbf{E}}^\textbf{D}, \textbf{D'}) - I(\widehat{\textbf{E}}^\textbf{D}, \textbf{D'} \mid \textbf{R})] \\
&= [I(\widehat{\textbf{E}}^\textbf{D}, \textbf{D'} \mid \textbf{R}) - I(\textbf{D}, \textbf{D'} \mid \textbf{R})] \\
&=[H(\textbf{D'} \mid \textbf{R}) - H(\textbf{D'}|\textbf{R})]-[H(\textbf{D'} \mid \textbf{R})-H(\textbf{D'} \mid \textbf{D}, \textbf{R})]\\
&= H(\textbf{D'} \mid \textbf{D}, \textbf{R}) - H(\textbf{D'} \mid \widehat{\textbf{E}}^{\textbf{D}}, \textbf{R}) \\
&= I(\widehat{\textbf{E}}^{\textbf{D}}, \textbf{D'} \mid \textbf{D}, \textbf{R}) + H(\textbf{D'} \mid \textbf{D}, \widehat{\textbf{E}}^{\textbf{D}}, \textbf{R}) \\
&\quad - I(\textbf{D}, \textbf{D'} \mid \widehat{\textbf{E}}^{\textbf{D}}, \textbf{R}) + H(\textbf{D'} \mid \textbf{D}, \widehat{\textbf{E}}^{\textbf{D}}, \textbf{R}) \\
&= I(\widehat{\textbf{E}}^{\textbf{D}}, \textbf{D'} \mid \textbf{D}, \textbf{R}) - I(\textbf{D}, \textbf{D'} \mid \widehat{\textbf{E}}^{\textbf{D}}, \textbf{R}) \\
&= I(\widehat{E}^{\textbf{D}}, \textbf{D'} \mid \textbf{D}, \textbf{R}) \\
&= 0.
\vspace{-5pt}
\nonumber
\end{aligned}
\end{equation}

In the same way, we could obtain $I(\textbf{D}, \textbf{D'}, \textbf{R}) -I(\widetilde{\textbf{E}}^{\textbf{D}}, \textbf{D'}, \textbf{R}) = 0$. Thus, we could have $I(\widehat{\textbf{E}}^{\textbf{D}}, \textbf{D}, \textbf{R}) = I(\textbf{D}, \textbf{D'}, \textbf{R}) = I(\widetilde{\textbf{E}}^{\textbf{D}},\textbf{D'}, \textbf{R})$.

Besides, according to Eq.\eqref{le2_1} and Eq.\eqref{le2_2}, we have:
\begin{equation}
\begin{aligned}
&\quad H(\widehat{\textbf{E}}^{\textbf{D}}) - H(\widetilde{\textbf{E}}^{\textbf{D}}) - H(\widehat{\textbf{E}}^{\textbf{D}} \mid \textbf{D'}) + H(\widetilde{\textbf{E}}^{\textbf{D}} \mid \textbf{D'})\\ 
&= H(\widehat{\textbf{E}}^{\textbf{D}}) - H(\widetilde{\textbf{E}}^{\textbf{D}}) - H(\widehat{\textbf{E}}^{\textbf{D'}}) + H(\textbf{D'}) + H(\widetilde{\textbf{E}}^{\textbf{D}}, \textbf{D'}) - H(\textbf{D'})\\
&= H(\widehat{\textbf{E}}^{\textbf{D}}) - H(\widetilde{\textbf{E}}^{\textbf{D}}) - H(\widehat{\textbf{E}}^{\textbf{D}}, \textbf{D'}) + H(\widetilde{\textbf{E}}^{\textbf{D}}, \textbf{D'}) \\
&= H(\widehat{\textbf{E}}^{\textbf{D}}) - H(\widetilde{\textbf{E}}^{\textbf{D}}) + H(\widehat{\textbf{E}}^{\textbf{D}}) + H(\widehat{\textbf{E}}^{\textbf{D}} \mid \textbf{D'}) + \widetilde{\textbf{E}}^{\textbf{D}} - H(\widetilde{\textbf{E}}^{\textbf{D}} \mid \textbf{D'})\\
&= H(\widehat{\textbf{E}}^{\textbf{D}} \mid \textbf{D'}) - H(\widetilde{\textbf{E}}^{\textbf{D}} \mid \textbf{D'}) \\
&= H(\widehat{\textbf{E}}^{\textbf{D}}) - I(\widehat{\textbf{E}}^{\textbf{D}}, \textbf{D'}) - H(\widehat{\textbf{E}}^{\textbf{D}}) + I(\widetilde{\textbf{E}}^{\textbf{D}}, \textbf{D'}) \\
&= 0.
\nonumber
\vspace{-5pt}
\end{aligned}
\end{equation}
Therefore, based on the above proof, we could obtain $H(\widehat{\textbf{E}}^{\textbf{D}}) - H(\widetilde{\textbf{E}}^{\textbf{D}}) = H(\widehat{\textbf{E}}^{\textbf{D}} \mid \textbf{D'}) - H(\widetilde{\textbf{E}}^{\textbf{D}} \mid \textbf{D'})$. We could divide Theorem~\ref{theorem_better} into two components. We proof the first as follows. We use the complement information of the representation extracted by our designed method and previous method as $I(\widehat{\textbf{E}}^{\textbf{D}}, \textbf{R}|\textbf{D'})$ and $I(\widetilde{\textbf{E}}^{\textbf{D}}, \textbf{R}|\textbf{D'})$. Since we split the representations into two components and we perform the structure alignment in shared part, we have $I(\widehat{\textbf{E}}^{\textbf{D}}, \textbf{R}|\textbf{D'}) \ge I(\widetilde{\textbf{E}}^{\textbf{D}}, \textbf{R}|\textbf{D'})$. Thus, we have $I(\widehat{\textbf{E}}^{\textbf{D}}, \textbf{R}) = I(\widehat{\textbf{E}}^{\textbf{D}}, \textbf{R}, \textbf{D'}) + I(\widehat{\textbf{E}}, \textbf{R}|\textbf{D'})$.


With Lemma.\ref{le_2}, we could have:
\begin{equation}
\begin{aligned}
 I(\widehat{\textbf{E}}^{\textbf{D}}, \textbf{R}) &= I(\widetilde{\textbf{E}}^{\textbf{D}}, \textbf{R}, \textbf{D'}) + I(\widehat{\textbf{E}}, \textbf{R}|\textbf{D'}),\\
 &=I(\widetilde{\textbf{E}}^{\textbf{D}}, \textbf{R}) - I(\widetilde{\textbf{E}}^\textbf{D},\textbf{R}|\textbf{D'}) + I(\widehat{\textbf{E}}^{\textbf{D}}, \textbf{R}|\textbf{D'})
\end{aligned}
\end{equation}

Moreover, $I(\widehat{\textbf{E}}^{\textbf{D}}, \textbf{R}|\textbf{D'}) \ge I(\widetilde{\textbf{E}}^{\textbf{D}}, \textbf{R}|\textbf{D'})$, we have $I(\widehat{\textbf{E}}^{\textbf{D}}, \textbf{R}) \ge I(\widetilde{\textbf{E}}^{\textbf{D}}, \textbf{R})$. After that, we use $H(\widehat{\textbf{E}}^{\textbf{D}}|\textbf{D'}, \textbf{R})$ and $H(\widetilde{\textbf{E}}^{\textbf{D}}|\textbf{D'}, \textbf{R})$ as the noisy information of the representations aligned by our method and previous method. Since we split the representation into specific and shared components. We only align with the shared representations. We have $H(\widehat{\textbf{E}}^{\textbf{D}}|\textbf{D'}, \textbf{R}) \le H(\widetilde{\textbf{E}}^{\textbf{D}}|\textbf{D'}, \textbf{R})$. According to Lemma.\ref{le_2}, we have:

\begin{equation}
\begin{aligned}
&H(\widehat{E}^{\textbf{D}}|\textbf{R}) = H(\widehat{\textbf{E}}^{\textbf{D}}) - I(\widehat{H}^{\textbf{D}}, T)\\
&= H(\widehat{\textbf{E}}^{\textbf{D}}) - [I(\widehat{\textbf{E}}^{\textbf{D}}, \textbf{R}, {\textbf{D'}}) + I(\widehat{\textbf{E}}^{\textbf{D}}, \textbf{R}|{\textbf{D'}})]\\
&= H(\widehat{\textbf{E}}^{\textbf{D}}) - [I(\widetilde{\textbf{E}}^{\textbf{D}}, \textbf{R}, {\textbf{D'}}) - I(\widehat{\textbf{E}}^{\textbf{D}}, \textbf{R}| {\textbf{D'}})]\\
&= H(\widehat{\textbf{E}}^{\textbf{D}}) - I(\widetilde{\textbf{E}}^{\textbf{D}}, \textbf{R}) + I(\widetilde{\textbf{E}}^{\textbf{D}}, \textbf{R}|{\textbf{D'}}) - I(\widehat{\textbf{E}}^{\textbf{D}}, \textbf{R}|{\textbf{D'}})\\
&= H(\widehat{\textbf{E}}^{\textbf{D}}) - [H(\widetilde{\textbf{E}}^{\textbf{D}}) - H(\widetilde{\textbf{E}}^{\textbf{D}}|\textbf{R})] + I(\widetilde{\textbf{E}}^{\textbf{D}}, \textbf{R}| {\textbf{D'}}) - I(\widehat{\textbf{E}}^{\textbf{D}}, \textbf{R}| {\textbf{D'}})\\
&= H(\widetilde{\textbf{E}}^{\textbf{D}}|\textbf{R}) + H(\widehat{\textbf{E}}^{\textbf{D}}) - H(\widetilde{\textbf{E}}^{\textbf{D}}) + I(\widetilde{\textbf{E}}^{\textbf{D}}, \textbf{R}| {\textbf{D'}}) - I(\widehat{\textbf{E}}^{\textbf{D}}, \textbf{R}|{\textbf{D'}})\\
&= H(\widetilde{\textbf{E}}^{\textbf{D}}|\textbf{R}) + H(\widehat{\textbf{E}}^{\textbf{D}}) - H(\widetilde{\textbf{E}}^{\textbf{D}}) + H(\widetilde{\textbf{E}}^{\textbf{D}}|{\textbf{D'}})
- H(\widetilde{\textbf{E}}^{\textbf{D}}|\textbf{D'}, \textbf{R})\\ 
&- H(\widehat{\textbf{E}}^{\textbf{D}}| {\textbf{D'}}) + H(\widehat{\textbf{E}}^{\textbf{D}}| {\textbf{D'}}, \textbf{R})\\
&= H(\widetilde{\textbf{E}}^{\textbf{D}}|\textbf{R}) - H(\widetilde{\textbf{E}}^{\textbf{D}}|\textbf{D'}, \textbf{R}) + H(\widehat{\textbf{E}}^{\textbf{D}}| {\textbf{D'}}, \textbf{R})
\nonumber
\end{aligned}
\end{equation}

Based on $H(\widehat{\textbf{E}}^{\textbf{D}}|\textbf{D'}, \textbf{R}) \le H(\widetilde{\textbf{E}}^{\textbf{D}}|\textbf{D'}, \textbf{R})$, we have $H(\widehat{\textbf{E}}^{\textbf{D}}|\textbf{R}) \le H(\widetilde{\textbf{E}}^{\textbf{D}}|\textbf{R})$. Therefore, we have completed the proof.

\end{proof}

\newpage

\bibliographystyle{IEEEtran}
\bibliography{main}

\end{document}